\documentclass[a4paper,11pt]{article}
\usepackage{heppub}
\usepackage{subcaption}

\pdfoutput=1
\graphicspath{{plots/}}

\newcommand{\ord}[1]{\mathcal{O}(#1)}

\newcommand{\df}{\mathrm{d}}
\newcommand{\img}{\mathrm{i}}
\newcommand{\f}{\frac}
\newcommand{\mufo}{\mu_\text{FO}}

\newcommand{\arccosh}{\mathrm{arccosh}}

\newcommand{\GeV}{\,\mathrm{GeV}}
\newcommand{\TeV}{\,\mathrm{TeV}}

\newcommand{\nn}{\nonumber}

\newcommand{\bn}{{\bar n}}
\newcommand{\ptveto}{p_T^{\mathrm{cut}}}
\newcommand{\ptv}{p_T^{\mathrm{cut}}}
\newcommand{\ptvd}{p_T^\text{2nd}}
\newcommand{\Rc}{R_{\mathrm{cut}}}
\newcommand{\Rj}{R_J}
\newcommand{\ptj}{p_T^J}

\newcommand{\cI}{{\mathcal I}}

\newcommand{\as}{\alpha_s}
\newcommand{\eps}{\epsilon}
\newcommand{\dely}{\Delta y}
\newcommand{\jetmeas}{\mathcal{M}}
\newcommand{\cusp}{\mathrm{cusp}}
\newcommand{\SC}{\mathcal{S}^R}
\newcommand{\SCs}{\mathcal{S}}
\newcommand{\ST}{S^T} 
\newcommand{\SCNG}{\mathcal{S}^\text{NG}}
\newcommand{\la}{\lambda}

\newcommand{\LSC}{L_R}

\allowdisplaybreaks[2]

\title{\boldmath Jet veto resummation for STXS $H+$1-jet bins at aNNLL$'$+NNLO}

\author[a]{Pedro Cal,}
\author[b]{Matthew A.~Lim,}
\author[c]{Darren J.~Scott,}
\author[a]{Frank J.~Tackmann}
\author[d,e]{\\and Wouter J.~Waalewijn}

\affiliation[a]{Deutsches Elektronen-Synchrotron DESY, Notkestr. 85, 22607 Hamburg, Germany}
\affiliation[b]{Department of Physics and Astronomy, University of Sussex, Sussex House, Brighton, BN1 9RH, UK\vspace{0.5ex}}
\affiliation[c]{Max Planck Institute for Physics, Boltzmannstr. 8  D-85748 Garching, Germany}
\affiliation[d]{Institute for Theoretical Physics Amsterdam and Delta Institute for Theoretical Physics, University of Amsterdam, Science Park 904, 1098 XH Amsterdam, The Netherlands}
\affiliation[e]{Nikhef, Theory Group, Science Park 105, 1098 XG, Amsterdam, The Netherlands}

\emailAdd{pedro.cal@desy.de}
\emailAdd{m.a.lim@sussex.ac.uk}
\emailAdd{dscott@mpp.mpg.de}
\emailAdd{frank.tackmann@desy.de}
\emailAdd{w.j.waalewijn@uva.nl}

\abstract{
Measurements of Higgs boson processes by the ATLAS and CMS experiments at the
LHC use Simplified Template Cross Sections (STXS) as a common framework for the
combination of measurements in different decay channels and their further
interpretation, e.g. to measure Higgs couplings. The different Higgs production
processes are measured in predefined kinematic regions -- the STXS bins
-- requiring precise theory predictions for each individual bin.
In gluon-fusion Higgs production a main division is into 0-jet, 1-jet, and
$\geq 2$-jet bins, which are further subdivided in bins of the Higgs transverse
momentum $p_T^H$. Requiring a fixed number of jets induces logarithms
$\ln\ptveto/Q$ in the cross section where $\ptveto$ is the jet-$p_T$ threshold and
$Q\sim p_T^H\sim m_H$ the hard-interaction scale. These jet-veto logarithms can
be resummed to all orders in perturbation theory to achieve the highest possible
perturbative precision. We provide state-of-the art predictions for the $p_T^H$
spectrum in exclusive $H+$1-jet production and the corresponding $H+$1-jet STXS
bins in the kinematic regime $\ptveto \ll p_T^H\sim m_H$. We carry out the
resummation at NNLL$'$ accuracy, using theory nuisance parameters to account for
the few unknown ingredients at this order, and match to full NNLO.
We revisit the jet-veto
factorization for this process and find that it requires refactorizing the total
soft function into a global and soft-collinear contribution in order to fully
account for logarithms of the signal jet radius. The leading nonglobal
logarithms are also included, though they are numerically small for the region
of phenomenological interest.}

\preprint{\vbox{%
\hbox{DESY-24-128}
\hbox{MPP-2024-105}}
}

\begin{document}

\maketitle

\clearpage

\section{Introduction}
\label{sec:intro}

Since the discovery of the Higgs boson \cite{ATLAS:2012yve, CMS:2012qbp}, a
major goal of the LHC physics programme has been to measure its properties by
leveraging all available production and decay channels. For this purpose, the
ATLAS and CMS experiments have adopted Simplified Template Cross Sections
(STXS)~\cite{Berger:2019wnu, Andersen:2016qtm,
LHCHiggsCrossSectionWorkingGroup:2016ypw} as a common framework. It is designed
to maximize the experimental sensitivity by allowing the combination of
measurements in different decay channels and experiments. At the same time it
aims to minimize the theory or model dependence in the measurements by dividing
the different Higgs production cross sections into predefined and mutually
exclusive kinematic regions called the STXS bins. For the latest STXS
measurements from Run 2 see e.g.\ \refscite{ATLAS:2022vkf, CMS:2021ugl, CMS:2021kom}.

For the dominant Higgs production channel (via gluon fusion), the cross section is first
divided into $p_T^H < 200$ GeV and $p_T^H > 200$ GeV bins, where $p_T^H$ is the
transverse momentum of the Higgs boson. The $p_T^H > 200$ GeV bin primarily
targets boosted Higgs production which is sensitive to possible BSM effects.
Most of the cross section lies in the $p_T^H < 200$ GeV bin, which is split into
exclusive 0-jet, exclusive 1-jet, and inclusive $\geq2$-jets bins. The 0-jet and
1-jet bins are further split into $p_T^H$ bins. To make maximal use of the STXS
measurements, precise theory predictions for the cross section in each bin
are required.

Requiring a fixed number of jets, as is the case for the 0-jet and 1-jet bins,
means that additional jets above some transverse momentum threshold $\ptveto$
are vetoed. For typical values of $\ptveto \sim 25-30$ GeV, the jet veto
constrains the phase space for additional emissions quite strongly, which
induces logarithms $\ln\ptveto/Q$ in the cross section where $Q\sim p_T^H\sim
m_H$ is a typical hard-interaction scale. The best possible perturbative
precision is then achieved by resumming the jet-veto logarithms to all orders in
perturbation theory.

The resummation for colour-singlet processes such as Drell-Yan or gluon-fusion
Higgs production in the 0-jet limit was first derived in
\refscite{Stewart:2009yx, Stewart:2010pd, Berger:2010xi} using the beam thrust
event shape. The factorization and resummation for the leading-jet $p_T$ was
derived in \refscite{Banfi:2012yh, Becher:2012qa, Tackmann:2012bt,
Banfi:2012jm}, and resummed results for the corresponding 0-jet cross section up
to NNLL$'$ order matched to either fixed NNLO or N$^3$LO are
available~\cite{Stewart:2013faa, Becher:2013xia, Banfi:2015pju}. The 0-jet
resummation has also been carried out for other colour-singlet
processes~\cite{Shao:2013uba, Li:2014ria, Moult:2014pja, Monni:2014zra,
Wang:2015mvz, Tackmann:2016jyb, Ebert:2016idf, Dawson:2016ysj, Arpino:2019fmo,
Kallweit:2020gva, Campbell:2023cha},
and also including an additional cut on the jet
rapidity~\cite{Michel:2018hui} or a smooth weighting with the jet
rapidity~\cite{Gangal:2014qda, Gangal:2020qik}. In ref.~\cite{Gavardi:2023aco}, the 0-jet resummation for the $WW$ process was used to match NNLO predictions to a parton shower algorithm using the \textsc{Geneva} framework~\cite{Alioli:2012fc, Alioli:2015toa}.

In this paper, we consider gluon-fusion Higgs production in association with one
hard (signal) jet and a jet-$p_T$ veto on additional jets, which we will refer
to simply as $H+$1-jet production from here on. This is directly relevant for
the gluon-fusion STXS 1-jet bins. In this case, the resummation is currently
only available up to NLL$'$~\cite{Liu:2012sz, Liu:2013hba, Boughezal:2013oha}.

We extend the resummation of jet-veto logarithms for $H+$1-jet to
NNLL$'$ order matched to NNLO. A few of the required perturbative
ingredients at NNLL$'$ are not yet available. They are parametrized in terms of
theory nuisance parameters (TNPs)~\cite{Tackmann:2024kci} and are treated as an additional
source of uncertainty. Despite this, the resulting approximate resummation, denoted as aNNLL$'$, still allows us to reduce the perturbative uncertainty at aNNLL$'+$NNLO substantially compared to NLL$'+$NLO and to the pure NNLO. Here, aNNLL$’$ refers to the global jet veto logarithms -- in the $H+$1-jet cross section there are additional nonglobal logarithms and clustering effects that will be discussed below. The uncertainties can be reduced further
once the few remaining ingredients become known. Due to the lack of a public code for the NNLO calculation, we have obtained it indirectly from publicly available numerical results~\cite{Chen:2016zka, Becker:2020rjp}. In practice this is not a limitation, but a public code would of course facilitate the matching procedure.

Our resummation setup is valid in the kinematic regime where $\ptveto$ is
considered parametrically small compared to any other scale $\ptveto \ll p_T^J
\sim p_T^H\sim m_H$, where $p_T^J$ is the $p_T$ of the leading jet. For $\ptveto
= 30$ GeV, this allows us to obtain state-of-the-art resummed predictions for
the $p_T^H$ spectrum above $p_T^H\gtrsim 60$ GeV and the corresponding two STXS
1-jet bins with $p_T^H\in[60, 120]$ GeV and $p_T^H\in[120, 200]$ GeV. On the
other hand, for smaller $p_T^H$ the more appropriate parametric hierarchy is
$\ptveto \sim p_T^H \sim p_T^J \ll m_H$. The resummation in this scenario requires a rather
different setup, which we leave to future work.%
\footnote{In principle, our results can be extended to smaller $p_T^H$ at fixed
order in a manner similar to \refcite{Boughezal:2013oha}.}

Compared to the 0-jet case, the presence of the additional signal jet in the
final state substantially complicates the factorization structure. We therefore
reconsider the derivation of the factorized cross section, focussing in
particular on the all-order factorization of the measurement. Although the
experimental analyses typically use the same jet radius for both signal and
vetoed jets, we find it important to distinguish between them, as it helps to
clarify the structure of the logarithms and their factorization. We denote the
jet radius of the signal jet by $\Rj$ and that of the vetoed jets by $\Rc$. The
factorization is formally derived in the narrow-jet limit for both, i.e.~$\Rj^2,
\Rc^2 \ll 1$. We also discuss how to include numerically important $\ord{\Rj^2}$
power corrections that are enhanced by jet-veto logarithms. Our resummation
fully accounts for all logarithms of the jet radius $\Rj$, which compared to
\refscite{Liu:2012sz, Liu:2013hba} requires a further factorization of the total
soft function into a global and soft-collinear contribution. The fact that
emissions outside the signal jet are constrained whilst inside the jet they are
unconstrained also leads to the appearance of nonglobal logarithms, that are
not captured by the jet-veto factorization. Their numerical impact (at fixed
order) was already observed to be minor in \refcite{Liu:2013hba}. Here, we also
include a resummation of the leading nonglobal logarithms, finding that it has only a small numerical effect for the region of phenomenological interest.

Concerning the dependence on the jet-veto radius $\Rc$, the situation is
analogous to the 0-jet case. The resummation of $\Rc$ logarithms (so-called
clustering logarithms) is a more delicate issue as explained in
\refcite{Tackmann:2012bt}. Their leading resummation was achieved via numerical
methods in \refscite{Dasgupta:2014yra, Banfi:2015pju}, and it was shown to have
a very small effect. For this reason, we do not attempt to include these effects
here and only include the $\Rc$ logarithms at fixed order. In principle, the
results of \refcite{Banfi:2015pju} could be included in our formalism via a
suitable modification of the anomalous dimensions. 

This work is organized as follows. In \sec{factorization}, we discuss the
factorization and resummation necessary to achieve NNLL$'$+NNLO  accuracy for
$H+$1-jet production. We begin by reviewing the Born kinematics in
\sec{kinematics}, present the factorized cross section and corresponding
building blocks in \sec{factorization formula} and discuss the factorization of
the measurement in \sec{factorization analysis}. We give all required
perturbative ingredients in \sec{perturbative ingredients}, and in
\sec{validation} we validate the singular structure of the factorized cross
section, highlighting the importance of including power corrections in $\Rj^2$
in the soft function. In \secs{resummation}{TNP}, we discuss the details of the
NNLL$'$ resummation and how we assess the perturbative uncertainties. We present our numerical results in \sec{results} and conclude in
\sec{conclusions}.

\section{Factorization}
\label{sec:factorization}

\subsection{Kinematics}
\label{sec:kinematics}

At Born level, the $H+$1-jet production process is a $2\to2$ scattering of the form
\begin{align}
p_a + p_b \to p_J+ p_H\, ,
\end{align}
where $p_a, p_b$ are the momenta of the incoming partons and $p_J$ and $p_H$ are the momenta of the signal jet and the Higgs boson. The Born kinematics are sufficient to describe the hard scattering processes, since the veto on additional jets prohibits additional \emph{hard} real radiation. The kinematics are fully specified by the transverse momentum and rapidity of the signal jet, $p_T^J$ and $y_J$, as well as the total rapidity of the system $Y$. The Mandelstam variables can be written in terms of these variables as
\begin{align}
\label{eq:invariants}
s_{ab}\equiv(p_a +p_b)^2 & = Q^2\,  ,    \nn \\
s_{aJ}\equiv(p_a - p_J)^2   & = - \ptj  \, Q e^{Y-y_J}\, ,    \nn \\
s_{bJ}\equiv(p_b - p_J)^2   & = - \ptj  \, Q e^{y_J-Y}
\,, 
\end{align}
where we have used the partonic centre-of-mass energy $Q$, which is itself expressed in terms of $p_T^J$, $Y$, and $y_J$ as
\begin{equation}
\label{eq:Qdef}
Q(p_T^J, Y, y_J) = p_T^J \cosh(y_J-Y) + \sqrt{(p_T^J)^2 \cosh^2(y_J-Y) + m_H^2} \, .
\end{equation}
We will also make use of the initial-state parton momentum fractions $x_a$ and $x_b$, and the large light-cone momenta associated with each of these, $\omega_a$ and $\omega_b$. In terms of $p_T^J$, $Y$, and $y_J$ (and equivalently $Q$) these are given by
\begin{equation}
\label{eq:xdef}
x_a = \frac{\omega_a}{E_{\rm cm}} = \frac{Q}{E_{\rm cm}}\,e^Y \, , \qquad   x_b = \frac{\omega_b}{E_{\rm cm}} = \frac{Q}{E_{\rm cm}}\, e^{-Y} \, .
\end{equation}

The parametrization in terms of $p_T^J$, $Y$, and $y_J$ is a natural one since the soft function depends on the rapidity of the jet (see \sec{soft functions}), which specifies the direction of its associated Wilson line. However, we are often interested in predictions for cross sections defined in terms of kinematic variables of the Higgs boson, rather than the jet. The change of variables from jet transverse momentum and rapidity to those of the Higgs, $(p_T^J,y_J) \to (p_T^H, y_H)$, can be performed in a straightforward manner. Due to the veto on additional jets, the transverse momentum of the Higgs boson is necessarily identical to that of the jet it recoils against (up to power corrections):
\begin{equation}
p_T^H = p_T^J
\,.
\end{equation}
The relation between $y_H$ and $(p_T^J, Y, y_J)$ is given by
\begin{equation}
y_H = y_J + \ln \bigg(\frac{Q e^{Y-y_J}-p_T^J}{E_T^H}\bigg) \, ,
\end{equation}
where $E_T^H = \sqrt{(p_T^H)^2+m_H^2}$ denotes the transverse energy of the Higgs boson. 

For a given $p_T^J$, the kinematically allowed range for $Y$ is 
\begin{equation}
\label{eq:Y_constraint}
|Y| \leq \ln \Biggl(\frac{E_{\rm cm}}{p_T^J + \sqrt{(p_T^J)^2 + m_H^2}}\Biggr) \, ,
\end{equation}
where the bound is saturated when $y_J = Y$. Similarly, for a given $p_T^J$ and $Y$, 
\begin{equation}
\label{eq:yJ_constraint}
|y_J - Y| \leq \arccosh \biggl( \frac{e^{-|Y|}\bigl(E_{\rm cm}^2-e^{2|Y|}m_H^2\bigr)}{2 E_{\rm cm}\,  p_T^J} \biggr) \, .
\end{equation}
The limit on $Y$ in \eq{Y_constraint} ensures that the argument of the hyperbolic cosine in \eq{yJ_constraint} is always greater than or equal to one, such that $|y_J-Y|$ is always bounded by a positive real number. Likewise, when using Higgs kinematics, a similar set of constraints holds:
\begin{align}
|Y| & \leq \ln \Biggl(\frac{E_{\rm cm}}{p_T^H + E_T^H}\Biggr), \nn \\
|y_H-Y| &\leq  \arccosh \biggl( \frac{e^{-|Y|}\bigl(E_{\rm cm}^2+e^{2|Y|}m_H^2\bigr)}{2 E_{\rm cm}\,  E_T^H} \biggr)  \, .
\end{align} 
%

\subsection{Factorization formula}
\label{sec:factorization formula}
\begin{table}[t]
\centering
\begin{tabular}{c|c|c}
\hline\hline
 Mode ($n$)                 &  $p^\mu \sim (+, -, \perp)$ 		     &  $p^2$                           \\ \hline
  Beam ($n_a, n_b$)    &   $Q(\la^2, 1, \la) $           		     &  $Q^2 \la^2$                  \\ 
 Jet  ($n_J$)                  &   $Q(R_J^2, 1, R_J) $      	              &  $Q^2 R_J^2 $              \\
 Soft	                             &   $Q\la(1, 1, 1) $			              &  $Q^2 \la^2$                  \\
 Soft-collinear ($n_J$)   &   $Q\la(R_J^2, 1, R_J) $		      &  $ Q^2 \la^2 R_J^2$
\\ \hline\hline
\end{tabular}
\caption{Power counting of the modes present in the factorization formula \eq{factorization}.  Here $Q$ is the hard scale, $\lambda$ is the small expansion parameter $\la\equiv\ptv/Q \ll 1$, while $R_J$ is the radius of the signal jet $\Rj\ll 1$. Where applicable, the collinear direction associated to each mode as specified in \eq{lc_dir} is given in parentheses.\label{tab:modes}}
\end{table}

In this section we detail the effective field theory and factorization formula used to describe $pp \to H+$1 jet in the presence of a jet veto. The veto is implemented by requiring additional jets to have transverse momentum $p_T<\ptveto$. We define our jets using the anti-$k_T$ clustering algorithm. We distinguish between the jet radius of the signal jet $\Rj$ and that of vetoed jets $\Rc$ to elucidate how each enters into our calculation. The typical experimental ranges of the jet-$p_T$ veto are $\ptveto \sim 25 - 30$ GeV, meaning that we consider the hierarchy $\ptv \ll Q$ and the resummation of logarithms of $\ptv/Q$ (where $Q$ is defined in \eq{Qdef}). We also work in the narrow-jet limit, $\Rj, \Rc \ll 1$. The limit $\Rj\ll 1$ is necessary to factorize the signal jet as we will see momentarily. As discussed in \refcite{Tackmann:2012bt}, the condition $\Rc \ll1$ is needed to derive a valid all-order factorization formula in the presence of the jet-veto measurement. We focus on the regime in which the signal jet is hard, $p_T^J \sim m_H \sim Q $, and well-separated in rapidity from the beam directions. In this way we are able to avoid introducing additional scale hierarchies into the problem.

Having identified our expansion parameters $\lambda \equiv \ptv/Q\ll1$ and $\Rj, \Rc \ll 1$, the next step is to establish the relevant degrees of freedom or modes in Soft-Collinear Effective Theory (SCET)~\cite{Bauer:2000yr,Bauer:2001ct,Bauer:2001yt,Bauer:2002nz,Beneke:2002ph}. To this end, we introduce three light-like reference vectors $n_a$, $n_b$, and $n_J$ for the three collinear directions associated with the three hard partons, as well as conjugate vectors $\bn_a$, $\bn_b$, and $\bn_J$,
\begin{align}
n_a^\mu &= (1,0,0,1), &
n_b^\mu &= (1,0,0,-1), &
n_J^\mu &= (\cosh y_J, 1, 0, \sinh y_J)
\, , \nn \\
\bn_a^\mu &= (1,0,0,-1), &
\bn_b^\mu &= (1,0,0,1), &
\bn_J^\mu &= (\cosh y_J, -1, 0, -\sinh y_J) \, ,
\label{eq:lc_dir}
\end{align}
where we choose as usual the beam direction to be the $z$ axis and the normalization $n_i \cdot \bar{n}_i = 2$. Any four-momentum  can then be decomposed into components along the light-cone vectors $n_i^\mu$ and $\bn_i^\mu$, and the components transverse to these $p_{\perp i}^\mu$,
\begin{equation}
p^\mu = p^+\,\frac{\bar{n}_i^\mu}{2} + p^-\,\frac{n_i^\mu}{2} + p_{\perp i}^\mu \, \qquad p^+ =  n_i \cdot p\,, \qquad p^- = \bar{n}_i \cdot p\,.
\end{equation}
Expressed in terms of these reference vectors, the requirement that the signal jet is well-separated in rapidity from the beam directions is given by $n_a\cdot n_J\sim n_b\cdot n_J \sim 1$.

We now describe the various modes in SCET, which are summarized in~\tab{modes}. There are collinear modes associated with the beams and jet, which carry the large momentum flowing in these directions. The power counting of these modes is most naturally expressed in the corresponding light-cone coordinates. The jet-veto measurement sets the power counting for the transverse momentum of collinear radiation along the beams. Due to its large energy, the jet collinear radiation must be confined to the jet, so its transverse momentum \emph{with respect to the jet axis} should be of order $p_T^J\Rj$. The typical virtuality scale of collinear emissions describing the jet is thus $p_T^J \Rj \sim Q \Rj$, which for the factorization to hold must be parametrically smaller than the hard scale $Q$. This is why we have to require $\Rj\ll 1$. The wide-angle soft radiation does not resolve the small angular scale $\Rj$ of the jet and the jet veto sets its transverse momentum to be of order $\ptv$. It is isotropic and so independent of which light-cone coordinates are used. Finally, there is soft-collinear radiation, whose power counting is set by the requirement that it resolves the jet, i.e.~its typical angular scale is $R_J$, and which is sensitive to the jet veto measurement, so that its energy is of order $\ptv$. Hence, having to require $\Rj\ll 1$ inevitably
leads to the appearance of the additional soft-collinear scale $\ptv\Rj$ with the same parametric distinction from the soft scale $\ptv$ as that between the jet and hard scales.

This mode picture leads to the following form of the factorized cross section:
\begin{align}\label{eq:factorization}
\frac{\df \sigma }{\df p_T^J \df Y \df y_J} (\ptv)  &= \sum_{\kappa} H_{\kappa} (p_T^J, Y, y_J; \mu) B_{a} \Bigl(\ptv, x_a, \Rc; \mu, \frac{\nu}{\omega_a}\Bigr)\,  B_{b}\Bigl(\ptv, x_b, \Rc; \mu, \frac{\nu}{\omega_b}\Bigr) \nn \\
&\quad \times J_{j}(p_T^J R_J; \mu) \,
S_{\kappa}(\ptv, \Rc, y_J; \mu, \nu) \, \SC_{j}(\ptv R_J; \mu)\, \SCNG_{j}\biggl(\frac{\ptv}{p_T^J}\biggr) \nn \\
&\quad \times \biggl[1+ \mathcal O\Big(\f{\ptv}{Q}, \Rc^2, \Rj^2\Big) \biggr].
\end{align}
The sum over $\kappa = \{a, b, j\}$ runs over the possible flavours $a,b,j \in \{g, u, \bar u, \dots\}$ of the incoming and outgoing partons. $H_{\kappa}$ denotes the hard function, which encodes the hard scattering process $ab \to Hj$ (and though written as a function of both $Y$ and $y_J$, depends only on the Lorentz-invariant difference $Y-y_J$). The beam functions $B$ describe initial-state collinear radiation from the incoming partons (with $x_i$ and $\omega_i$ defined in \eq{xdef}), while the global soft function $S$ describes isotropic soft radiation.  Hard-collinear radiation along the jet direction is described by the jet function $J$.

The soft-collinear function $\SC_j$ is associated with the soft-collinear mode described above. Such soft-collinear modes were introduced in a similar context in \refcite{Chien:2015cka} to resum the logarithms of $\Rj$ in the soft sector, which arise from the ratio of the soft and soft-collinear scales $\ptv$ and $\ptv \Rj$. These are distinct in origin from logarithms of $\Rj$ arising from the ratio of the hard and jet scales $p^J_T$ and $p^J_T \Rj$, whose resummation is addressed by the factorization between the jet and hard function. While we are able to resum all logarithms of $\Rj$, there are logarithms of $\Rc$ which remain unresummed (as discussed in ref.~\cite{Tackmann:2012bt}). For work on these clustering logarithms beyond their contribution to the NNLO cross section, see refs.~\cite{Dasgupta:2014yra,Banfi:2015pju,Alioli:2013hba}.
Finally, since the in-jet region is unconstrained and the out-of-jet region is sensitive to the $p_T$-veto measurement, there are nonglobal logarithms (NGLs) of $\ptveto/p_T^J$, arising
from the separation of the soft-collinear and $n_J$-collinear modes~\cite{Larkoski:2015zka, Becher:2015hka}.
As discussed in \sec{soft functions}, we will include the leading nonglobal logarithms in the function $\SCNG_j$, which formally contribute at NLL order to the cross section but which are in practice numerically small.

Before moving on to a detailed analysis of the validity of the factorization in~\eq{factorization}, it is worth commenting on the differences with respect to the work of~\refcite{Liu:2012sz} where the resummation of veto logarithms for $pp \to H+$1 jet was originally studied. The main difference between the factorization formula presented in that work and~\eq{factorization} is the additional refactorization of the soft sector into global soft, soft-collinear and nonglobal pieces. Compared to~\refcite{Liu:2012sz} which treated only logarithms of $\Rj$ arising from the ratio of the jet and hard scales, this allows us also to deal with equally large logarithms of the ratio of soft and soft-collinear scales. Having supplemented this with the resummation of the leading nonglobal logarithms and the inclusion in the soft sector of subleading terms in $\Rj$ (see~\sec{validation}), we have been able to address many of the issues identified by the authors of~\refcite{Liu:2012sz} as topics for future study. 

\subsection{Factorization analysis}
\label{sec:factorization analysis}

The all-order validity of the leading-power factorization formula in \eq{factorization} rests on the factorization properties of the measurement, which we will now discuss in some detail.
The measurement function acting on the complete final state is given by
\begin{equation} \label{eq:meas_full}
\jetmeas(\ptveto, \Rc, \Rj) = \prod_{j\in \mathcal{J}(\Rc, \Rj)} \Theta(|\vec{p}_{Tj}| < \ptveto)
\,,\end{equation}
where $\mathcal{J}(\Rc, \Rj)$ denotes the set of all additional jets beyond the signal jet.
More precisely, we first run the jet algorithm with $\Rj$ to identify the leading
signal jet and remove all its particles from the final state. We then run
the jet algorithm again with $\Rc$ on this reduced final state and denote its
outcome by $\mathcal{J}(\Rc, \Rj)$. To establish \eq{factorization}, we must be able
to separate the full measurement in \eq{meas_full} into a product of measurements acting independently on the soft, beam-collinear, jet-collinear, and soft-collinear sectors. That is,
we wish to make the decomposition
\begin{equation} \label{eq:meas_full_fact}
\jetmeas =  \jetmeas_a\, \jetmeas_b\, \jetmeas_J\, \jetmeas_s\, \jetmeas_{sc} + \delta\jetmeas
\,,\end{equation}
where the measurement applied independently within each individual sector is
denoted by $\jetmeas_i$, implicitly defining a remainder term
$\delta\jetmeas(\ptveto)$. Achieving factorization amounts to demonstrating that
this remainder function (more precisely its contribution to the cross section)
is power-suppressed, or in other words that the measurement does not mix or
entangle the constraints on different sectors to leading power.%
\footnote{To be precise, the measurement entangling the constraints on individual
emissions in different sectors does not necessarily preclude any resummation.
It does prevent a simple (in our case multiplicative)
all-order factorization structure of the cross section, from which the resummation
to any resummation order would follow.}
Assuming the jet algorithm does not cluster emissions from different sectors,
the product form in \eq{meas_full_fact} immediately follows from the product
of $\Theta$ functions in \eq{meas_full} with $\delta\jetmeas = 0$~\cite{Tackmann:2012bt}.

In \refcite{Tackmann:2012bt}, the properties of jet-veto resummations were studied for the class
of processes $pp\to V$, where $V$ represents any colour-singlet final state. It was shown that the
contribution from mixing of soft and $n_{a,b}$-collinear sectors entering in
$\delta\jetmeas(\ptveto)$ is indeed power-suppressed by $\mathcal{O}(\Rc^2)$.%
\footnote{As discussed in detail in \refcite{Tackmann:2012bt}, due to the presence of such mixing in the measurement function the all-order factorization and resummation at $\ord{\Rc^2}$ is presently unknown, since the standard leading-power arguments become insufficient at $\ord{\Rc^2}$. The
mixing contributions first appear at fixed $\ord{\alpha_s^2}$. There are different schemes to incorporate them into the final resummed result~\cite{Banfi:2012jm, Stewart:2013faa, Becher:2013xia}, which are formally equivalent at NNLL as discussed in detail in \refcite{Gangal:2016kuo}.}
It remains for us to demonstrate that the additional $n_J$-collinear and
soft-collinear sectors in our case do not yield any further, unsuppressed mixing
terms.

Firstly, $n_{a,b}$-collinear and $n_J$-collinear particles cannot be clustered
at leading power, just as the clustering of $n_a$-collinear and $n_b$-collinear
particles is not allowed.
Next, we consider the case in which a $n_J$-collinear particle is clustered
with a soft particle. This is power suppressed by $\mathcal{O}(\Rj^2)$ for the global soft sector, since it describes wide-angle radiation and does not resolve the jet boundary. Soft-collinear radiation, however, does resolve the jet boundary. This can in principle lead to clustering effects -- in \app{cscclust}, we demonstrate that these contribute at the same level as subleading nonglobal logarithms. Since we include only the leading nonglobal terms in this work (see \sec{ngls}), we can thus use a static boundary in calculating the soft-collinear function and ignore the effect of clustering with individual $n_J$-collinear emissions. At this point, we can conclude that the full measurement factorizes as
\begin{align} \label{eq:meas_full_fact_2}
\jetmeas(\ptveto, \Rc, \Rj)
&= \jetmeas_a(\ptveto, \Rc) \jetmeas_b(\ptveto, \Rc) \jetmeas_J(\Rj)
\jetmeas_\mathrm{soft}(\ptveto, \Rc, \Rj)
\nn\\ & \quad \times
\bigl[1 + \ord{\Rj^2, \Rc^2}\bigr]
\,.\end{align}
where
\begin{equation}
\jetmeas_{a,b}(\ptveto, \Rc) = \prod_{j\in \mathcal{J}_{a,b}(\Rc)} \Theta(|\vec{p}_{Tj}| < \ptveto)
\,.\end{equation}
is the standard jet-veto measurement in the $n_{a,b}$-collinear sectors with
$\mathcal{J}_{a,b}(\Rc)$ denoting the set of all jets returned by the jet
algorithm acting solely on the $n_{a,b}$-collinear final state. In particular,
for the same reason $n_{a,b}$-collinear emissions cannot be clustered with
$n_J$-collinear ones, the $\jetmeas_{a,b}$ do not depend at all on the presence
of the signal jet and $\Rj$.

The $n_J$-collinear
measurement $\jetmeas_J(\Rj)$ simply constrains all $n_J$-collinear emissions to
lie inside the signal jet without any further constraints inside the jet.%
\footnote{This means that, in the terminology of \refcite{Ellis:2010rwa}, we are considering an unmeasured jet function. The relationship between measured and unmeasured jet functions is further discussed in \refcite{Chien:2015cka}.}
In particular, it does not actually depend on the jet-veto measurement and
$\ptveto$. To see this, assume there is an $n_J$-collinear emission outside the
signal jet. It would be clustered into its own jet with $p_T \sim p_T^J$, which
would then be subject to the jet-veto requirement $p_T^J < \ptveto$, which
however violates the power counting $\ptveto \ll p_T^J$. Hence, the jet veto
outside the signal jet effectively forces all $n_J$-collinear emissions to be
clustered into the signal jet.

The last measurement in \eq{meas_full_fact_2} is given by
\begin{equation} \label{eq:meas_soft}
\jetmeas_\mathrm{soft}(\ptveto, \Rj, \Rc)
= \prod_{j\in \mathcal{J}_{\rm soft}(\Rc, \Rj)} \Theta(|\vec{p}_{Tj}| < \ptveto)
\,.\end{equation}
It acts on the total soft sector comprised of both soft and soft-collinear
emissions, where $\mathcal{J}_{\rm soft}(\Rc, \Rj)$ denotes the result of the
jet algorithm acting on both soft and soft-collinear emissions outside a fixed
cone of $\Rj$ around the jet direction $n_J$. The remaining step is to consider the factorization of
\begin{equation} \label{eq:meas_soft_fact}
\jetmeas_\mathrm{soft}(\ptveto, \Rc, \Rj)
= \jetmeas_s(\ptveto, \Rc)\, \jetmeas_{sc}(\ptveto, \Rc, \Rj)
\end{equation}
into measurements acting separately on the global (i.e.\ overall) soft and the
soft-collinear sectors, and in the process determine the precise
form of the soft-collinear measurement function $\jetmeas_{sc}$.
With no particles in the final state, we trivially have
\begin{equation}
\jetmeas^{[0]}_{\rm soft} = 1 = 1 \times 1 \equiv \jetmeas_s^{[0]} \jetmeas_{sc}^{[0]}
\,,\end{equation}
where the superscript $[n]$ denotes the measurement acting on the $n$-particle
final state.
This result corresponds to $S^{(0)}{\mathcal{S}}^{R,(0)}$.

For a single emission,
\begin{align} \label{eq:meas_fact_1}
\jetmeas_{\rm soft}^{[1]}
&= {\bf 1}_{\rm in} + {\bf 1}_{\rm out} {\bf 1}_{\rm cut}
= {\bf 1}_{\rm in}({\bf 1}_{\rm cut} + {\bf \bar 1}_{\rm cut}) + {\bf 1}_{\rm out} {\bf 1}_{\rm cut}
\nn\\
&= \underbrace{{\bf 1}_{\rm cut}}_{\jetmeas_s^{[1]}} + \underbrace{{\bf 1}_{\rm in} \bar {\bf 1}_{\rm cut}}_{\jetmeas_{sc}^{[1]}}
\equiv \jetmeas_s^{[1]}\jetmeas_{sc}^{[0]} + \jetmeas_s^{[0]}\jetmeas_{sc}^{[1]}
= (\jetmeas_s \jetmeas_{sc})^{[1]}
\,,\end{align}
where ${\bf 1}_{\rm in}$ and ${\bf 1}_{\rm out}$ denote the restriction (i.e.\ a
suitable set of theta functions) for particle 1 to be inside and outside the
signal jet respectively. Similarly, ${\bf 1}_{\rm cut}$ denotes the restriction
that the $p_T$ of particle 1 is below $p_T^{\rm cut}$. The bar denotes the
complement of a restriction, e.g.~${\bf 1}_{\rm out} = \bar {\bf 1}_{\rm in} = 1
- {\bf 1}_{\rm in}$. The first equality simply encodes the full measurement on
the soft emission, namely that its $p_T$ is constrained when outside the jet and
unconstrained when inside the jet, which can be rewritten as shown. In the
second line we can identify the measurement $\jetmeas_s^{[1]}$ acting on the
global soft radiation, which does not resolve the jet and therefore the emission
is constrained by $\ptveto$ everywhere. The form of the soft-collinear
measurement $\jetmeas_{sc}^{[1]} = {\bf 1}_{\rm in} \bar {\bf 1}_{\rm cut}$
is perhaps not intuitive. It can be interpreted as correcting the global soft
measurement inside the jet by adding back the soft emissions that are removed by
the global soft measurement, or equivalently as $\jetmeas_{sc}^{[1]} = {\bf
1}_{\rm in} - {\bf 1}_{\rm in} {\bf 1}_{\rm cut}$ as subtracting the overlap
with the global soft measurement to avoid double counting.

Next, for two soft emissions we have
\begin{align} \label{eq:meas_soft_2}
\jetmeas_{\rm soft}^{[2]} &=
  {\bf 1}_{\rm in}   {\bf 2}_{\rm in} +   {\bf 1}_{\rm in} {\bf 2}_{\rm out} {\bf 2}_{\rm cut}
  + {\bf 1}_{\rm out} {\bf 1}_{\rm cut}  {\bf 2 }_{\rm in} 
  + {\bf 1}_{\rm out}  {\bf 2}_{\rm out}
  (\{{\bf 12}\}_{\rm clust} \{{\bf 12}\}_{\rm cut}
  +\overline{\{{\bf 12}\}}_{\rm clust} {\bf 1}_{\rm cut} {\bf 2}_{\rm cut})
  \nn \\
  &= \underbrace{( {\bf 1}_{\rm cut} + {\bf 1}_{\rm in} \bar {\bf 1}_{\rm cut})
  ( {\bf 2}_{\rm cut} + {\bf 2}_{\rm in} \bar {\bf 2}_{\rm cut})}_{\jetmeas_\text{soft,no clust}^{[2]}}
  +  \underbrace{ {\bf 1}_{\rm out} {\bf 2}_{\rm out}   
  \{{\bf 12}\}_{\rm clust} 
  (\{{\bf 12}\}_{\rm cut} - {\bf 1}_{\rm cut} {\bf 2}_{\rm cut})}_{\jetmeas_\text{soft,clust}^{[2]}}
\end{align}
where $\{{\bf 12}\}_{\rm clust}$ encodes the condition that particle 1 and 2 are clustered and  $\{{\bf 12}\}_{\rm cut}$ denotes the $p_T$ cut on their combined momenta. In the first term, $\jetmeas_\text{soft,no clust}^{[2]}$, no clustering happens so each emission is treated as its own jet. It is clear that these contributions factorize, being of the form
\begin{equation}
\jetmeas_\text{soft,no clust}^{[2]}
\equiv \Bigl(\jetmeas_{s}^{[2]} + \jetmeas_{s}^{[1]}\jetmeas_{sc}^{[1]} + \jetmeas_{sc}^{[2]}\Bigr)_\text{no clust}
= (\jetmeas_{s} \jetmeas_{sc})^{[2]}_\text{no clust}
\,.\end{equation}
Note that when written as a product of separate measurement operators acting on
separate sectors in the final state, the product $\jetmeas_{s}^{[1]}\jetmeas_{sc}^{[1]}$
accounts for both cross terms ${\bf 1}_{\rm cut} {\bf 2}_{\rm in} \bar {\bf 2}_{\rm cut}$
and ${\bf 2}_{\rm cut} {\bf 1}_{\rm in} \bar {\bf 1}_{\rm cut}$ appearing in
\eq{meas_soft_2}, depending on which of the final-state particles is
soft or soft-collinear.

The second term in \eq{meas_soft_2}, $\jetmeas_\text{soft,clust}^{[2]}$, contains the clustering correction, which we can rewrite as
\begin{equation}
\jetmeas_\text{soft,clust}^{[2]}
=  \underbrace{\{{\bf 12}\}_{\rm clust}
  (\{{\bf 12}\}_{\rm cut} - {\bf 1}_{\rm cut} {\bf 2}_{\rm cut})}_{\jetmeas_{s,{\rm clust}}^{[2]}}
  + \underbrace{({\bf 1}_{\rm out} {\bf 2}_{\rm out} - 1)   \{{\bf 12}\}_{\rm clust} 
  (\{{\bf 12}\}_{\rm cut} - {\bf 1}_{\rm cut} {\bf 2}_{\rm cut})}_{\jetmeas_{sc,{\rm clust}}^{[2]}}
.\end{equation}
$\jetmeas_{s,{\rm clust}}^{[2]}$ corresponds to the global soft contribution, which does not involve the jet boundary, and $\jetmeas_{sc,{\rm clust}}^{[2]}$ is defined by the difference between $\jetmeas_{\rm soft,clust}^{[2]}$ and $\jetmeas_{s,{\rm clust}}^{[2]}$. We can now argue that $\jetmeas_{sc,{\rm clust}}^{[2]}$ is fully accounted for by the soft-collinear function, because the constraint $({\bf 1}_{\rm out} {\bf 2}_{\rm out} - 1)$ forces at least one of the two emissions to be inside the jet, whilst the clustering condition $\{{\bf 12}\}_{\rm clust}$ forces the other emission to have separation $\Rc$ to it. Hence, for the case $\Rc \sim R_J \ll 1$ that we consider, both emissions are forced to have the same angular scale and thus belong to the soft-collinear sector.%
\footnote{%
On the other hand, in case of a hierarchy $R_J \ll \Rc \sim 1$, one of the emissions could be a wide-angle soft emission outside the jet, implying a mixing between the soft and soft-collinear measurement constraints, which would prevent us from establishing a simple (multiplicative) measurement factorization using only power-counting arguments.}
Combining the two contributions we have
\begin{equation}
\jetmeas_{\rm soft}^{[2]}
= (\jetmeas_{s} \jetmeas_{sc})^{[2]}_\text{no clust}
+ \jetmeas_{s,\rm clust}^{[2]} + \jetmeas_{sc,\rm clust}^{[2]}
= (\jetmeas_{s} \jetmeas_{sc})^{[2]}
\,.\end{equation}

It should be clear that the above pattern extends to any number of soft emissions.
First, without any clustering, we immediately have
\begin{equation} \label{eq:meas_soft_noclust_n}
\jetmeas_\text{soft,no clust}^{[n]}
= \prod_{\bf k = 1}^{\bf n} ( {\bf k}_{\rm cut} + {\bf k}_{\rm in} \bar {\bf k}_{\rm cut})
= (\jetmeas_s \jetmeas_{sc})^{[n]}_\text{no clust}
\,.\end{equation}
Including clustering effects, we define the soft measurement as the jet-veto
measurement acting on the soft sector without any reference to the signal jet,
\begin{equation} \label{eq:jetmeas_s}
\jetmeas_s(\ptveto, \Rc) = \prod_{j\in \mathcal{J}_s(\Rc)} \Theta(|\vec{p}_{Tj}| < \ptveto)
\,,\end{equation}
where $\mathcal{J}_s(\Rc)$ is the result of the jet algorithm acting on the soft
final state. The resulting soft function by definition only depends on the scale
$\ptveto$ but not on $\Rj$. Starting at $\ord{\as^2}$, it will also include
clustering logarithms of $\Rc$, which here we do not aim to resum but include at
fixed order, as discussed earlier.

The soft-collinear measurement is then defined for any number of emissions as the necessary
correction with respect to the full measurement including all clustering effects.
That is, we use \eq{meas_soft_fact} to solve for $\jetmeas_{sc}$ with
$\jetmeas_\mathrm{soft}$ and $\jetmeas_{s}$ given in \eqs{meas_soft}{jetmeas_s},
analogous to what we did for one and two emissions above.
Symbolically, we can write
\begin{equation}
\jetmeas_{sc}(\ptveto, \Rc, \Rj) = \jetmeas_\mathrm{soft}(\ptveto, \Rc, \Rj)
\jetmeas_s^{-1}(\ptveto, \Rc)
\,.\end{equation}
The last step is to argue that this remaining constraint indeed forces all
involved emissions to be soft-collinear, i.e., to have the same small angular
scale given by $\Rj\sim\Rc$, analogous to what we found for
$\jetmeas^{[2]}_{sc,\rm clust}$ in case of two emissions above. First, we can
always rewrite the total soft measurement including clustering for $n$ emissions
analogously to the second line of \eq{meas_soft_2} as the no-clustering
contribution \eq{meas_soft_noclust_n} plus a clustering correction, which in
general is a sum of products of clustering constraints of the form
\begin{equation}
{\bf 1}_{\rm out}{\bf 2}_{\rm out}\dotsb {\bf k}_{\rm out}\{\bf \bf 12 \dotsb k\}_{\rm clust}(\ldots)
\equiv \{{\bf k}\}_{\rm out} \{{\bf k}\}_{\rm clust}(\dotsb)
\,,\end{equation}
where $k$ particles are clustered together in a jet and the parantheses
only contain differences of $\ptveto$ constraints to appropriately correct
from the unclustered to the clustered case. In particular, since the jet algorithm
for $\jetmeas_{\rm soft}$ only acts outside the signal jet any clustering correction
for $k$ particles must have the constraint $\{\bf k\}_{\rm out} = {\bf 1}_{\rm out}\dotsb {\bf k}_{\rm out}$
as shown, forcing all particles to be outside the jet.
The global soft measurement can be written in exactly the same way, except that
this constrain is absent. For a single clustering term like the above, the resulting
remainder appearing in $\jetmeas_{sc}^{[k]}$ is then given by
\begin{equation}
\underbrace{(\{{\bf k}\}_{\rm out} - 1)\{\bf k\}_{\rm clust}(\ldots)}_{\in\jetmeas_\text{sc,clust}^{[k]}}
= \underbrace{\{{\bf k}\}_{\rm out} \{\bf k\}_{\rm clust}(\ldots)}_{\in\jetmeas_\text{soft,clust}^{[k]}}
- \underbrace{\{\bf k\}_{\rm clust}(\ldots)}_{\in\jetmeas_\text{s,clust}^{[k]}}
\,,\end{equation}
which forces all involved $k$ emissions to be soft-collinear because it requires
at least one of the emissions to be inside the signal jet and all others to be clustered with it.
For a product of such clustering terms, say for $k_i$ and $k_j$ particles that are clustered into separate jets, their constraints to be outside the signal jet can always be rearranged as
\begin{align}
\underbrace{\{{\bf k_i}\}_{\rm out} \{{\bf k_j}\}_{\rm out}}_{\in\jetmeas_\text{soft,clust}^{[k_i+k_j]}}
&= (\{{\bf k_i}\}_{\rm out} - 1 + 1) (\{{\bf k_j}\}_{\rm out} - 1 + 1)
\\\nn
&= \underbrace{(\{{\bf k_i}\}_{\rm out} -1) (\{{\bf k_j}\}_{\rm out} - 1)}_{\in \jetmeas^{[k_i+k_j]}_{sc,\rm clust}}
+ \underbrace{(\{{\bf k_i}\}_{\rm out} - 1)\times 1}_{\in \jetmeas_{sc,\rm clust}^{[k_i]} \jetmeas_{s,\rm clust}^{[k_j]}}
+ \underbrace{1\times (\{{\bf k_j}\}_{\rm out} -1)}_{\in\jetmeas_{s,\rm clust}^{[k_i]}\jetmeas_{sc,\rm clust}^{[k_j]}}
+ \underbrace{1}_{\in \jetmeas_{s,\rm clust}^{[k_i+k_j]}}
\,,\end{align}
where for brevity we have dropped the overall factor of $\{\bf k_i\}_{\rm clust}\{\bf k_j\}_{\rm clust}$ on both sides. The first term on the right-hand side again forces both sets
of particles to be soft-collinear. As indicated, the cross terms correspond to cross terms in
$[\jetmeas_s \jetmeas_{sc}]^{[k_i+k_j]}_{\rm clust}$.
Finally, whenever clustering terms appear multiplied by additional unclustered
emissions, these are accounted for by cross terms between lower-emission clustering and
no-clustering contributions.

It may not be obvious where the nonglobal logarithms, included through $\SCNG$ in \eq{factorization}, appear in this discussion. These leading nonglobal logarithms arise from the soft-collinear sector i.e., the matrix element of $\jetmeas_{sc}$ gives rise to the product $\SC_{j} \SCNG_{j}$ in \eq{factorization}. Specifically,
they are not caused by a failure of the measurement to factorize, but are due to correlated soft-collinear emissions involving both the constrained and unconstrained regions of phase space. In practice, the jet scale $p_T^J$ limits the radiation in the unconstrained region, leading to nonglobal logarithms of $p_T^J/\ptv$. Subleading nonglobal logarithms are sensitive to collinear emissions inside the jet -- their resummation is more complicated, and cannot be achieved using a factorisation formula of the kind studied in this work. Resummation has nevertheless been achieved up to NLL using both coherent branching and SCET approaches~\cite{Banfi:2021owj,Becher:2023vrh}. At the same order the effect of soft-collinear clustering also enters, described in \app{cscclust}.

\section{Resummation at \texorpdfstring{NNLL$'+$NNLO}{NNLL'+NNLO}}

We begin this section by presenting the necessary perturbative ingredients in \sec{perturbative ingredients}. We use these to validate the singular structure of the cross section at next-to-leading order in \sec{validation}. This validation reveals relatively large power corrections in the jet radius, arising from the factorization of the total soft function into a global and soft-collinear part (discussed in \sec{soft functions}). The resummation is discussed in \sec{resummation}, and the treatment of the missing ingredients needed to reach (global) NNLL$'$ accuracy using theory nuisance parameters (TNPs) is presented in \sec{TNP}.

\subsection{Perturbative ingredients}
\label{sec:perturbative ingredients}

In this section we provide (most of) the perturbative ingredients necessary to attain (global) NNLL$'$ accuracy. We will express these in terms of the cusp and noncusp anomalous dimensions, using the following convention for their expansion in $\alpha_s$:
\begin{align} \label{eq:Gamma_exp}
\Gamma^i_\text{cusp} (\as)=&\sum_{n=0}^\infty \Gamma^i_{n} \Big(\f{\as}{4\pi} \Big)^{n+1} , \hspace{1cm}  \gamma_F^i(\as)=\sum_{n=0}^\infty \gamma^i_{F\, n} \Big(\f{\as}{4\pi} \Big)^{n+1}.
\end{align}
Here $F$ stands for any function in the factorized cross section in \eq{factorization}, i.e.~hard, beam, soft, soft-collinear or jet function, and $\gamma_F$ its noncusp anomalous dimension. The superscript $i \in \{q,g\}$ specifies the representation of the colour algebra. It will occasionally be convenient to use the colour-stripped coefficients of the cusp anomalous dimension $\Gamma_n = \Gamma_n^i /C_i$ for $n\leq2$, with the overall colour factor $C_i = C_A$ ($C_F$) for $i=g$ ($i=q$) made explicit. Similarly, the $\beta$-function is expanded as
\begin{align} \label{eq:beta_exp}
\beta(\as)=&-2\as \sum_{n=0}^\infty \beta_{n} \Big(\f{\as}{4\pi} \Big)^{n+1}.
\end{align}
The coefficients in \eqs{Gamma_exp}{beta_exp} needed up to NNLL$'$ are given in \eqs{beta_coef}{Gamma_coef}.
The boundary condition for a function $F$ in the factorized cross section is expanded as
\begin{align} \label{eq:function expansion}
F=\sum^\infty_{n=0} \Big(\f{\as}{4\pi}\Big)^n F^{(n)}
\,.\end{align}
%

\subsubsection{Hard function}
\label{sec:hardfn}

We work in the heavy-top limit in which the top-quark loop coupling the Higgs boson to gluons has been integrated out, resulting in the effective Lagrangian
\begin{align}
\mathcal{L}_{\mathrm{eff}} = C_t \frac{\alpha_s}{12\pi}\frac{H}{v}G^a_{\mu\nu}G^{a\,\mu\nu}
\,,  \end{align}
where the Wilson coefficient $C_t$ has an expansion in $\alpha_s$. In this limit, the relevant hard functions can be extracted from the helicity amplitudes~\cite{Becher:2013vva,Moult:2015aoa}, which were computed at 1-loop order in \refcite{Schmidt:1997wr} and at 2-loops in \refcite{Gehrmann:2011aa}. The squared amplitudes can be written as
\begin{align}
  |\mathcal{M}_{ggg}^{+++}(p_1,p_2,p_3)|^2 &= \frac{M_H^8}{2\varsigma_{12}\varsigma_{23}\varsigma_{13}} |\alpha|^2,\,\nn\\
  |\mathcal{M}_{ggg}^{++-}(p_1,p_2,p_3)|^2 &= \frac{\varsigma_{12}^3}{2\varsigma_{23}\varsigma_{13}} |\beta|^2,\,\nn\\
  |\mathcal{M}_{q\bar{q}g}^{-++}(p_1,p_2,p_3)|^2 &= \frac{\varsigma_{23}^2}{2\varsigma_{12}} |\gamma|^2,\,
\end{align}
where $\varsigma_{ij}=2p_i\cdot p_j$ and the functions $\alpha$, $\beta$ and $\gamma$ admit a perturbative expansion in $\alpha_s(\mu)$.\footnote{In the following we will suppress the $\mu$ dependence in our notation.} The remaining helicity amplitudes can be deduced by exploiting parity and charge conjugation symmetries. We are interested in the case in which the momenta $p_1=p_a$ and $p_2=p_b$ are incoming and $p_3=p_J$ is outgoing. Using the definitions in \eq{invariants} and further defining
\begin{align}
t=\frac{s_{aJ}}{s_{ab}}
\,,\qquad
u=\frac{s_{bJ}}{s_{ab}}
\,,\qquad
v=\frac{M_H^2}{s_{ab}}
\,,\end{align}
the hard functions can be written as
\begin{align}
  H_{ggg}(s_{ab},s_{bJ},s_{aJ}) &= 2 \biggl[ \frac{M_H^8}{2s_{ab}s_{bJ}s_{aJ}} |\alpha_2(t,v)|^2 + \frac{s_{ab}^3}{2s_{bJ}s_{aJ}} |\beta_2(t,v)|^2 
  \nn \\ & \qquad
  + \frac{s_{bJ}^3}{2s_{ab}s_{aJ}} |\beta_4(t,v)|^2 + \frac{s_{aJ}^3}{2s_{ab}s_{bJ}} |\beta_4(u,v)|^2\biggr],\,\nn\\
  H_{q\bar{q}g}(s_{ab},s_{bJ},s_{aJ}) &= 2 \biggl[\frac{s_{bJ}^2}{2s_{ab}} |\gamma_2(u,v)|^2 + \frac{s_{aJ}^2}{2s_{ab}} |\gamma_2(t,v)|^2 \biggr],\,\nn\\
  H_{qgq}(s_{ab},s_{bJ},s_{aJ}) &= 2 \biggl[\frac{s_{bJ}^2}{2s_{aJ}} |\gamma_3(u,v)|^2 + \frac{s_{ab}^2}{2s_{aJ}} |\gamma_4(u,v)|^2 \biggr]\,,
\end{align}
where the subscript $n=\{2,3,4\}$ on the functions $\alpha, \beta, \gamma$ denotes the kinematic region in which they are to be evaluated, following the conventions of \refcite{Becher:2013vva}. We take the perturbative coefficients of these functions from that reference, where the translation from the Catani scheme in which they were originally computed to the $\overline{\mathrm{MS}}$ scheme conventionally used in SCET has already been performed.\footnote{We note that the original results for the $q\bar{q}\to Hg$ channel presented in \refcite{Gehrmann:2011aa} contained a typographical error in the ancillary files. This was remedied in the later results of \refcite{Becher:2013vva}.} Since their expressions, even at one loop, are rather involved, we refrain from writing them out explicitly here.

Lastly, the renormalization group equation (RGE) for the hard function is given by
\begin{align} \label{eq:hardRGE}
\mu \frac{\df}{\df \mu}  \ln H_{\kappa} (\ptj, Y, y_J; \mu)
&= \gamma^{\kappa}_H  (\ptj, Y, y_J; \mu)
\end{align}
where the hard anomalous dimension is given to all orders by~\cite{Gardi:2009qi}
\begin{equation}
\gamma^\kappa_H(s_{ab}, s_{aJ}, s_{bJ};\mu)
= \sum_{(klm)} \Bigl\{\Gamma^m_\cusp[\as(\mu)] - \Gamma^k_\cusp[\as(\mu)] - \Gamma^l_\cusp[\as(\mu)] \Bigr\} \ln\frac{\lvert s_{kl}\rvert}{\mu^2}
+ \gamma^\kappa_H[\as(\mu)]
\label{eq:hard_noncusp}
\,,\end{equation}
where the sum runs over the three permutations $(klm) = (abJ), (aJb), (bJa)$.
Up to two loops the noncusp anomalous dimension is given by
\begin{align}
\gamma^\kappa_H(\as) &= 2\gamma_C^a(\as) + 2\gamma_C^b(\as) + 2\gamma_C^j(\as) + \ord{\as^3}
\,,\nn \\
\gamma^q_{C\,0} &= -3 C_F
\,,\nn\\
\gamma^q_{C\,1}
&= - C_F \biggl[
   C_A \Bigl(\frac{41}{9} - 26\zeta_3\Bigr)
   + C_F \Bigl(\frac{3}{2} - 2 \pi^2 + 24 \zeta_3 \Bigr)
   + \beta_0 \Bigl(\frac{65}{18} + \frac{\pi^2}{2} \Bigr)
\biggr]
\,,\nn \\
\gamma_{C\,0}^g &= -\beta_0
\,,\nn\\
\gamma_{C\,1}^g
&= C_A \biggl[
   C_A \Bigl(-\frac{59}{9} + 2\zeta_3 \Bigr)
   + \beta_0 \Bigl(-\frac{19}{9} + \frac{\pi^2}{6} \Bigr)
\biggr]
 - \beta_1
\,.\end{align}

\subsubsection{Jet function}

The jet function RGE is given by
\begin{align} \label{eq:jetRGE}
\mu \frac{\df}{\df \mu}  \ln J_i(\ptj \Rj; \mu)&= \gamma^{i}_J (\ptj \Rj; \mu)\, ,
\end{align}
where the jet anomalous dimension is
\begin{align}
\gamma^i_J(\ptj \Rj; \mu) &=   2 \Gamma^i_\text{cusp}[\as(\mu)] \ln\f{\mu }{\ptj \Rj}
+ \gamma^i_{J}[\as(\mu)]
\,,\end{align}
with the one- and two-loop noncusp anomalous dimensions given by
\begin{align}
\gamma_{J \,0}^q &= 6 C_F\,, \nn \\
\gamma_{J \,0}^g &= 2 \beta_0\,,  \nn \\
\gamma_{J \,1}^q &= 11.17(5) C_F^2 -181.30(6) C_F C_A -7.916(5) C_F n_f T_F\,, \nn \\
\gamma_{J \,1}^g &= 52.898C_A^2 - 215.20C_AC_F - 10.045C_F^2 - 5.1089C_An_fT_F - 0.37279C_Fn_fT_F\, .
\end{align}
The two-loop quark noncusp anomalous dimension $\gamma_{J \,1}^q$ was computed in~\refcite{Liu:2021xzi}, while we obtain the gluon case $\gamma_{J \,1}^g $ from consistency relations, as discussed in  \sec{consistency}.

Expanding the jet function $J_i$ according to \eq{function expansion}, we have 
\begin{align}
J^{(0)}_i(p_T R; \mu)&=1
\,,\nn \\
J^{(1)}_i(p_T R; \mu)&= \Gamma^i_0 L_J^2 + \gamma_{J \,0}^i  L_J + j_{i}^{(1)}
\,, \nn \\
J_i^{(2)}(p_T R; \mu)&=  \f{\Gamma^{i\,2}_0}{2} L_J^4 +\biggl(\f{2 \beta_0 \Gamma^i_0}{3} + \Gamma^i_0 \gamma_{J \,0}^i  \biggr) L_J^3 + \bigg( \Gamma^i_1+ \Gamma^i_0  j_{i}^{(1)} +\beta_0 \gamma_{J \,0}^i  +\f{\gamma_{J \,0}^{i \, 2}}{2} \bigg) L_J^2
\, \nn \\ & \quad
+ \Bigl[ (2\beta_0 +\gamma_{J \,0}^i  ) j_{i}^{(1)}+ \gamma_{J\, 1}^i\Bigr]  L_J + j_{i}^{(2)}
\,, \end{align}
where $L_J\equiv \ln [\mu / (\ptj \Rj) ]$, and the NLO and NNLO constant terms are given by
\begin{align}
j^{(1)}_q&=  C_F \bigg(13 -\f{3 \pi^2}{2} \bigg),   \nn \\
j^{(1)}_g&=  C_A \bigg( \f{5}{6} -\f{3 \pi^2}{2} \bigg) +\f{23}{6} \beta_0,  \nn \\
j^{(2)}_q&= 4 \times [ -1.78(2) C_F^2 - 106.87(2)C_A C_F + 14.072(2) C_F n_f T_F],   \nn \\
j^{(2)}_g&\propto  \theta_2
\, .\end{align}
The NLO anti-$k_T$ jet functions were calculated in~\refcite{Ellis:2010rwa}, while the NNLO quark jet function has been computed in~\refcite{Liu:2021xzi}. The NNLO gluon jet function remains unknown, and its constant term $j^{(2)}$ is therefore treated as a theory nuisance parameter $\theta_2$ and included as part of our perturbative uncertainty, as discussed in \sec{TNP}.

\subsubsection{Beam function}

The beam function virtuality and rapidity RGEs are given by
\begin{align} \label{eq:BeamRGEs}
\mu \frac{\df}{\df \mu}  \ln B_i\Bigl(\ptv, x, \Rc; \mu, \frac{\nu}{\omega}\Bigr)&= \gamma^{i}_B
\Bigl(\mu, \frac{\nu}{\omega}\Bigr)\, , \nn \\
\nu \frac{\df}{\df \nu}  \ln B_i\Bigl(\ptv, x, \Rc; \mu, \frac{\nu}{\omega}\Bigr)&= \gamma^{i}_{\nu,B} (\ptv,\Rc; \mu )\, ,
\end{align}
where the anomalous dimensions are
\begin{align}\label{eq:anom_dim_B}
\gamma^{i}_{B} \Bigl(\mu, \frac{\nu}{\omega}\Bigr)&= 2 \Gamma_\text{cusp}^i[\as(\mu)] \, \ln \frac{\nu}{\omega} + \gamma^{i}_{B}[\as(\mu)]
\,,\nn\\
\gamma^{i}_{\nu,B}(\ptv, \Rc; \mu)&= 2\eta_\Gamma^i(\ptv;\mu) + \gamma^{i}_{\nu,B}[\as(\ptv),\Rc]
\,,\end{align}
with 
\begin{align}\label{eq:anom_dim_b_explicit}
\gamma_{B \,0}^g &= 2\beta_0\,,  \\
\gamma_{B \,1}^g &= 2\beta_1 + 8C_A \bigg[\Bigl(-\frac{5}{4}+2(1+\pi^2)\ln 2-6\zeta_3\Bigr)C_A + \Big(\frac{5}{24}-\frac{\pi^2}{3}+\frac{10}{3}\ln2\Big)\beta_0\bigg]\,, \nn \\
\gamma_{B \,0}^q &= 6C_F\,,  \nn \\
\gamma_{B\,1}^q &= C_F\bigg[ (3-4\pi^2+48\zeta_3 )C_F+(-14+16(1+\pi^2)\ln 2-96\zeta_3)C_A
\nn \\ & \qquad \quad
+\bigg(\frac{19}{3}-\frac{4\pi^2}{3}+\frac{80}{3}\ln 2\bigg)\beta_0\bigg] \,, \nn \\
\gamma^{i}_{\nu,B\, 0}(\Rc) &= 0 \,, \nn \\
\gamma^{i}_{\nu,B\, 1}(\Rc) &= 8C_i\biggl[\biggl( \frac{17}{9}-\bigl(1+\pi^2\bigr)\ln2 + \zeta_3 \biggr)C_A + \biggl(\frac{4}{9}+\frac{\pi^2}{12}-\frac{5}{3}\ln2\biggr)\beta_0\biggr] -\frac{1}{2}C_i C_2(\Rc) \,,\nn
\end{align}
where $C_i=C_A$ for $i=g$, $C_i=C_F$ for $i=q$, and the $\Rc$ dependent part of the rapidity anomalous dimension is given by \cite{Stewart:2013faa, Abreu:2022zgo}
\begin{align}
C_2(\Rc) =& \, 2\ln \Rc^2\biggl[\biggl(1-\frac{8\pi^2}{3}\biggr)C_A + \biggl(\frac{23}{3}-8\ln2\biggr)\beta_0\biggr]  + 15.62C_A-9.17\beta_0 \nn\\
& + 54.21 \Rc^2 - 4.40 \Rc^4 + 0.11 \Rc^6 + \mathcal{O}\bigl(\Rc^8\bigr) \, 
\,,\end{align}
where the terms without an explicit colour factor assume $n_f = 5$. 

As discussed in \sec{factorization analysis},
placing a jet veto can mix the constraints imposed by the measurement on the soft and $n_{a,b}$-collinear sectors leading to the presence of soft-collinear mixing terms at $\ord{\Rc^2}$~\cite{Tackmann:2012bt}. Various ways for treating these subleading terms have been proposed in the literature. Since we will use the results of \refcite{Abreu:2022sdc} for the 2-loop beam functions, we follow their prescription and exponentiate these terms rather than treating them as an additional additive contribution. The rapidity anomalous dimension $\gamma_{\nu,B\,1}$ then takes the form shown above.

The beam function can be written as a convolution between perturbative matching kernels $\mathcal{I}_{ij}(\ptv,\Rc,\omega,z;\mu,\nu)$ and the standard PDFs $f_j(x,\mu)$,
\begin{align}
B_i\Bigl(\ptv, x, \Rc; \mu, \frac{\nu}{\omega}\Bigr) = \sum_j \int_x^1 \frac{\mathrm{d}z}{z}\mathcal{I}_{ij}\Bigl(\ptv, z, \Rc; \mu, \frac{\nu}{\omega}\Bigr)\, f_j\Bigl(\frac{x}{z},\mu\Bigr).
\end{align}
The matching kernels can be perturbatively expanded as
\begin{align}
  \mathcal{I}_{ij}=\delta_{ij}\delta(1-z) + \frac{\as(\mu)}{4\pi}\mathcal{I}_{ij}^{(1)} + \Bigl(\frac{\as(\mu)}{4\pi}\Bigr)^2\mathcal{I}_{ij}^{(2)}+\mathcal{O}(\as^3)
\,.\end{align}
Suppressing most arguments for brevity and using
\begin{equation}
L_B^\mu = \ln \frac{\mu}{\ptv}
\,, \quad
L_B^\nu = \ln \frac{\nu}{Q}
\,,\end{equation}
the one- and two-loop matching kernels are given by
\begin{align} \label{eq:I_standard}
\cI_{ij}^{(1)}(z) &=
\delta_{ij} \delta(1-z) \, L^\mu_B (2 \Gamma^i_0 L^\nu_B+\gamma^i_{B \,0})
-2 L^\mu_B P_{ij}^{(0)}(z)
+I_{ij}^{(1)}(z)
\,,\nn \\
\cI_{ij}^{(2)}(z) &=
\delta_{ij} \delta(1-z) \biggl\{
   (L^\mu_B)^2 \Bigl[ 2 (\Gamma^i_0)^2 (L^\nu_B)^2 + L^\nu_B (2 \beta_0 \Gamma^i_0+2 \Gamma^i_0 \gamma^i_{B \,0})+\beta_0 \gamma^i_{B \,0}+\frac{(\gamma^i_{B \,0})^2}{2} \Bigr]
\nn\\[-0.25em] &\quad\quad\quad\quad\quad\quad
   +L^\mu_B \Bigl[ 2 \Gamma^i_1 L^\nu_B+\gamma^i_{B \,1} \Bigr]
   -\frac{1}{2} \gamma^i_{\nu\,1}(\Rc) L^\nu_B
   \biggr\}
\nn \\[0.25em] &\quad
+ P_{ij}^{(0)}(z) \, (L^\mu_B)^2 \Bigl[ -4 \Gamma^i_0 L^\nu_B -2 \beta_0-2 \gamma^i_{B \,0} \Bigr]
+ I_{ij}^{(1)}(z) \, L^\mu_B \Bigl[ 2 \Gamma^i_0 L^\nu_B+ 2 \beta_0+\gamma^i_{B \,0} \Bigr]
\nn \\[0.25em] &\quad
-2 L^\mu_B \sum_k I_{ik}^{(1)}(z) \otimes_z P_{kj}^{(0)}(z)
-2 L^\mu_B P_{ij}^{(1)}(z)
+2 (L^\mu_B)^2 \sum_k P_{ik}^{(0)}(z) \otimes_z P_{kj}^{(0)}(z)
\nn \\ &\quad
+I_{ij}^{(2)}(\Rc, z)
\,,\end{align}
The splitting functions and respective convolutions can e.g.~be found in~\refcite{Michel:2018hui}.
The finite parts are given by~\cite{Becher:2012qa,Becher:2013xia,Stewart:2013faa}
\begin{align}
  I_{gq}^{(1)}(z)&=2C_Fz\, , \nn \\
  I_{gg}^{(1)}(z)&= 0\, , \nn \\
  I_{qq}^{(1)}(z)&=2C_F(1-z)\, ,\nn \\
  I_{qg}^{(1)}(z)&=4T_Fz(1-z)\, .
\end{align}
At NNLO, the quark and gluon beam functions were computed via a numerical approach in~\refcite{Bell:2022nrj,Bell:2024epn}; semi-analytic expressions for both quark and gluon cases later appeared in~\refcite{Abreu:2022zgo}. We have implemented these expressions up to $\mathcal{O}(\Rc^4)$ in \textsc{SCETlib}. The effect of higher order terms $\mathcal{O}(\Rc^6)$ is negligible for phenomenologial applications~\cite{Gavardi:2023aco}.

\subsubsection{Soft and soft-collinear functions}
\label{sec:soft functions}

 \begin{figure}[t]
 \centering
 \begin{subfigure}{0.3\textwidth}
 \centering
 \includegraphics[width=0.9\textwidth]{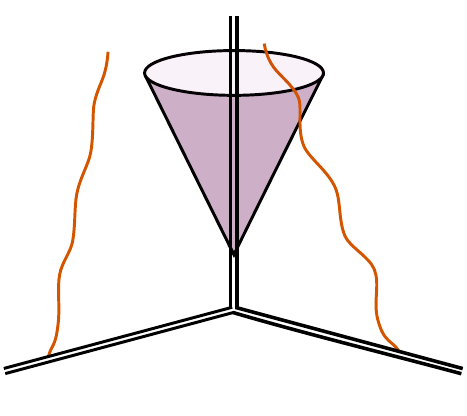}
 \caption{Total-soft $S^T$} \label{fig:totalsoft}
 \end{subfigure} \hfill
 \begin{subfigure}{0.3\textwidth}
 \centering
 \includegraphics[width=0.9\textwidth]{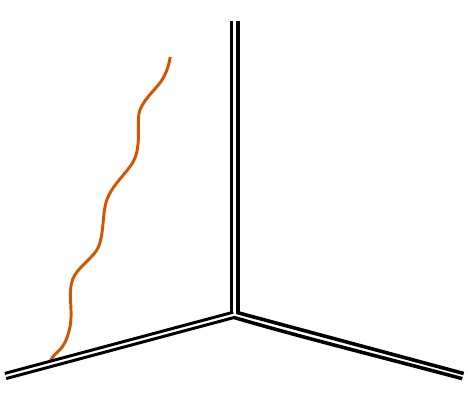}
 \caption{Global-soft $S_\kappa$}\label{fig:globalsoft}
 \end{subfigure} \hfill
  \begin{subfigure}{0.3\textwidth}
 \centering
 \includegraphics[width=0.78\textwidth]{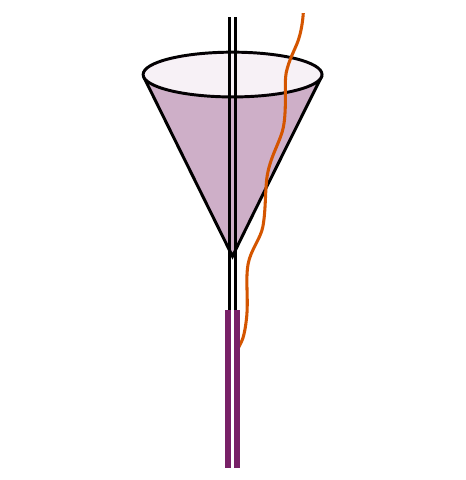}
 \caption{Soft-collinear $\mathcal{S}^R$} \label{fig:soft-collinear}
 \end{subfigure}\hfill
\caption{Depiction of the total soft function $\ST$ (a) present in the $R_J \sim 1$ limit, and of its refactorization into the global soft function $S_\kappa$ (b) and soft-collinear function $\mathcal{S}^R$ (c) in the $\Rj\ll1$ limit, the latter two appearing in \eq{factorization}.  } \label{fig:SoftSector}
 \end{figure}
 
In this section we discuss the global soft ($S_\kappa$) and soft-collinear ($\SC$) functions appearing in the factorized cross section in \eq{factorization}. To understand their origin, we start by discussing the soft sector in its entirety before taking the $R_J\ll 1$ limit. The total soft function, denoted $\ST$, is given by the squared matrix element of a set of Wilson lines which can act as sources of soft(-collinear) gluons, where the gauge field has the appropriate soft or soft-collinear momentum scaling. Its definition is given by
\begin{align}
S^T_\kappa  &= \frac{1}{N_\kappa}\ \sum_{X_{s}} \big\langle 0 \big\vert  \bar{\bf T}[Y^\dagger(\{n_i\},\kappa)] \big\vert X_{s}\big\rangle \big\langle X_{s}\big\vert {\bf T}[Y(\{n_i\},\kappa) ] \big\vert 0 \big\rangle\, \mathcal{M}_{\rm full}(X_{s})
\, ,\end{align}
where $Y(\{n_i\},\kappa)$ collects the set of three soft Wilson lines, one for each external parton. The normalization factor $N_\kappa$ is chosen such that the tree-level soft function is the identity matrix in colour space. The individual soft Wilson lines are defined as 
\begin{equation}
Y^i_{n_i}(x) = {\bf P}\exp\biggl\{\img g_s\int_{-\infty}^{0} \! \df s \, n_i\cdot A_s^i(x+sn_i)\biggr\} \, ,
\end{equation}
for a given incoming parton along the $n_i$ direction, and the representation of the gauge field $A_s^i$
depends on whether the associated particle $i$ is a (anti-)quark or gluon.
This total soft function corresponds to the soft function of \refcite{Liu:2013hba}. For $\ST$, depicted in \fig{totalsoft}, radiation clusters outside the jet are required to have transverse momentum $k_T<\ptv$, while radiation inside the jet is unconstrained.

An explicit NLO computation of $\ST$ is presented in \app{R2-corrections}, where we see that it depends on two scales: $\ptv$ and $\ptv \Rj$. In the $\Rj \sim 1$ limit, these scales are parametrically of the same size and $\ln [\mu/\ptv] $ and $\ln [\mu/(\ptv \Rj)]$ can be simultaneously minimized by choosing $\mu \sim \ptv$. However, the factorization of the jet-collinear sector requires us to employ the narrow jet limit $\Rj\ll1$. This makes the two scales parametrically different and therefore requires the factorization of the total soft function into global-soft and soft-collinear functions,
\begin{align} \label{eq:ST}
S_{\kappa}^T(\ptv, \Rc, \Rj, y_J; \mu, \nu) = S_{\kappa}(\ptv, \Rc, y_J; \mu, \nu) \SC_j(\ptv \Rj; \mu) + \mathcal{O}(\Rj^2)
\,,\end{align}
each only depending on a single scale $\ptv$ or $\ptv \Rj$.

The global-soft function $S$ describes wide-angle radiation and  does not probe the (narrow) jet boundary, so it simply requires all clusters to have transverse momentum below $\ptv$, as depicted in \fig{globalsoft}. Its definition in terms of Wilson lines is therefore similar to that of the total soft function, only with a different measurement function 
\begin{equation}
S_\kappa = \sum_{X_{s}} \big\langle 0 \big\vert  \bar{\bf T}[Y^\dagger(\{n_i\},\kappa)] \big\vert X_{s}\big\rangle \big\langle X_{s}\big\vert {\bf T}[Y(\{n_i\},\kappa) ] \big\vert 0 \big\rangle\, \mathcal{M}_{s}(X_{s})
\, ,
\end{equation}
where $\mathcal{M}_s$ is now the global measurement, see \eq{jetmeas_s}, and is independent of $R_J$.
Its virtuality and rapidity RGEs are given by 
\begin{align} \label{eq:globalRGEs}
\mu \frac{\df}{\df \mu}  \ln S_{\kappa}(\ptv, \Rc, y_J; \mu, \nu)&= \gamma^{\kappa}_S (\ptv, y_J;  \mu, \nu)\,, \nn \\
\nu \frac{\df}{\df \nu}  \ln S_{\kappa}(\ptv, \Rc, y_J; \mu, \nu) &= \gamma^{\kappa}_{\nu,S} (\ptv, \Rc;\mu ) \, .
\end{align}
The anomalous dimensions are 
\begin{align}
\gamma^{\kappa}_{S}(\ptv, y_J;  \mu, \nu) &=  2 \Gamma^{a}_\text{cusp}[\as(\mu)]  \ln\Big( \f{\mu}{\nu e^{-y_J} }\Big) + 2 \Gamma^{b}_\text{cusp}[\as(\mu)]  \ln \Big( \f{\mu}{\nu e^{y_J} } \Big)  \nn \\
&\quad + 2 \Gamma^{j}_\text{cusp}[\as(\mu)]  L_S^\mu  + \gamma^{\kappa}_{S}[\as(\mu)]\, \nn \\[2mm]
&=2 (\Gamma^{a}_\text{cusp}[\as(\mu)]+\Gamma^{b}_\text{cusp}[\as(\mu)]+\Gamma^{j}_\text{cusp}[\as(\mu)])L_S^\mu \nn \\
&\quad +2\Gamma^{a}_\text{cusp}[\as(\mu)] L_S^a +2\Gamma^{b}_\text{cusp}[\as(\mu)] L_S^b + \gamma^{\kappa}_{S}[\as(\mu)]\,, \nn \\[2mm]
\gamma^{\kappa}_{\nu,S}(\ptv, \Rc; \mu)&= - \gamma^a_{\nu, B}(\ptv, \Rc, \mu)-\gamma^b_{\nu, B}(\ptv, \Rc, \mu) \, ,
\end{align}
where the beam rapidity anomalous dimensions $\gamma_{\nu,B}$ are given in \eqs{anom_dim_B}{anom_dim_b_explicit}, and we use the shorthand
\begin{equation}
L_S^\mu = \ln \frac{\mu}{\ptv}\, , \qquad L_S^a = \ln \Bigl(\frac{e^{y_J}\ptv}{\nu}\Bigr)\, , \qquad L_S^b = \ln \Bigl(\frac{e^{-y_J}\ptv}{\nu}\Bigr) \, .
\end{equation}
RG consistency only fixes the sum of the anomalous dimension of the global soft function and the soft-collinear function. The individual terms are not known at two-loop order, and we therefore account for the splitting of the total soft anomalous dimension into the two pieces using a theory nuisance parameter, as discussed in \sec{TNP}.
The logarithmic terms of the global soft function at a given order are fixed by its RGE and lower-order constant terms. For definiteness, we still show how it can be directly calculated at NLO:
\begin{align}\label{eq:soft-global-measurement}
 S_{\kappa,\text{bare}}(\ptveto, \Rc, y_J) &= 1 \!-\!
 \frac{\alpha_s}{\pi^{2-\epsilon}} (e^{\gamma_E}\mu^2)^\eps
 \sum_{l<m}\mathbf{T}_l\cdot \mathbf{T}_m  
\int \mathrm{d}^d k \, \delta(k^2) \frac{n_l\cdot n_m}{ (n_l\cdot k ) (n_m\cdot k)}\, \Theta (\ptv\!-\!k_T),
\end{align}
with $l,m \in \{a,b,j\}$. Following the same procedure taken in the calculation of the total soft function (\app{R2-corrections}) this yields
\begin{align} \label{eq:soft-global-NLO}
S^{(1)}_{\kappa}(\ptveto, y_J;\mu,  \nu)&=(C_{a} + C_{b} +C_{j})\Gamma_0 \bigl(L_S^\mu\bigr)^2
+2\Gamma_0 \bigl[C_{a} L_S^a
+ C_{b} L_S^b\bigr]  L_S^\mu
+ s^{(1)}_{\kappa }
\, .
\end{align}
At NLO the constant term is
\begin{align} 
s^{(1)}_{\kappa }&=-(C_{a} + C_{b} +C_{j}) \f{\pi^2}{6}
\, .
\end{align}
The NNLO soft function is
\begin{align}  \label{eq:soft-global-NNLO}
S^{(2)}_{\kappa}(\ptveto,\Rc, y_J ;\mu, \nu) &=
\f12 (C_{a}+C_{b}+C_{j})^2 \Gamma_0^2 \bigl(L_S^\mu\bigr)^4  \nn \\
&\quad +\f23 (C_{a} + C_{b} + C_{j})\Gamma_0 \Bigl[\beta_0 + 3\Gamma_0 \bigl(C_{a} L_S^a + C_{b}  L_S^b\bigr)\Bigr] \bigl(L_S^\mu\bigr)^3  \nn \\
&\quad+ \Bigl\{(C_{a} + C_{b} + C_{j}) (\Gamma_0 s_{\kappa}^{(1)}+ \Gamma_1) +2 \Gamma_0 \bigl[  C_{a} L_S^a +  C_{b} L_S^b  \bigr] \nn \\
&\quad\quad\quad \times 
\Bigl[\beta_0 + \Gamma_0 \bigl(C_{a} L_S^a + C_{b} L_S^b\bigr) \Bigr] \Bigr\} \bigl(L_S^\mu\bigr)^2 \nn \\
&\quad+\Bigl\{   2 s_{\kappa}^{(1)} \beta_0   +\gamma^{\kappa}_{S \, 1} + 2\bigl[s_{\kappa}^{(1)} \Gamma_0+ \Gamma_1\bigr] \bigl[ C_{a} L_S^a + C_{b} L_S^b \bigr] \Bigr\}  L_S^\mu  \nn \\
&\quad+  \gamma_{\nu, S \, 1}^{\kappa} \ln \f{\nu}{\ptv}  + s^{(2)}_{\kappa}
\,,
\end{align}
where we have used $\gamma_{S\, 0}^{\kappa}=\gamma_{\nu, S \, 0}^{\kappa}=0$ to simplify the expression. The two-loop constant term $s^{(2)}_{\kappa}$ is unknown and will be parametrized in terms of a theory nuisance parameter, as explained in \sec{TNP}. Note that the dependence of the global soft function on $\Rc$ starts at two-loop order and enters through the rapidity anomalous dimension and the two-loop constant term.
In principle, the presence of logarithms of $e^{-y_J}$ in \eqs{soft-global-NLO}{soft-global-NNLO} implies that a further factorization is possible. However, 
here we always consider the case where the jet is well separated from the beams, $n_a \cdot n_J \sim n_b \cdot n_J \sim \mathcal{O}(1)$, such that the rapidity dependence can be treated in fixed-order perturbation theory. Moreover, these logarithms vanish for the most phenomenologically relevant channel, $gg\to Hg$.

We now turn to the soft-collinear function $\SC$, which is depicted in \fig{soft-collinear}. As explained in \sec{factorization analysis}, the soft-collinear function arises as a correction to the global soft function. The latter applies the $p_T$-veto to soft radiation everywhere, including inside the jet region. However, radiation inside the jet should not be subject to a veto. The collinear nature of soft-collinear radiation allows it to resolve the jet boundary and correct for this. Its RGE is given by
\begin{align} \label{eq:softcollinearRGE}
\mu \frac{\df}{\df \mu}  \ln \SC_i(\ptv \Rj; \mu) &= \gamma^{i}_\mathcal{S} (\ptv \Rj; \mu)\, 
\end{align}
where the anomalous dimension is 
\begin{align}
\gamma^i_\mathcal{S}(\ptv \Rj; \mu) &= -  2 \Gamma^i_\text{cusp}[\as(\mu)] \ln\Bigl(\f{\mu }{\ptv \Rj}\Bigr)  + \gamma^i_{\mathcal{S}}[\as(\mu)] \, ,
\end{align}
with
\begin{align}
\gamma_{\SCs \,0}^i&= 0 \, .
\end{align}
$\gamma_{\SCs \,1}^i$ depends on the unknown split of the total soft anomalous dimension into global and soft-collinear pieces, which is parameterized by a nuisance parameter.

The definition of the soft-collinear function is
\begin{align}
\mathcal{S}^R_{i}(\ptv  R_J)  &= \frac{1}{d_{i}(N_c)}\sum_{X_{sc}}  \text{Tr}\big[ \big\langle 0 \big\vert  \bar{T}[V^\dagger_{n_J}  X_{n_J}] \big\vert X_{sc}\big\rangle \big\langle X_{sc}\big\vert T[ X^\dagger_{n_J} V_{n_J}]\big\vert 0 \big\rangle \big] \mathcal{M}_{sc}(X_{sc})
\, ,\end{align}
where the representation of the Wilson lines depends on the flavour of the parton initiating the jet and 
$d_{j}(N_c)$ denotes the dimension of this representation, i.e. $d_{q}=N_c$ and $d_{g}=N_c^2-1$. $X^\dagger_n$ and $V^\dagger_n$ are soft-collinear Wilson lines defined as 
\begin{align}
X^\dagger_n (x) &= {\bf P}  \bigg\{ \exp \bigg[ -\img g \int^{\infty}_0 \! \df s \, n\cdot A_{sc} (x+sn)   \bigg]   \bigg\} \,,  \nn \\
V^\dagger_n (x) &= {\bf P}  \bigg\{ \exp \bigg[ -\img g  \int^{\infty}_0 \!  \df s\, \bn\cdot A_{sc} (x+s\bn)    \bigg]  \bigg\} \,,
\end{align}
where $A_{sc}$ is the soft-collinear gluon field. The subscript $n$ does not refer to the direction of the Wilson line, but indicates that these soft-collinear gluons are collinear to the $n$ direction.

At NLO, where there is only a single real emission with momentum $k_{sc}$, the soft-collinear function is given by
\begin{align}
\mathcal{S}^R_{i,  \text{bare}}(\ptv  R_J)  &=  1-
 \frac{\alpha_s}{\pi^{2-\epsilon}} (e^{\gamma_E}\mu^2)^\eps
\, C_{i}
\int \mathrm{d}^d k \, \delta(k^2)\, \frac{n_J\cdot \bn_J}{ (n_J\cdot k ) (\bn_J\cdot k)}\,  \mathcal{M}_{sc}(X_{sc})
\, ,\end{align}
where the one-loop measurement function is (see \eq{meas_fact_1})
\begin{align}
\mathcal{M}^{(1)}_{sc}(k_{sc})&= \Theta(k^{sc}_T-\ptv) \Theta(R_J-\Delta R^{sc})\nn \\
&=\left[1-\Theta(\ptv-k^{sc}_T)\right] \Theta(R_J-\Delta R^{sc}) \,,
\end{align}
and the light-like vectors $n_J$ and $\bn_J$ are defined in \eq{lc_dir}.
This yields the NLO expression
\begin{align} \label{eq:soft-collinear-1loop}
\mathcal{S}^{R\,, (1)}_j(\ptv\Rj; \mu) &= - \Gamma_0^i \LSC^2 + \gamma^i_{\mathcal{S}\, 0} \LSC +s^{R \,,(1)}_i \, ,
\end{align}
where $\LSC=\ln (\mu/(\ptv \Rj))$ and the NLO constant piece is
\begin{align} 
s^{R\,, (1)}_i&= C_{i}\, \f{\pi^2}{6}
\, . \end{align}
At NNLO we have
 \begin{align}
\mathcal{S}^{R \,,(2)}_j(\ptveto \Rj; \mu) &=  
 \f{\Gamma^{j\,2}_0}{2} \LSC^4 
 -\Gamma^{j}_0 \Bigl(\f{2}{3} \beta_0 +\gamma^j_{\mathcal{S}\,0} \Bigr)\,   \LSC^3 
 +\biggl[\beta_0 \gamma^j_{\mathcal{S}\,0}+  \f{\gamma^{j\, 2}_{\mathcal{S}\,0}}{2} -\Gamma^{j}_0 s^{R \,,(1)} _{j}  - \Gamma_1^j \biggr]  \LSC^2 \nn \\
 & \quad 
 + \Bigl[ (2 \beta_0 + \gamma^j_{\mathcal{S}\,0}) s^{R\,,(1)}_{j} + \gamma^j_{\mathcal{S}\,1} \Bigr] \LSC
  + s^{R \,, (2)}_{j} 
\, ,\end{align}
where the NNLO constant will be treated as a theory nuisance parameter, as described in \sec{TNP}.

\subsubsection{Nonglobal logarithms}
\label{sec:ngls}

As explained in \sec{factorization}, the function  $S_\text{NG}(\ptv/\ptj)$ encodes the nonglobal logarithms arising from correlated emissions inside and outside the signal jet. After a boost in the jet direction (exploiting type-III reparametrization invariance), the inside and outside of the jet can be turned into complementary hemispheres, implying that the leading NGLs are given by the same universal function as for the hemisphere masses. We will work in the large $N_c$ approximation (the leading NGLs without this approximation have been studied in~\refcite{Hatta:2013iba}). Rather than using the fit of~\refcite{Dasgupta:2001sh}, we employ the solution to the BMS equation~\cite{Banfi:2002hw} up to five-loop order~\cite{Schwartz:2014wha},
\begin{align} \label{eq:S_NG}
 \mathcal{S}_{q}^\text{NG}\bigg( \f{\ptv}{\ptj}\bigg) = 1 - \frac{\pi^2}{24} \widehat L^2 + \frac{\zeta_3}{12} \widehat L^3 + \frac{\pi^4}{34560} \widehat L^4+ \Big(-\frac{\pi^2 \zeta_3}{360} + \frac{17\zeta_5}{480}\Big) \widehat L^5 + \ord{L^6}
\,,\end{align}
where
\begin{align}
\widehat L= \frac{\as N_c}{\pi} \ln \frac{\ptj}{\ptv}
\,.
\end{align}

\begin{figure}[t]
    \centering
    \includegraphics[width=0.48\textwidth]{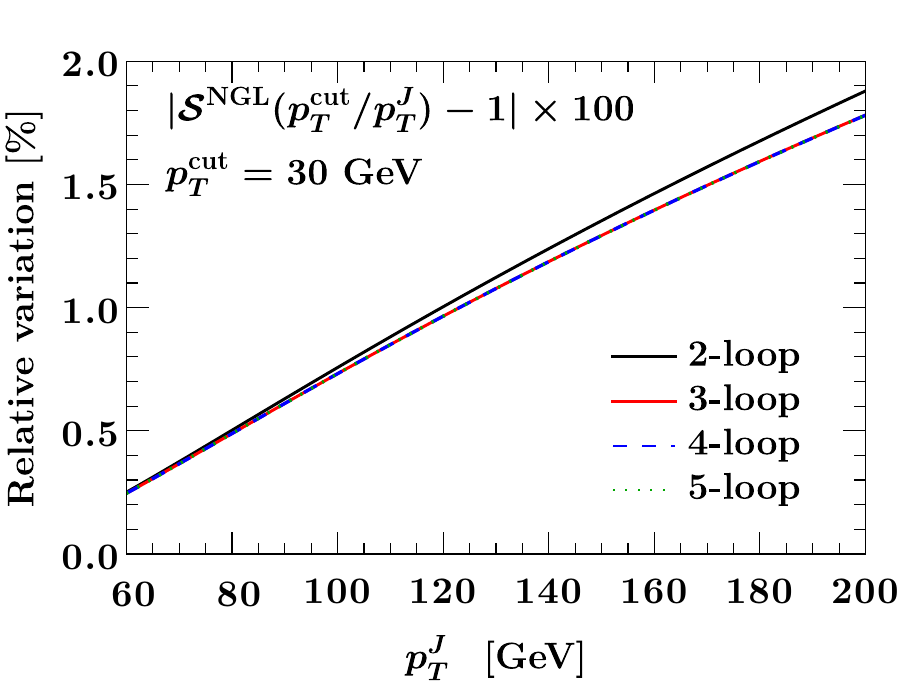}
    \caption{The relative correction from nonglobal logarithms up to 2-loop, 3-loop, 4-loop and 5-loop order, using $\alpha_s(M_Z) = 0.118$ as a representative value. For the region of interest the 2-loop thus affects the cross section at the percent level, the 3-loop at the permille level, and beyond it is negligible.}
    \label{fig:NGL convergence}
\end{figure}

The convergence of this expression for the kinematic region we are interested in can be seen in \fig{NGL convergence}.
For the indicative values, $p_T^J=120$ GeV, $\ptv=30$ GeV, $R_J = 0.4$, we have
$\alpha_s(\mu = \ptv R_J)=0.171$ and $\widehat L = 0.226$, and the expansion reads
\begin{align}
\mathcal{S}_{q}^\text{NG} = 1 - 2.10 \times 10^{-2} + 1.15 \times 10^{-3} + 7.34 \times 10^{-6} + 2.22 \times 10^{-6} + \dotsb \,.
\end{align}

Given the excellent convergence beyond three-loop order, we employ the five-loop result in \eq{S_NG} as a proxy for the resummed tower of leading NGLs.  We therefore want to keep all terms in \eq{S_NG} in the resummation region, while in the fixed-order region the nonglobal logs should be expanded in $\as$ to the same order as the rest of the cross section. A simple way to achieve this is by defining
\begin{align}\label{eq:ModifiedNGLs}
\SCs_{i}^\text{NG}(\ptv, \ptj ;\mu_\SCs, \mu_J) =  \SCs^\text{NG}_{i\, \text{resum}}\Bigl( \f{\mu_\SCs}{\mu_J}\Bigr)\, \SCs^\text{NG}_{i\, \text{FO}}\Bigl( \f{\ptv}{\mu_\SCs} \f{\mu_J}{\ptj} \Bigr)
\, ,\end{align} 
where both $\SCs^\text{NG}_{i\, \text{resum}}$ and $\SCs^\text{NG}_{i\, \text{FO}}$ are obtained from \eq{S_NG}, with the former treated as an evolution kernel (and the full five-loop expansion used), while the latter is considered as a boundary condition and is therefore expanded to order $\as^2$. The scales $\mu_\SCs$ and $\mu_J$ take different values in the resummation and fixed-order regions: in the resummation region, they are chosen to minimize the size of large logarithms as $\mu_\SCs=\ptv \Rj$ and $\mu_J=\ptj\Rj$, while in the fixed-order region they take the common value $\mu_\SCs=\mu_J=\mu_\text{FO}$. This ensures that \eq{ModifiedNGLs} behaves appropriately in the two limits, since in each we have:

 \begin{align}
 \text{Resummation region:}&\hspace{.5cm} \SCs_{i}^\text{NG}(\ptv, \ptj ;\mu_\SCs=\ptv \Rj, \mu_J=\ptj \Rj)=
 \SCs^\text{NG}_{i\, \text{resum}}\bigg( \f{\ptv}{\ptj}\bigg)\, , \nn \\  
  \text{Fixed order region:}& \hspace{.5cm}\SCs_{i}^\text{NG}(\ptv, \ptj ;\mu_\SCs=\mufo, \mu_J=\mufo)=
 \SCs^\text{NG}_{i\, \text{FO}}\bigg( \f{\ptv}{\ptj}\bigg) 
 \, .\end{align}
 We will thus use \eq{ModifiedNGLs} in our numerical implementation.

\subsection{Validation of the singular structure}
\label{sec:validation}

\begin{figure}[!b]
\centering
 \includegraphics[width=0.32\textwidth]{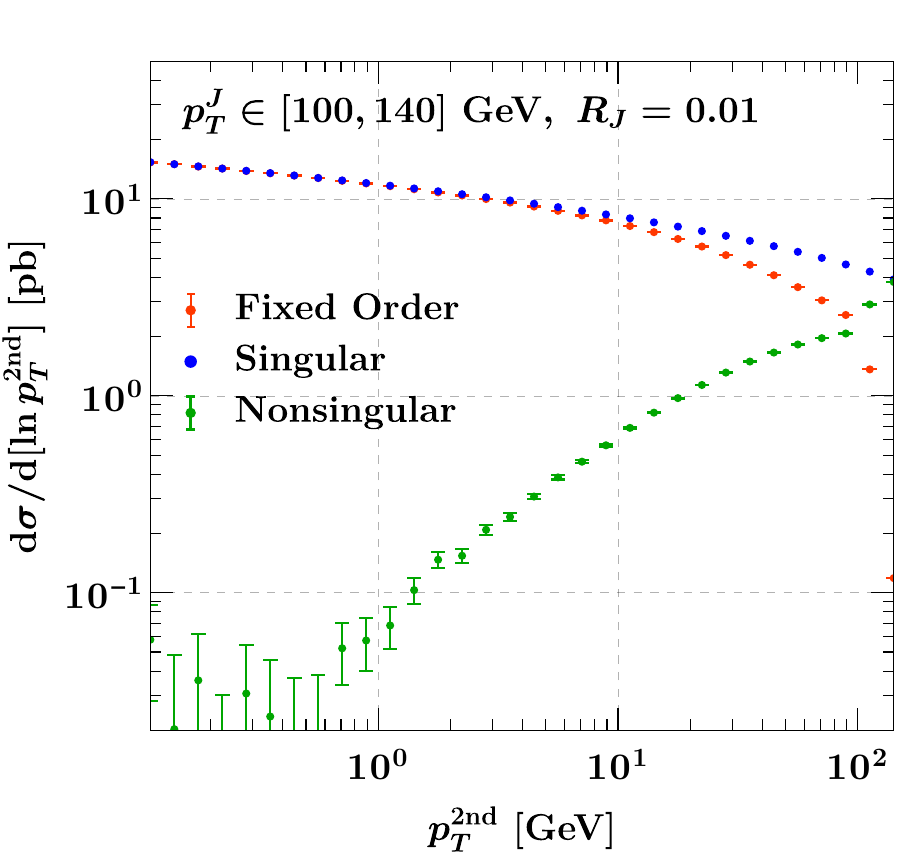}
 \includegraphics[width=0.32\textwidth]{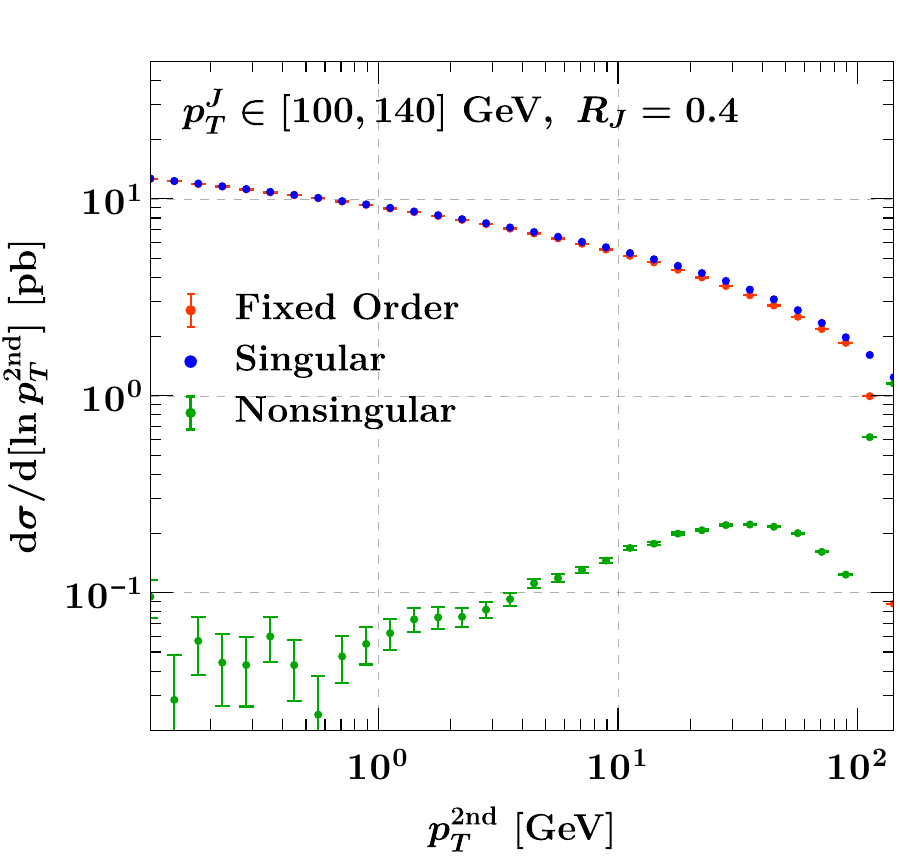}
 \includegraphics[width=0.32\textwidth]{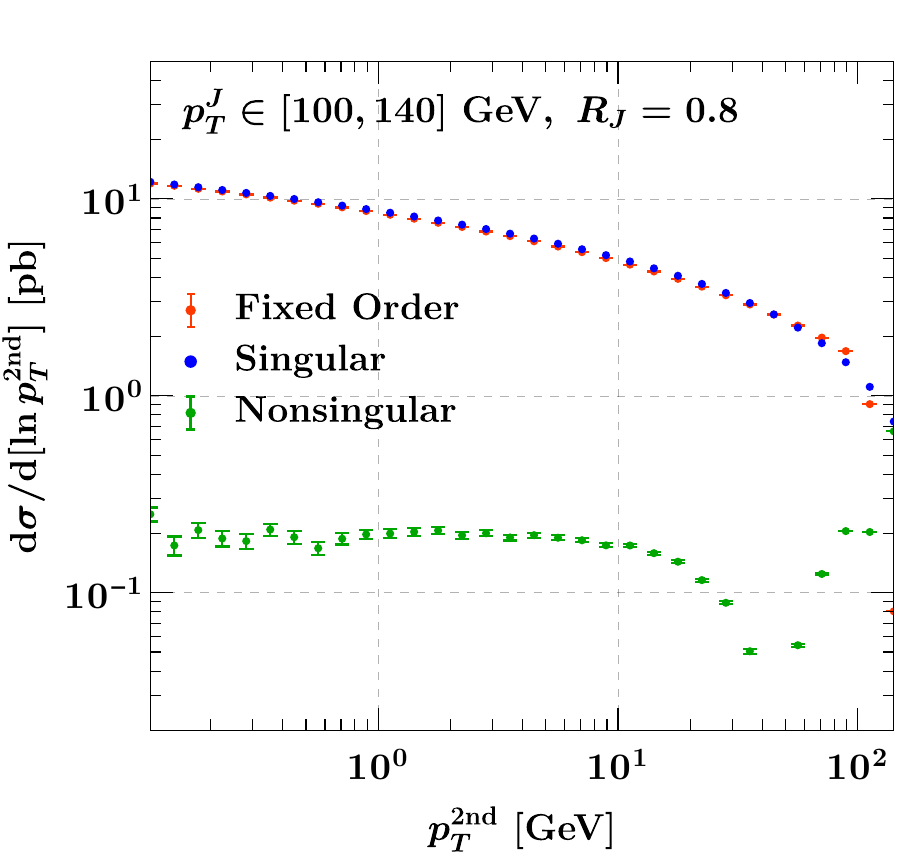}
  \caption{Full fixed order (red), singular (blue) and nonsingular (green) at NLO, differential in $\ln \ptvd$, for $\Rj=0.01, 0.4$ and $0.8$. Wee see that the small $\ptvd$ behaviour of the nonsingular is linear for $\Rj=0.01$ and constant for large $\Rj$, which implies the presence of $\Rj^2 \ln \ptv$ power corrections.
   }\label{fig:SNS-bad}
 \end{figure}
 
To check whether the factorization formula in \eq{factorization} reproduces all the leading-power logarithms of the fixed-order result, we examine the behaviour of the nonsingular cross section differential in $\ln \ptvd$ in the limit $\ptvd \to 0$. (We use $\ptv$ for the integrated cross section  and $\ptvd$ for the corresponding  differential cross section.) The nonsingular cross section is defined as
\begin{align}
\f{\df \sigma^\text{ns}}{\df\ln \ptvd}\equiv \f{\df \sigma^\text{FO}}{\df\ln \ptvd} - \f{\df \sigma^\text{sing}}{\df\ln \ptvd},
\end{align}
where $\df \sigma^\text{FO} / \df\ln \ptvd$ is the full fixed-order cross section and $\df \sigma^\text{sing}/ \df \ln \ptvd$ is given by the factorization formula in \eq{factorization} made differential, which reproduces the singular terms in $\ptvd$ of the fixed-order result. At order $\as$ these leading-power singular terms take the form 
\begin{align}\label{eq:singular-structure}
  \frac{\df \sigma^\text{FO}}{\df \ptvd} &= \as\biggl(c_{-1} \,\delta(\ptvd) +  c_{0} \frac{1}{[\ptvd]_+} + c_{1} \bigg[\frac{\ln \ptvd}{\ptvd}\bigg]_+ + \text{p.c.}\biggr) \nn \\
  & \implies \frac{\df \sigma^\text{FO}}{\df \ln \ptvd} = \as\bigl( c_{0} + c_{1} \ln \ptvd + \text{p.c.} \bigr),
\end{align}
where $c_{-1}$, $c_0$ and $c_1$ are constants that do not depend on $\ptvd$ and p.c.~denotes power corrections.  If the singular piece correctly captures the $c_0$ and $c_1$ terms above, these will not be present in the nonsingular difference, and we therefore expect $\df \sigma^\text{ns}/\df \ln \ptvd\to 0$ as $\ptvd \to 0$.

In \fig{SNS-bad}, we show the nonsingular, singular and full fixed-order results for various values of the jet radius $\Rj$, where the fixed NLO calculation for Higgs+jet used to obtain the nonsingular is provided by the Monte Carlo event generator \textsc{Geneva}~\cite{Alioli:2023har}. We observe that for very small values of the jet radius (e.g. $R_J=0.01$) the nonsingular cross section goes to zero as $\ptv\to0$, as expected. In fact, its linear behaviour at small $\ptvd$ in the log-log plot is consistent with the presence of a $\ptv \ln \ptv$ power correction in $\sigma(\ptv)$. For larger values of the jet radius (e.g. $\Rj=0.8$), however, we see that the nonsingular tends to a constant in the small $\ptvd$ limit, implying that the singular cross section is insufficient to fully reproduce the $c_{0}$ term in \eq{singular-structure} for larger values of $\Rj$. The worsening of the behaviour for increasing $\Rj$ is due to the presence of terms of $\ord{\Rj^2}$ and $\ord{\Rj^2 \ln \ptv}$ in the nonsingular cross section $\sigma(\ptv)$. Since our factorization formula assumes $\Rj\ll1$, these terms are formally a power correction, which cannot be expected to be reproduced by the singular. However, \fig{SNS-bad} shows that they can be numerically sizeable even for intermediate values of $\Rj=0.4$, since they are only suppresed by $\Rj^2$ but are effectively singular in $\ptv$, and we would therefore like to include them explicitly.

\begin{figure}[t]
    \centering
    \includegraphics[width=0.32\textwidth]{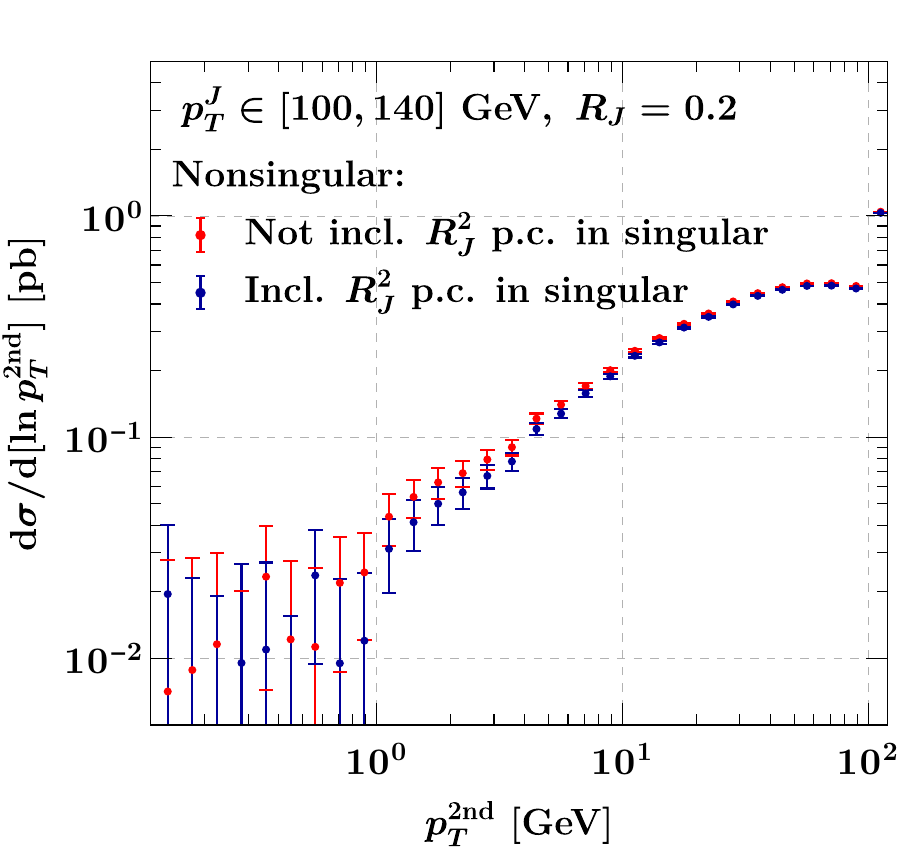}
    \includegraphics[width=0.32\textwidth]{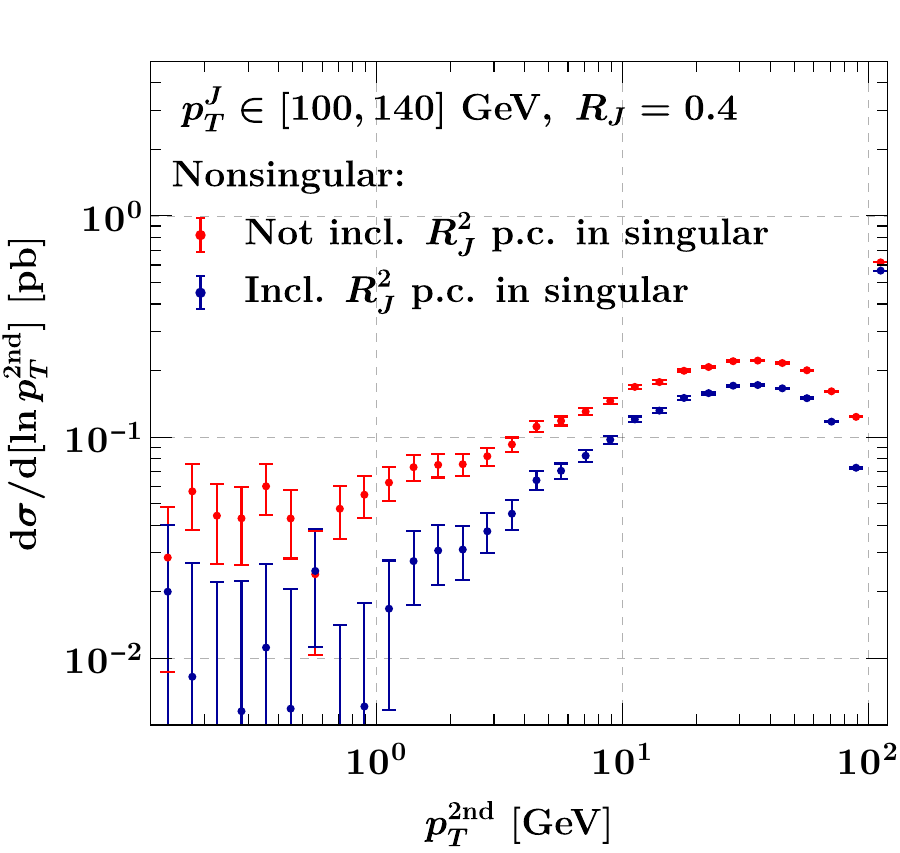}
    \includegraphics[width=0.32\textwidth]{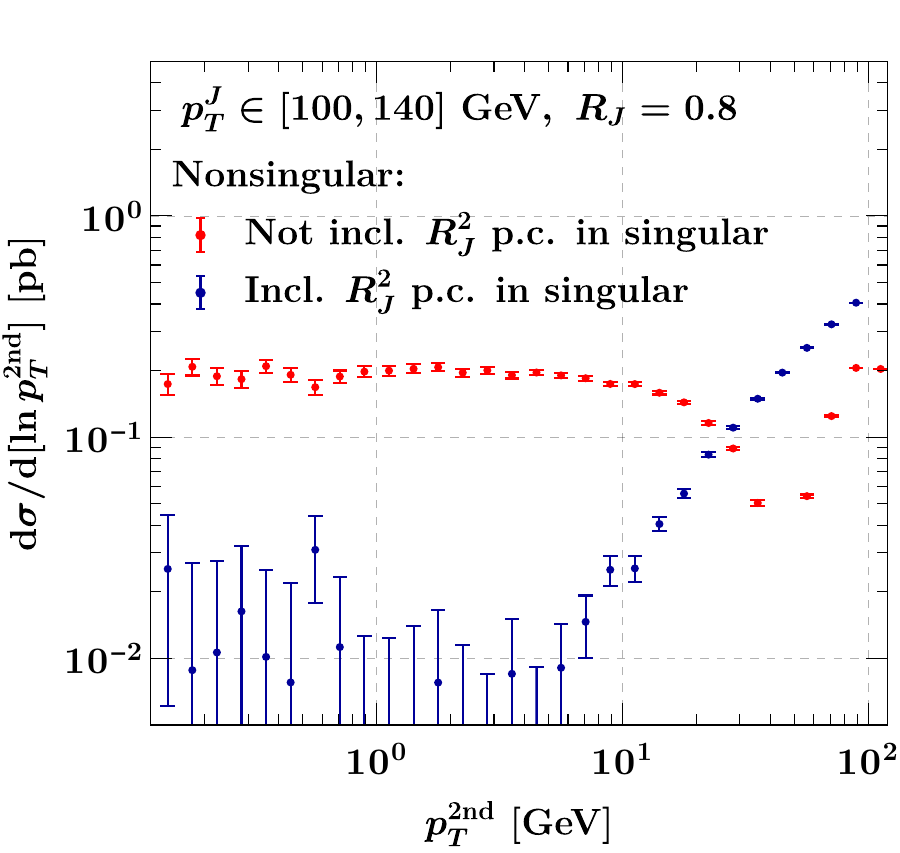} \\
    \includegraphics[width=0.32\textwidth]{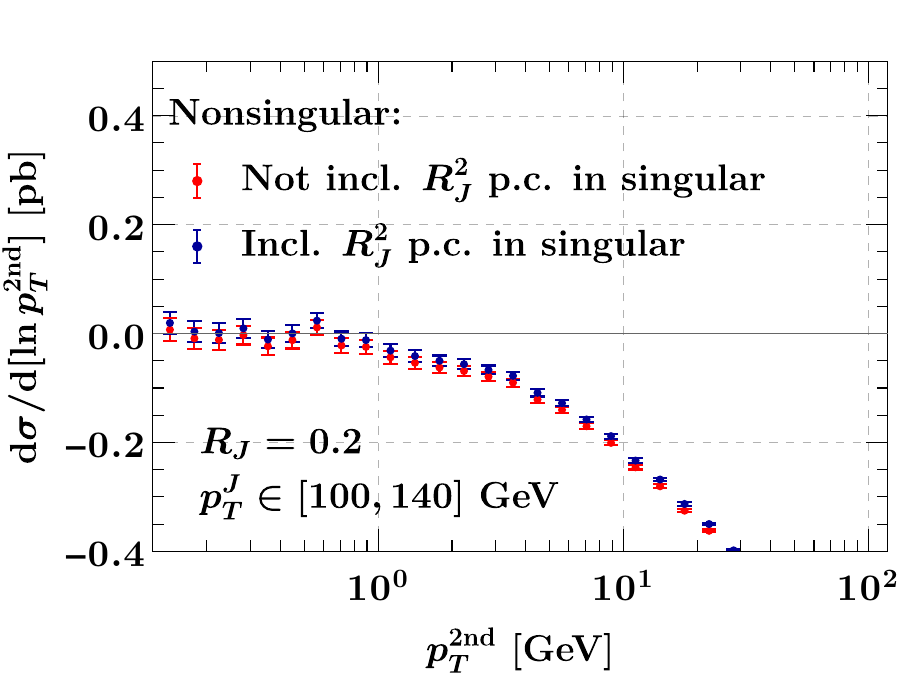}
    \includegraphics[width=0.32\textwidth]{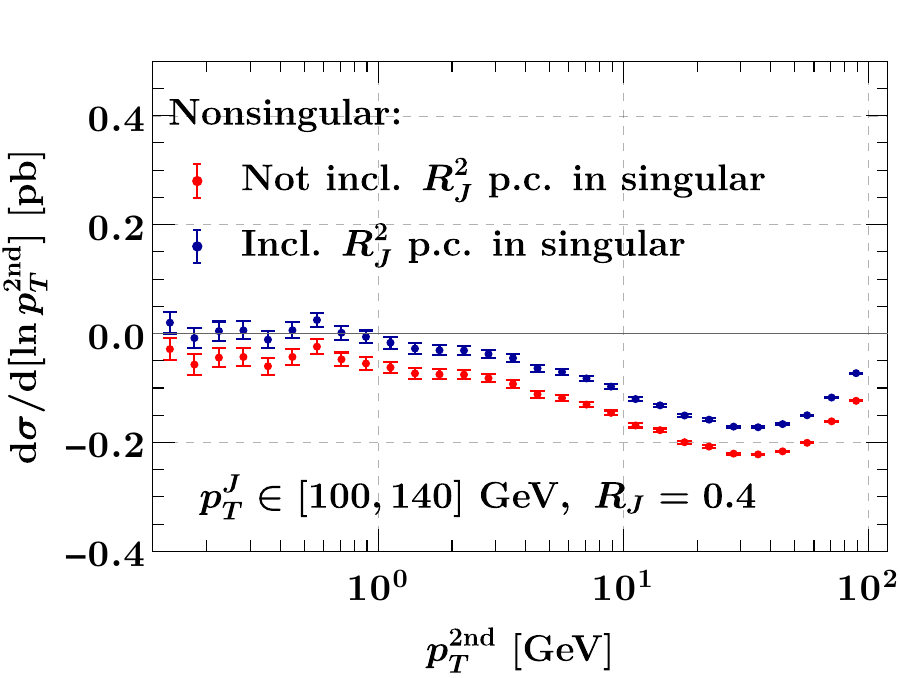}
    \includegraphics[width=0.32\textwidth]{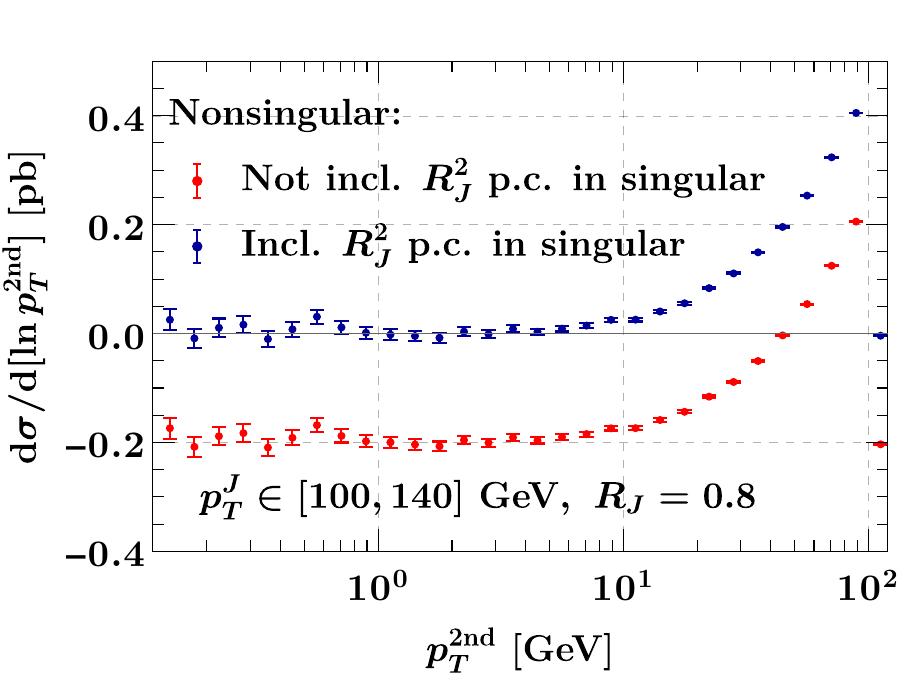}
    \caption{Nonsingular with (blue) and without (red)  $\Rj^2 \ln \ptv$ power corrections for $\Rj=0.2$ (left), 0.4 (middle) and (right), using log-log axes (top) and log-linear axes (bottom).}
    \label{fig:SNSgood}
\end{figure}

The factorization of the soft sector into the global soft $S_{\kappa}(\ptv; \mu, \nu)$ and the soft-collinear $\SC(\ptv \Rj; \mu)$ functions relies on the limit $\Rj\ll 1$ and is valid up to $\mathcal{O}(\Rj^2)$ terms, which are precisely the $\Rj^2$ power corrections we wish to retain. To compute them we therefore consider the total soft function in \eq{ST} including $\Rj^2$ corrections.
The full calculation is laid out in \app{R2-corrections}, and the NLO result is
\begin{align} \label{eq:soft-sector-1loop}
S^{T, \, (1)}_{\kappa}(\ptveto,\Rj,\mu) &=  (C_{a}+C_{b})\bigg[-4\ln^2\Bigl(\frac{\mu}{\ptveto}\Bigr)-8\ln\Bigl(\frac{\mu}{\ptveto}\Bigr)\ln\Bigl(\frac{\nu}{\mu}\Bigr)-\frac{\pi^2}{6}  \\
&\quad \color{red}+ 2 \Rj^2 \ln \bigg(\f{\mu}{\ptv R_J} \bigg) + \Rj^2 \color{black} \bigg] + 8 y_J(C_{a}-C_{b})\ln\Bigl(\frac{\mu}{\ptveto}\Bigr) \nn \\
&\quad + C_{j}\biggl[4\ln^2\Bigl(\frac{\mu}{\ptveto}\Bigr) - 4\ln^2\Bigl(\frac{\mu}{\ptveto \Rj}\Bigr) \color{red} - \Rj^2 \ln \bigg(\f{\mu}{\ptv \Rj} \bigg) + \f{\Rj^2}{6}\color{black} \biggr].
\nn\end{align}
Note that besides the $\Rj^2$ power corrections (highlighted in red), the above expression also contains the full analytic dependence in $\ln \Rj$, which was missing in ref.~\cite{Liu:2013hba}. The effects of including the $\Rj^2$ power corrections are displayed in \fig{SNSgood}. As is clear, including the power corrections remove the constant offset at small $\ptveto$ that was visible in \fig{SNS-bad}. 

Given that the power corrections in the soft sector are accompanied by logarithms of $\mu/(\ptv \Rj)$, it makes sense to absorb them into the soft-collinear function by making the replacements
\begin{align} \label{eq:SC_R2}
\gamma_{\SCs \,0}^{j} \to \tilde{\gamma}_{\SCs \,0}^{\kappa}&=\gamma_{\SCs \,0}^{j} + \Rj^2\,(2C_{a}+2C_{b}-C_{j}) \, , \nn \\
s^{R, (1)}_j \to \tilde{s}^{R, (1)}_j &=s^{R, (1)}_j + \Big(C_{a} +C_{b}+ \f{1}{6}C_{j}\Big)\Rj^2
\, .\end{align}
The inclusion of $\Rj^2$ power corrections in the soft-collinear anomalous dimension introduces a scale dependence that must be cancelled elsewhere to maintain RGE consistency. Since the only other function that exhibits $\Rj$ dependence is the jet function, we also make the replacement
\begin{align} \label{eq:gammaJ_R2}
\gamma_{J \,0}^{j} \to \tilde{\gamma}_{J \,0}^{\kappa}&=\gamma_{J \,0}^{j} - \Rj^2\,(2C_{a}+2C_{b}-C_{j})
\,.\end{align}
We stress that although we are able to make these replacements to
absorb $\Rj^2$ terms into the leading-power (with respect to $\Rj$) resummation, we do not claim that this necessarily correctly resums the $\ln(\ptv/\ptj)$ logarithms at the subleading $\ord{\Rj^2}$ at either NLL or NLL$'$.

\subsection{Resummation}
\label{sec:resummation}

The ingredients necessary to obtain a given resummed accuracy are shown in \tab{orders}. To reach NNLL$'$, we require the 3-loop cusp anomalous dimension and the 2-loop noncusp anomalous dimensions and 2-loop boundary conditions for the various functions.

\begin{table}
\centering
\renewcommand{\tabcolsep}{2ex}
\begin{tabular}{l | c c c | c}
\hline\hline
& Boundary & \multicolumn{2}{c|}{Anomalous dimensions} & FO matching
\\
Order & conditions & $\gamma_i$ (noncusp) & $\Gamma_\mathrm{cusp}$, $\beta$ & (nonsingular)
\\ \hline\hline
LL             & $1$  & - & 1-loop & -
\\
NLL            & $1$  & 1-loop & 2-loop & -
\\[0.25ex]
\hline
NLL$'$ $(+$NLO$)$  & $\alpha_s$ & 1-loop & 2-loop & $\alpha_s$ \\
NNLL $(+$NLO$)$ & $ \alpha_s$ & 2-loop & 3-loop & $\alpha_s$
\\[0.25ex]
\hline
NNLL$'$ $(+$NNLO$)$ & $\alpha_s^2$ & 2-loop & 3-loop & $\alpha_s^2$ \\
N$^3$LL $(+$NNLO$)$ & $\alpha_s^2$ & 3-loop & 4-loop & $\alpha_s^2$
\\\hline\hline
\end{tabular}
\caption{Definition of resummation orders.
The $(+$N$^n$LO$)$ in the order refers to whether or not the nonsingular
$\ord{\alpha_s^n}$ corrections in the last column are included.}
\label{tab:orders}
\end{table}

\subsubsection{Anomalous dimensions from consistency relations}
\label{sec:consistency}

Some of the perturbative ingredients at $\mathcal{O}(\alpha_s^2)$ that are needed
at NNLL$'$ remain, at present, unknown. Specifically, these are the 2-loop
noncusp anomalous dimensions and constant terms of the global soft, soft-collinear and gluon jet function.
We can deduce most of the missing anomalous dimensions by demanding RGE consistency, which implies that
\begin{align}
  \gamma_H^{\kappa} + \gamma_{S}^{\kappa} + \gamma_{\mathcal{S}}^{j} + \gamma_{J}^{j} + \gamma_{B}^{a} + \gamma_{B}^{b} = 0.
\end{align}
However, no such constraints hold for the unknown 2-loop constant terms. The remaining
independent unknowns will be treated as theory nuisance parameters, which we vary in order to estimate the uncertainties due to these missing higher order corrections, as discussed in \sec{TNP}.

We first ignore the factorization of the total soft function
into global and soft-collinear parts. The consistency relations must hold separately
for cusp and noncusp parts, where the latter are in particular free of any
kinematic dependence.
Considering the $qg\to Hq$ channel, we have
\begin{align}
\label{eq:qgqconsistency}
    \gamma_{S,T}^{qgq}+\gamma_B^q + \gamma_B^g + \gamma_H^{qgq}+\gamma_J^q=0\,,
\end{align}
where the total soft anomalous dimension $\gamma_{S,T}$ is given by
\begin{align}
  \gamma_{S,T}^{qgq} = \gamma_{S}^{qgq} + \gamma_{\mathcal{S}}^{q}\,.
\end{align}
The two-loop quark jet anomalous dimension $\gamma_{J\,1}^{q}$ is known~\cite{Liu:2021xzi},
as are the relevant hard and beam anomalous dimensions.
Using \eqref{eq:qgqconsistency}, we are then able to deduce $\gamma_{S,T\, 1}^{qgq}$. 

Together with the fact that $\gamma_{S,T\,1}^{qgq}=\gamma_{S,T\,1}^{q\bar{q}g}$, we can then extract $\gamma_{J\,1}^{g}$ from the $q \bar q \to Hg$ channel
\begin{align}
    \gamma_J^g=-\bigl(\gamma_{S,T}^{q\bar{q}g}+2\gamma_B^q + \gamma_H^{q\bar{q}g}\bigr)\,.
\end{align}
By considering $gg \to Hg$, we can then also obtain the total soft noncusp anomalous dimension for the $ggg$ channel,
\begin{align}
    \gamma_{S,T}^{ggg}= -\bigl(2\gamma_B^g + \gamma_H^{ggg}+\gamma_J^g\bigr)\,.
\end{align}

We are thus able to determine all anomalous dimensions at 2-loop order so far.
However, considering the further factorization of the total soft function into global soft and soft-collinear functions, we do not have sufficient constraints to determine how the total
soft noncusp anomalous dimension is split into the global soft and soft-collinear ones,
for which we resort to using a theory nuisance parameter as discussed in \sec{TNP}.

\subsubsection{Renormalization group evolution}
\label{sec:RGE}

We minimize the size of the logarithms associated with each of the hard, soft, soft-collinear, beam and jet functions by evaluating them at their canonical scales
\begin{align} \label{eq:canonical}
\mu_H &= \sqrt{(p_T^H)^2+m_H^2}, \qquad \mu_J = \ptj \Rj, \qquad \mu_\mathcal{S}= \ptv \Rj, \qquad \mu_S = \mu_{B} = \ptv,
\nn \\[1ex]
 \nu_{S} &= \ptv, \qquad \nu_{B \, a,b}= \omega_{a,b}\,.
\end{align}
We then RG-evolve the individual functions to common $\mu$ and $\nu$ scales via the following solutions to their RGEs in eqs.~\eqref{eq:hardRGE}, \eqref{eq:jetRGE}, \eqref{eq:BeamRGEs}, \eqref{eq:globalRGEs} and \eqref{eq:softcollinearRGE}. The solution to the virtuality evolution equations can be written as $F(\mu) = F(\mu_0)\, U_F(\mu_0,\mu)$, where the evolution kernels $U_F$ are given by
\begin{align}
U_J^{j}(\ptj\Rj; \mu_0, \mu)
&= \exp\Bigl[2 K^{j}_\Gamma(\mu_0, \mu) + K^{j}_{\gamma_J}(\mu_0, \mu) \Bigr]\, \Bigl ( \f{\mu_0}{\ptj \Rj} \Bigr)^{2 \eta^{j}_\Gamma(\mu_0, \mu)}
\,, \nn \\
U^{j}_{\SC}(\ptv \Rj; \mu_0, \mu)
&= \exp\Bigl[-2 K^j_\Gamma(\mu_0, \mu) + K^{j}_{\gamma_\mathcal{S}}(\mu_0, \mu) \Bigr] \Bigl(\f{\mu_0}{\ptv \Rj} \Bigr)^{-2 \eta^{j}_\Gamma(\mu_0, \mu)}
\,, \nn \\
U^{\kappa}_{S}(\ptv, y_J;  \nu; \mu_0, \mu)
&= \exp\Bigl[2K^{a}_\Gamma(\mu_0, \mu) + 2K^{b}_\Gamma(\mu_0, \mu) + 2K^{j}_\Gamma(\mu_0, \mu) + K^{\kappa}_{\gamma_S}(\mu_0, \mu) \Bigr]
\nn \\
& \quad \times  \Bigl(\f{\mu_0}{\nu e^{-y_J}}  \Bigr)^{2 \eta^{a}_\Gamma(\mu_0, \mu)} \Bigl(\f{\mu_0}{\nu e^{y_J}} \Bigr)^{2 \eta^{b}_\Gamma(\mu_0, \mu)}  \Bigl(\f{\mu_0}{\ptv} \Bigr)^{2 \eta^{j}_\Gamma(\mu_0, \mu)}
\,, \nn \\
U_{B}^{a}(\omega; \nu;  \mu_0, \mu)
&= \exp\Bigl[ K^{a}_{\gamma_B} (\mu_0, \mu)\Bigr] \Bigl ( \f{\nu}{\omega} \Bigr)^{2 \eta^{a}_\Gamma (\mu_0, \mu)}
\,, \nn \\
U^{\kappa}_H(\{s_{ik}\}, \mu_0, \mu)
&= \exp \Bigl[2K^a_{\Gamma}(\mu_0, \mu) + 2K^b_{\Gamma}(\mu_0, \mu) + 2K^j_{\Gamma}(\mu_0, \mu)
+ K_{\gamma_H}^{\kappa}(\mu_0, \mu) \Bigr]
\nn \\ & \quad \times
\Bigl(\frac{s_{bJ}\mu_0^2}{s_{ab}s_{aJ}}\Bigr)^{\eta^{a}_{\Gamma}(\mu_0, \mu)}
\Bigl(\frac{s_{aJ}\mu_0^2}{s_{ab}s_{bJ}}\Bigr)^{\eta^{b}_{\Gamma}(\mu_0, \mu)}
\Bigl(\frac{s_{ab}\mu_0^2}{s_{aJ}s_{bJ}}\Bigr)^{\eta^{J}_{\Gamma}(\mu_0, \mu)}
\,,\end{align}
where $K^i_\Gamma$, $K^i_\gamma$ and $\eta^i_\Gamma$ are given in \app{resummation}.
We remind the reader that we also include the $\Rj^2$ power corrections by making the replacement $\gamma_{\SCs,J} \rightarrow \tilde{\gamma}_{\SCs,J}$ in the above formulas, where $\tilde{\gamma}_{\SCs,J}$ are defined in \eqs{SC_R2}{gammaJ_R2}.

The rapidity evolution equations yield solutions of the form $F(\nu) = F(\nu_0)V(\nu_0,\nu)$ with the rapidity evolution kernels
\begin{align}
V^\kappa_S (\ptv;\mu_0;\nu_0, \nu)
&= \Bigl(\f{\nu}{\nu_0}\Bigr)^{-\gamma_{\nu,B}^{a}(\ptv, \Rc; \mu_0) - \gamma_{\nu,B}^{b}(\ptv, \Rc; \mu_0)}
\,, \nn \\
V^{i}_B (\ptv; \mu_0; \nu_0, \nu)
&=\Bigl(\f{\nu}{\nu_0}\Bigr)^{\gamma_{\nu,B}^{i}(\ptv, \Rc; \mu_0)}
\,.\end{align}

\subsubsection{Profile scales}
\label{sec:scales}

As we transition from the resummation region, where $\ptv \ll Q$, to the fixed-order region, where $\ptveto \sim Q$, the resummation needs to be turned off. This is accomplished by arranging a transition of the canonical scales in \eq{canonical} to a common fixed-order scale $\mufo$ using profile scales~\cite{Ligeti:2008ac,Abbate:2010xh}. Because the fixed-order scales can be varied, we define the central value as
\begin{equation} \label{eq:muFO_central}
\mufo^c = \sqrt{(p_T^H)^2+m_H^2} \, ,
\end{equation}
such that
\begin{equation}
\mufo = \mufo^c 2^{V_{\rm FO}} \, ,
\label{eq:scale_FO}
\end{equation}
where $V_{\rm FO} \in \{0,1,-1\}$ and implements the usual factor of two variations (we discuss the full set of scale variations in \sec{TNP}). The scales entering in the resummed cross section then take the form
\begin{align}
\mu_H &= \mu_{\rm FO} \, , \label{eq:scaleForm_H}\\
\mu_X &= \mu_{\rm FO} \, f_{\rm run}\bigg(\xi, \frac{\mu_X^{\rm canon}}{\mu_{\rm FO}^c}\bigg)f_{\rm vary}(\xi)^{V_X}\, , \label{eq:scaleForm_mu} \\
\nu_Y &= \mu_{\rm FO} \, f_{\rm run}\bigg(\xi, \frac{\nu_Y^{\rm canon}}{\mu_{\rm FO}^c}\bigg)f_{\rm vary}(\xi)^{W_Y} \, ,
\label{eq:scaleForm_nu}
\end{align}
where $X\in\{J,S,\mathcal{S},B\}$, $Y\in\{S,B\}$, the scales $\mu_X^{\rm canon}, \nu_X^{\rm canon}$ are given in~\eq{canonical} and the variation factors $V_X,W_Y\in\{0,1,-1\}$. The function $f_{\rm vary}$ is used to turn off the resummation scale variations gradually as we enter the fixed-order region of phase space.
The argument $\xi$ of the profile functions is a dimensionless function of $\ptv$ and the hard kinematics which measures how far along the transition a given phase space point lies. The $\xi$ dependence of the profile function $f_{\mathrm{run}}(\xi,y)$ factorizes as
\begin{align}
f_\text{run}(\xi, y)=   g_\text{run}(\xi) \, y +  [1-g_\text{run}(\xi)] \, ,
\end{align}
where
\begin{align} 
g_\text{run}(x) &= 
\begin{cases}
1 & 0 < x \leq x_1 \,, \\
1- \frac{(x-x_1)^2}{(x_2-x_1)(x_3-x_1)} & x_1 < x \leq x_2
\,, \\
\frac{(x-x_3)^2}{(x_3-x_1)(x_3-x_2)} & x_2 < x \leq x_3
\,, \\
0 & x_3 \leq x
\,,
\end{cases}
\label{eq:grun}
\end{align}
enabling a smooth transition of the scales in \eq{canonical} to the common scale $\mu_{\rm FO}$ in the fixed-order regime. 

\begin{figure}[t]
    \centering
    \begin{subfigure}{0.48\textwidth}
        \centering
        \includegraphics[width=\textwidth]{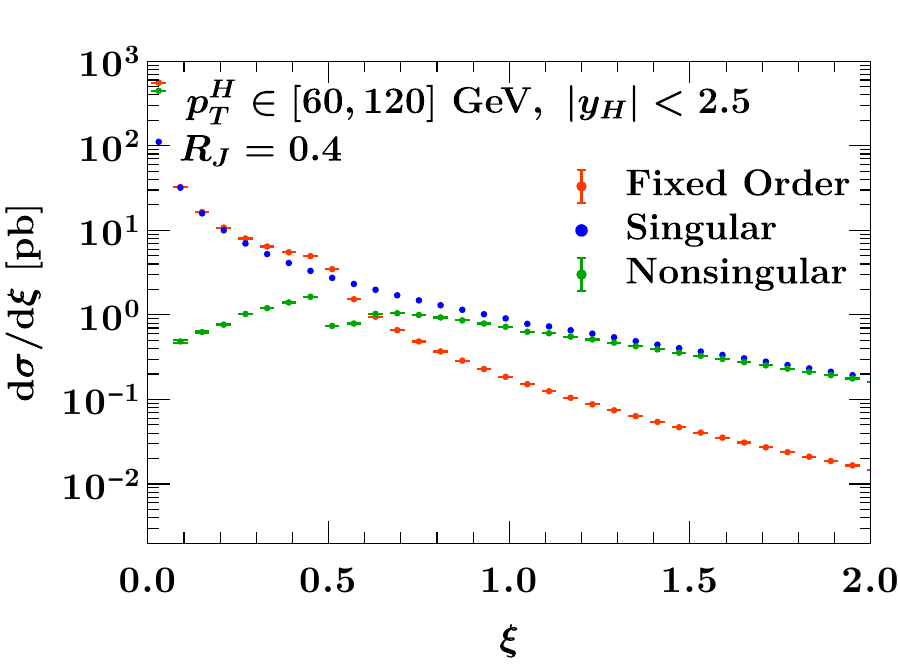}
    \end{subfigure}
    \hfill
    \begin{subfigure}{0.48\textwidth}
        \centering
        \includegraphics[width=\textwidth]{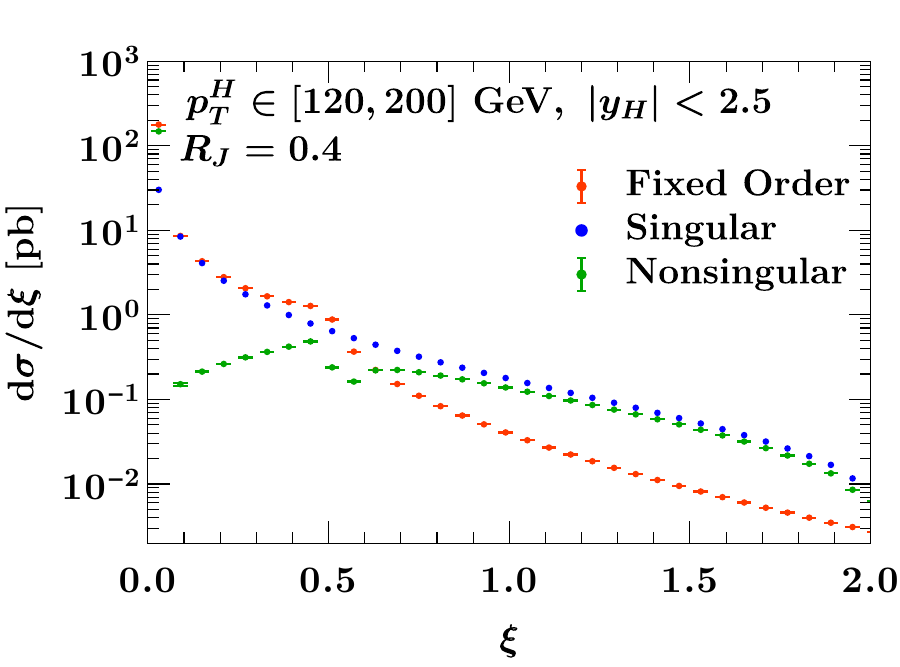}
    \end{subfigure}
    \hfill
    \caption{Decomposition of the full cross section (red) in terms of the singular (blue) and nonsingular (green) as function of $\xi=\ptv/p_T^H$. Our transition points are chosen on the basis of this plot. By using $\xi$, this is fairly independent of the hard kinematics, as can be seen by comparing the left ($p_T^H \in [60, 120]$ GeV) and right ($p_T^H \in [120, 200]$ GeV) panel.  \label{fig:transition}}
\end{figure}

For a $2\to 1$ process such as $gg\to H$, there is limited freedom in how one defines the transition parameter $\xi$ due to the simple kinematic dependence of the hard scattering. For a $2\to 2$ process, however, more possibilities arise -- for example, far from the singular limit $\ptj\neq p_T^H$ and $\xi$ could depend on either quantity (or conceivably both).
Given that we are interested in making predictions for $p_T^H$, we choose
\begin{equation}
\xi= \frac{\ptv}{p_T^H} \, .
\end{equation}
We have verified that this choice ensures the points of transition are roughly independent of the phase space bin being considered. This is illustrated in \fig{transition} for two different bins of the hard scattering kinematics.
We also use this plot to select the values of $x_1, x_2$,  and $x_3$ which appear in \eq{grun} and which are used to control precisely where and how quickly the scale merging occurs. We select the values $\{x_1, x_2, x_3\} = \{0.15, 0.4, 0.65\}$, which coincides with the choices made in ref.~\cite{Stewart:2013faa}.
%

\subsubsection{Fixed-order matching}

Our NLL$'$ resummed predictions for the Higgs $p_T^H$ spectrum in $H+$1-jet production are matched to fixed-order calculations obtained from \textsc{Geneva}~\cite{Alioli:2023har} at NLO$_1$ accuracy (where NLO$_i$ denotes a next-to-leading order calculation with $i$ partons in the final state). To match our approximate NNLL$'$ results consistently, we require a corresponding NNLO$_1$ calculation. Unfortunately, however, no publicly available code capable of producing such a prediction currently exists. We therefore reconstruct the NNLO$_1$ inclusive cross section as a function of $p_T^H$ by taking the NLO$_1$ prediction from \textsc{Geneva} with $\ptv\rightarrow\infty$ (retaining all rapidity cuts) and reweighting by a $K$-factor which is determined via examination of the results for the Higgs $p_T^H$ spectrum at NNLO$_1$ shown in~\refscite{Chen:2016zka, Becker:2020rjp}. This is viable since with the fixed-order
scale choice in \eq{muFO_central} (and also used in \refscite{Chen:2016zka, Becker:2020rjp}) the NNLO$_1$ correction is
approximately flat as a function of $p_T^H$ in our region of interest. While the results in \refscite{Chen:2016zka, Becker:2020rjp} do not include a cut on the Higgs rapidity,
we have verified that the effect of placing a rapidity cut $|y_H|<2.5$ has an almost identical effect on $p_T^H$ at LO$_1$ and NLO$_1$. This allows us to use the NNLO$_1$ $K$-factor obtained from  results which do not have this
additional rapidity cut. We obtain a central value of $K_{\rm NNLO_{1}} = 1.25$.

From this inclusive NNLO$_1$ result we then subtract the  contributions from the region of phase space where the transverse momentum of the second jet, $\ptvd$, is above our desired cut-off $\ptv$,
\begin{equation}
\frac{\df \sigma^{\rm NNLO}_{Hj}}{\df p_T^H}\bigg|_{\ptvd<\ptv}
= K_{\rm NNLO_1} \frac{\df \sigma^{\rm NLO}_{Hj}}{\df p_T^H}\bigg|_{\ptv \rightarrow \infty} - \frac{\df \sigma^{\rm NLO}_{Hjj}}{\df p_T^H}\bigg|_{\ptvd>\ptv} \, .
\label{eq:nnlo_construction}
\end{equation}
For distributions differential in $p_T^H$, this subtraction term (the NLO$_2$ cross section with $\ptvd > \ptv$) can be obtained directly from  \textsc{Geneva} by setting appropriate cuts on the second jet. 
This method allows us to reconstruct the NNLO$_1$ result down to values of $\ptv > 10$~GeV.

\subsection{Perturbative uncertainties and theory nuisance parameters}
\label{sec:TNP}

The theoretical uncertainties associated with our calculation have three distinct origins. The first relates to the fact that, at the order in resummed perturbation theory to which we are working, we have incomplete knowledge of the perturbative ingredients. We deal with this by introducing theory nuisance parameters~\cite{Tackmann:2024kci}, for which a central value is estimated and variations performed to assess the associated uncertainty due to the missing ingredient.
The second source of uncertainty arises from the freedom we have in choosing how the resummation is turned off in transitioning to the fixed-order region of phase space. This ``matching'' uncertainty is associated with our choice of the parameters $x_1, x_2$, and $x_3$ in \eq{grun}.
The third, more usual source is the lack of knowledge of terms which are of higher order in perturbation theory than the order to which we work and arise because of the truncation of the perturbative series in the first place. We first discuss how we quantify the latter uncertainties by varying the
profile scales before moving on to provide a more detailed description of uncertainties from
theory nuisance parameters.
%

\subsubsection{Profile scale variations}

As is usual, the contribution from missing higher order terms is estimated by a scale variation procedure. By independently varying the exact form of the profile functions $\mu_B,\,\mu_S,\,\mu_J,\,\mu_{\mathcal{S}},\,\nu_S,\,\nu_B$  in \eqs{scaleForm_mu}{scaleForm_nu} while keeping the scale $\mu_H$ fixed, we alter the arguments of the logarithms being resummed and hence obtain a resummation uncertainty. This probes the intrinsic uncertainty in the resummed logarithmic series. It is important, however, that the uncertainties generated from these scale variations are gradually turned off in the same manner as the resummation as one enters the fixed-order region of phase space.

The profile scale variations and how they are shut off are controlled by the parameters $V_X,W_Y \in \{0,1,-1\}$ and $f_{\rm vary}(\xi)$ in \eqs{scaleForm_mu}{scaleForm_nu} where
\begin{align} 
f_\text{vary}(x) &= \begin{cases}
2\left(1-\frac{x^2}{x_3^2}\right) & 0 \leq x < \frac{x_3}{2} \,, \\
1 + 2\left(1-\frac{x}{x_3}\right)^2 & \frac{x_3}{2} \leq x < x_3
\,, \\
1 & x_3 \leq x \,.
\end{cases}
\label{eq:fvary}
\end{align}
Thus $f_{\rm vary}(\xi)$ runs from $2$ in the resummation region to $1$ in the fixed-order region and acts as a multiplier of the scales around their central values, while $V_X$ and $W_Y$ control whether the variation is by a factor of $2$ or $1/2$ (or not at all). Our procedure for varying the scales follows that in Section III.C of ref.~\cite{Stewart:2013faa} and we outline it only briefly here. We highlight that varying each of the six scales $\mu_B,\,\mu_S,\,\mu_J,\,\mu_{\mathcal{S}},\,\nu_S,\,\nu_B$ independently through all possible combinations is undesirable, since it can lead to combinations which alter the arguments of logarithms by factors of four rather than two (such as the combination $\{2\nu_B,\nu_S/2\}$ for logarithms containing the ratio $\nu_B/\nu_S$). To avoid such situations we omit simultaneous scale variations which would lead to these factors of four. Furthermore, we vary the scales in the global soft and soft-collinear functions simultaneously by the same factor (though their central values are of course different). Our factorization of the total soft function into global and soft-collinear functions as outlined in \sec{soft functions} already introduces a number of nuisance parameters. Keeping their scale variations consistent avoids double counting the associated resummation uncertainty.

In summary, we impose the following conditions
\begin{align}
V_S = V_{\mathcal{S}} \, , \quad V_S V_B \geqslant 0 \, ,  \quad W_S W_B \geqslant 0 \, ,  \quad V_S W_S \geqslant 0 \, ,
\end{align}
where the last three conditions avoid generating factors of four in logarithms containing ratios of $\mu_S/\mu_B$, $\nu_S/\nu_B$, and $\mu_S/\nu_S$ respectively. This results in 122 separate variation combinations to consider. The resulting uncertainty is then given by constructing a symmetric envelope using the points which produce the largest magnitude of deviation across all variations, resulting in the resummation uncertainty $\Delta_{\rm resum}$.

In addition to these individual scale variations, we collectively vary all scales up and down. At large $\ptv$ this corresponds to the normal procedure followed in a fixed-order calculation, while at small $\ptv$ it preserves the arguments of all logarithms being resummed. We therefore refer to this as a fixed-order variation and it is achieved through the choice of $V_{\rm FO} \in \{0,1,-1\}$ in \eq{scale_FO}, which feeds into $\mu_H$ via \eq{scaleForm_H} and all other scales via \eqs{scaleForm_mu}{scaleForm_nu}. This produces only two new variations in addition to the central curve -- we again symmetrize by taking the maximum deviation to produce an uncertainty $\Delta_{\rm FO}$. When performing these variations we also consistently vary the scales entering in the nonsingular, adjusting the NNLO$_1$ K-factor appropriately. This ensures that we correctly reproduce the scale variation bands of the inclusive NNLO$_1$ calculation given in~\refscite{Chen:2016zka, Becker:2020rjp}.

The second source of uncertainty introduced at the start of this section originates from the arbitrariness in choosing the transition points $x_i$ which define the shape of the profile. Our central choice for these parameters is given by $\{x_1, x_2, x_3\} = \{0.15, 0.4, 0.65\}$ as discussed at the end of \sec{scales}, while variations from this are given by
\begin{align}
\{x_1,x_2,x_3\} &= \big\{\{0.1, 0.35, 0.6\}, \{0.1, 0.4, 0.7\},
\{0.2, 0.4, 0.6\}, \{0.2, 0.45, 0.7\}\big\} \, .
\end{align}
This corresponds to moving all transition points to larger and smaller values by $0.05$, as well as keeping the centre of the transition fixed and moving the start and end points simultaneously closer or further away from the central point again by $0.05$. The symmetrized envelope results in the matching uncertainty $\Delta_{\rm match}$.
%

\subsubsection{Theory nuisance parameters}

We now move on to discuss the parameterization of missing terms at the perturbative order to which we are working using theory nuisance parameters (TNPs)~\cite{Tackmann:2024kci}. The unknown ingredients include both two-loop anomalous dimensions and also NNLO constant (nonlogarithmic) terms. In \sec{consistency}, we obtained all anomalous dimensions from consistency relations with one exception -- the split between the two-loop noncusp anomalous dimensions for the global soft and soft-collinear functions. We manifest our ignorance by introducing a TNP
$\theta_1$ as
\begin{align}
    \gamma_{\mathcal{S}\,1}^{j}\propto C_{j}\,\theta_1
\label{eq:theta1}
\,,\end{align}
and obtain the two-loop coefficient of the global soft anomalous dimension from the difference $\gamma_{S\,1}^{\kappa}=\gamma_{S,T\,1}^{\kappa} - \gamma_{\mathcal{S}\,1}^{j}$.
We thus need only one TNP for the remaining missing anomalous dimensions.

A priori, we have a total of 6 unknown two-loop constant terms, namely
\begin{align}
    &j^{(2)}_{g},\, s^{(2)}_{ggg},\,s^{(2)}_{qgq},\,s^{(2)}_{q\bar{q}g},\,s^{R,(2)}_{g},\,s^{R,(2)}_{q}\,.
\end{align}
If we assume Casimir scaling for the soft and soft-collinear constants and also neglect any possible $\Rc$ dependence in them, this can be reduced to three unknowns. We therefore introduce three more TNPs,
\begin{align}
    j^{(2)}_{g} \propto \theta_2
    \,, \qquad
    s_{\kappa}^{(2)} \propto (C_{a}+C_{b}+C_{j})\,\theta_3
    \,, \qquad
    s^{R,(2)}_{i} \propto C_i\,\theta_4
\,,\label{eq:theta234}
\end{align}
where the proportionality symbol again reminds us that we have a certain amount of freedom in how we choose the central values (and does not necessarily imply strict proportionality). We are thus left with a total of four parameters $\theta_{1-4}$. 

It is worth noting that in principle the two-loop soft constant $s_{\kappa}^{(2)}$ may be a function of the Born kinematics of the process as well as the jet radius. One could therefore consider introducing further TNPs to parameterize these dependences to capture correlations in them. Since, however, we only consider one value of the jet radius and a single rapidity bin, this is unnecessary in our case and we are able to treat these terms as effective constants. For a further discussion see \refcite{Tackmann:2024kci}.

Having identified the role played by each nuisance parameter, we are able to make some educated guesses about the relations in \eqs{theta1}{theta234} based on the structure of each of the missing terms. For $\gamma_{\mathcal{S}}^{j}$, we pick the central value to correspond to the known two-loop noncusp anomalous dimension for the hemisphere soft function appearing in the resummation of thrust (see eq.~(3.56) of \refcite{Monni:2011gb}) and write
\begin{equation}
\gamma_{\mathcal{S}\,1}^{j} = -16 C_j\biggl[C_A \biggl(-\frac{101}{54} + \frac{11}{144}\pi^2 + \frac{7}{4}\zeta_3 \biggr) + T_F n_f\biggl(\frac{14}{27} - \frac{\pi^2}{36} \biggr)\biggr] \theta_1 \,.
\end{equation}
This can (only) be expected to give the correct typical size of $\gamma_{\mathcal{S}}^{j}$. We therefore take as the central value $\theta_1=1$ and vary it by $\pm 1$, as shown along with the other TNPs in~\tab{tnp_variations}.

For $j_g^{(2)}$ we pick the central value by Casimir scaling with respect to the known expression in the quark case, $j_q^{(2)}$,
\begin{equation} \label{eq:jg2-ansatz}
j_g^{(2)} = 4C_A \bigl(-1.78\, C_F -106.87\, C_A + 14.072\, T_F n_f \bigr) \theta_2\,.
\end{equation}
This can again be expected to give the correct typical size (but only that), which
we have checked at one loop and also for other jet functions that are known at two loops for
both quarks and gluons. We therefore take the central value to be $\theta_2=1$ and vary it by $\pm 1$.

The two remaining unknowns relate to constant terms in the NNLO soft functions. Since these are given by correlators of Wilson lines related to the external partons, we parametrize both functions by assuming each is related to products of single Wilson lines, which in particular allows us to obtain the appropriate Casimir scaling for the leading overall colour factor. Specifically, if we parametrize the contribution to the constant term in a soft function from a single Wilson line associated to parton $i$ as
\begin{equation}
Y_{i} = 1 + \frac{\alpha_s}{4\pi}C_{i} Y^{(1)} + \left(\frac{\alpha_s}{4\pi}\right)^2 C_{i} C_A Y^{(2)} + \dotsb
\,,\end{equation}
then we can model the constant terms in the global soft function as
\begin{align}
s_\kappa \approx Y_{a} Y_{b} Y_{j}
&=  1 + \frac{\alpha_s}{4\pi} \big(C_{a} + C_{b} + C_{j}\big) Y^{(1)}
\\\nonumber & \quad
+ \left(\frac{\alpha_s}{4\pi}\right)^2 \Bigl[\bigl(C_{a} C_{b} + C_{a} C_{j} + C_{b} C_{j}\bigr)\bigl(Y^{(1)}\bigr)^2 + C_A \bigl(C_{a} + C_{b} + C_{j}\bigr)Y^{(2)}\Bigr]
\,.\end{align}
The entire $\alpha_s^2$ term comprises our unknown $s_\kappa^{(2)}$ term, but constructing it in this manner gives us an expected known piece related to the square of the one-loop term $s_\kappa^{(1)}$ term. Matching the known NLO term requires that $Y^{(1)} = -\frac{\pi^2}{6}$. Substituting this in the $\mathcal{O}(\alpha_s^2)$ term we finally parametrize the unknown $s_\kappa^{(2)}$ as
\begin{equation}
s_\kappa^{(2)} = \big(C_{a} C_{b} + C_{a} C_{j} + C_{b} C_{j}\big)\frac{\pi^4}{36} + 16\,C_A \big(C_{a} + C_{b} + C_{j}\big)\theta_3
\,,\label{eq:s2-global-ansatz}
\end{equation}
where we have replaced $Y^{(2)}$ by $16\theta_3$. With this normalization convention,
the missing two-loop term can be expected to be of size $|\theta_3|\lesssim 1$~\cite{Tackmann:2024kci}.
Having no prior expectation of the sign of this term, we set the central value to
$\theta_3=0$ and vary it (conservatively) by $\pm 2$.

An analogous analysis for the missing two-loop constant for the soft-collinear function yields
\begin{equation}
s^{R,(2)}_{i} = \frac{\pi^4}{36} C_j^2 + 32\, C_A C_j \theta_4 \, ,
\label{eq:s2-col-ansatz}
\end{equation}
where as for $\theta_3$ we take $\theta_4=0 \pm 2$.

The uncertainty associated with these unknowns is quantified by varying the $\theta_i$ independently, one-at-a-time, around their central values by $\pm\Delta\theta_i$ as summarized in \tab{tnp_variations}.
\begin{table}[]
\centering
\begin{tabular}{c|ccc}
\hline\hline
           & Central & Upper & Lower \\ \hline
$\theta_1$ & $1$     & $2$   & $0$   \\
$\theta_2$ & $1$     & $2$   & $0$   \\
$\theta_3$ & $0$     & $2$   & $-2$  \\
$\theta_4$ & $0$     & $2$   & $-2$  \\ \hline\hline
\end{tabular}
\caption{Central values and variations used for the theory nuisance parameters.}
\label{tab:tnp_variations}
\end{table}
The TNPs we use are associated solely with unknown parts of the resummed calculation, while the fixed-order piece is, in principle, known exactly. It would therefore be undesirable to continue to vary them in the fixed-order region dominated by the fixed-order calculation, since this would result in an overestimate of the theoretical uncertainty. To avoid this, when performing the fixed-order matching we gradually shut off the TNP variation as a function of $\xi=\ptv/p_T^H$, using the same function $g_{\rm run}$ defined in \eq{grun},
\begin{equation}
\theta_i=\theta_i^{\rm central} \pm \Delta\theta_i \,  g_{\mathrm{run}}(\xi)
\,.
\end{equation}
In so doing, we ensure that the TNP variation is switched off in exactly the same way as the resummation itself. The uncertainty due to each TNP is again taken as the maximum absolute deviation between the up and down variations. Since each TNP parametrizes an uncorrelated source of uncertainty, their resulting individual uncertainties are added in quadrature to give a combined uncertainty estimate $\Delta_{\rm TNP}$.

\begin{figure}[t]
    \centering
    \begin{subfigure}{0.48\textwidth}
         \centering
         \includegraphics[width=\textwidth]{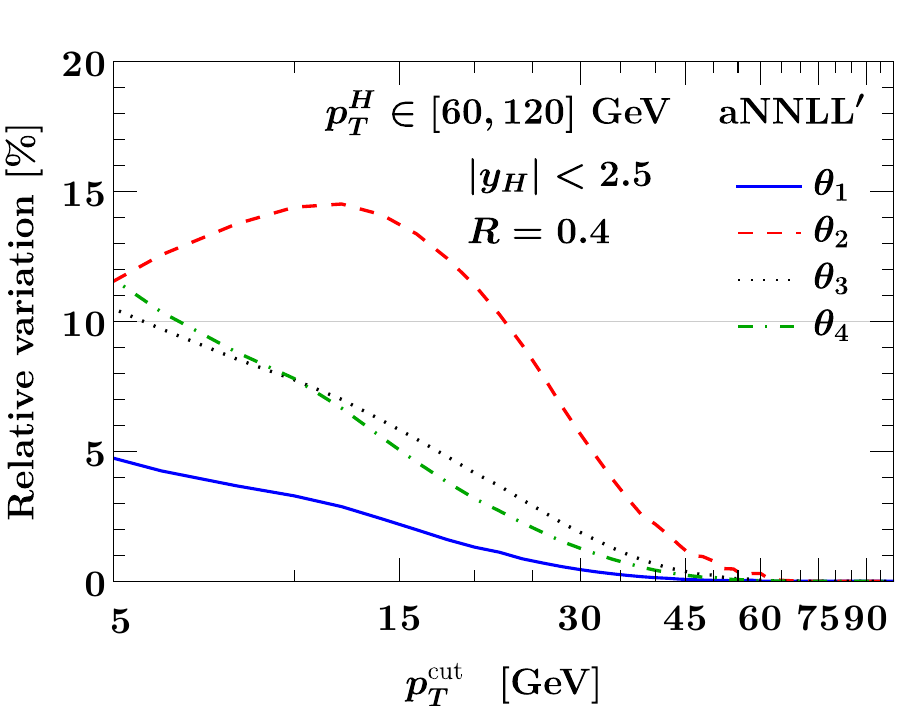}
     \end{subfigure}
     \hfill
     \begin{subfigure}{0.48\textwidth}
         \centering
         \includegraphics[width=\textwidth]{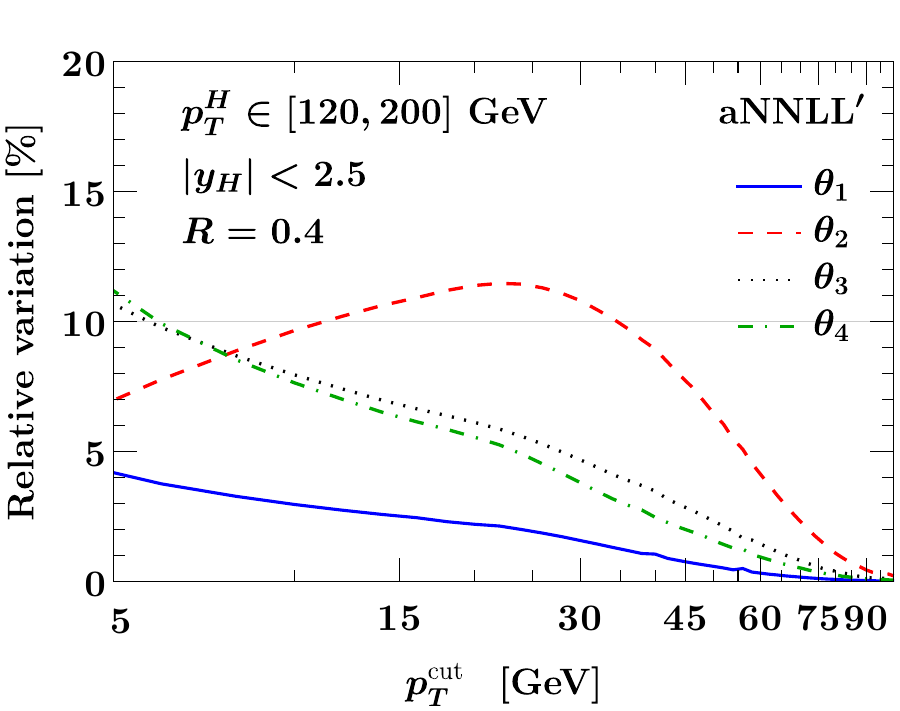}
     \end{subfigure}
    \hfill
\caption{Relative uncertainty from varying each theory nuisance parameter as a
function of $\ptv$ for two different STXS bins.}
\label{fig:impact-cumulant-nuisance}
\end{figure}

\begin{figure}[t]
    \centering
    \includegraphics[width=0.48\textwidth]{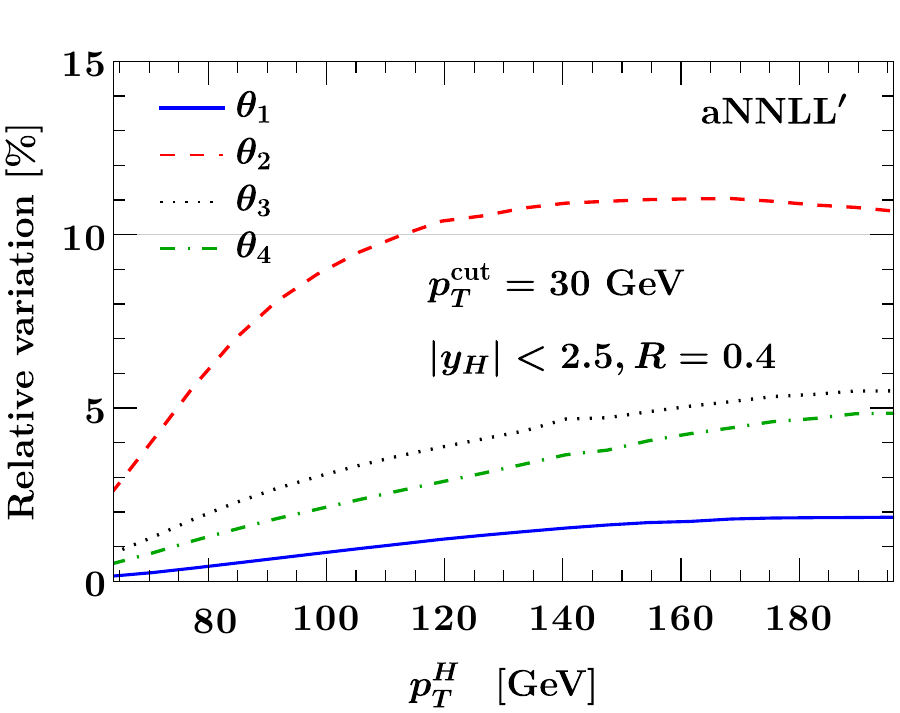}
    \caption{Relative uncertainty from varying each theory nuisance parameter for
    the $p_T^H$ spectrum with a jet veto at $\ptv = 30\GeV$.}
    \label{fig:impact-diffpTH-nuisance}
\end{figure}
In \figs{impact-cumulant-nuisance}{impact-diffpTH-nuisance} we show the relative impact due to the variation of each TNP for the $\ptv$ cumulant and $p_T^H$ distributions. It is clear that $\theta_2$ (associated with the unknown two-loop constant term in the gluon jet function) has the biggest impact, though $\theta_3$ and $\theta_4$ (which are associated with the missing two-loop constant terms in the global soft and soft-collinear functions) give increasing contributions as $\ptv$ decreases. The uncertainties associated with $\theta_3$ and $\theta_4$, while still less than $\theta_2$, grow as a function of $p_T^H$ for fixed $\ptv$.

One possible reason for the relatively large uncertainty generated by $\theta_2$ compared to say $\theta_3$ or $\theta_4$, which also represent unknown two-loop constant contributions, is the very large numerical value of $j_q^{(2)}$ on which our estimate of $j_g^{(2)}$ is based. The expression in \eq{jg2-ansatz} evaluates to
\begin{equation}
j_g^{(2)} = -3453.64\,\theta_2
\,.\end{equation}
The expressions for $s_\kappa^{(2)}$ and $s_i^{R,(2)}$ in \eqs{s2-global-ansatz}{s2-col-ansatz}
evaluate in the gluon-gluon channel to
\begin{equation}
s_{ggg}^{(2)}
= 73.06 + 432\, \theta_3
\quad \mathrm{and} \quad
s_g^{R,(2)}= 24.35 + 288\,\theta_4
\,.\end{equation}
Although $j_g^{(2)}$ only enters in contributions with a final state gluon and the soft terms will enter in all channels, the large numerical hierarchy between them for the gluon-gluon channel could already point to a reason for the significant difference in the uncertainty generated through variation of the associated TNPs.

The various sources of uncertainty $\Delta_i$ discussed in this section are combined in quadrature to obtain the final uncertainty estimate,
\begin{equation}
\Delta_{\rm total} = \sqrt{\Delta_{\rm FO}^2 + \Delta_{\rm resum}^2 + \Delta_{\rm match}^2 +\Delta_{\rm TNP}^2} \, .
\label{eq:total_unc}
\end{equation}
The uncertainty bands of a given distribution are then generated by $\df \sigma(x)\pm \Delta_{\rm total}(x)$ for $x\in \{\ptv,p_T^H\}$ depending on the distribution under consideration and $\df \sigma(x)$ represents the central curve. All our results are shown with uncertainties generated in this manner (where $\Delta_{\rm TNP}=0$ for results to NLL and NLL$'$ accuracy).

\section{Numerical results}
\label{sec:results}
In this section we present our numerical results. In the first set of plots we study the dependence on the jet veto for the two STXS 1-jet bins with $p_T^H \in [60,120]$ GeV and $p_T^H \in [120,200]$ GeV.
In the second set of plots we explore the dependence on $p_T^H$ for a fixed value of the jet veto $\ptv = 30$ GeV. We always include a cut on the Higgs rapidity $|y_H| \leq 2.5$. For the jet radius we always use $\Rj = \Rc = R = 0.4$. We always work at $E_{\rm cm} = 13\TeV$ using the \texttt{MSHT20nnlo} PDFs with $\alpha_s(m_Z) = 0.118$~\cite{Bailey:2020ooq}.

\begin{figure} [t]
     \centering
     \begin{subfigure}{0.48\textwidth}
         \centering
         \includegraphics[width=\textwidth]{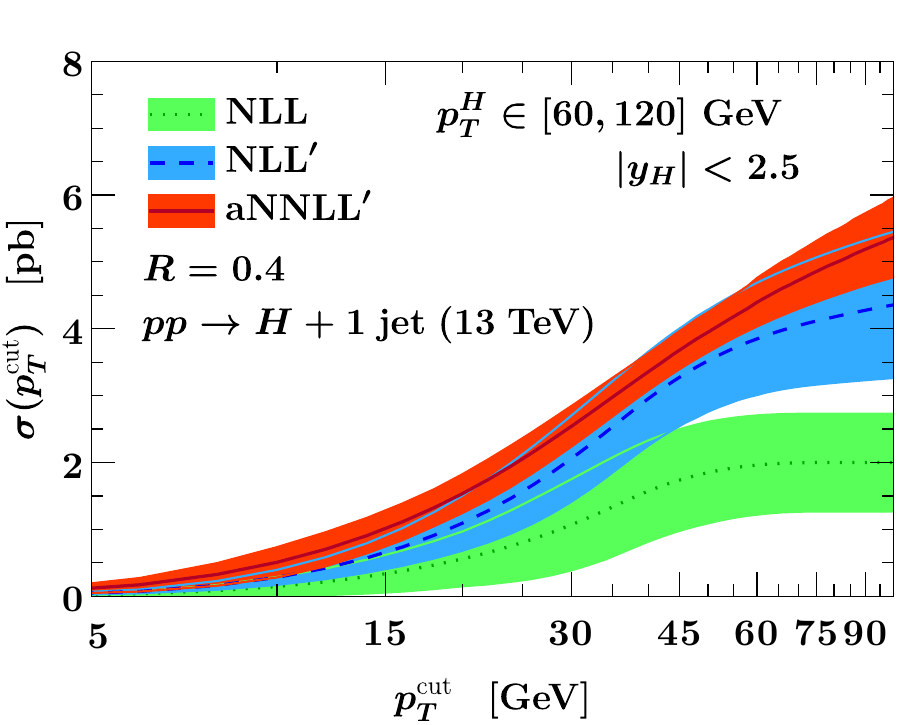}
     \end{subfigure}
     \hfill
     \begin{subfigure}{0.48\textwidth}
         \centering
         \includegraphics[width=\textwidth]{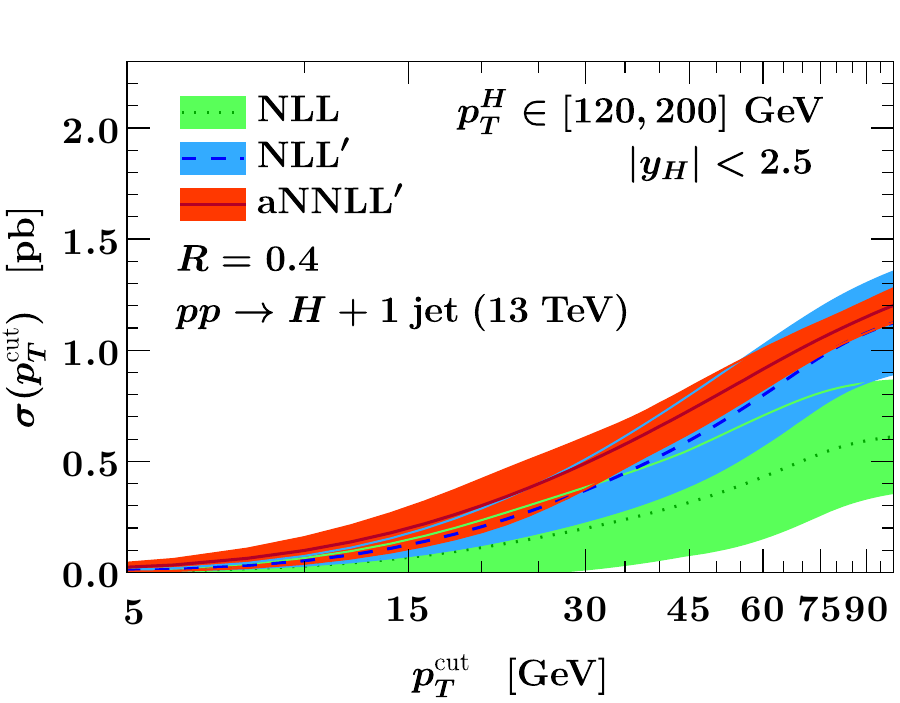}
     \end{subfigure}
    \hfill
\caption{Convergence of resummed perturbation theory as function of $\ptv$, comparing NLL (green dotted), NLL$'$ (blue dashed) and aNNLL$'$ (red solid) for two different STXS bins. The uncertainty bands are obtained as described in \sec{TNP}.}
\label{fig:resum_convergence}
\end{figure}

In \fig{resum_convergence} we show the convergence of the pure resummed results, comparing NLL, NLL$'$ and aNNLL$'$ for the two STXS bins. In the resummation region (small $\ptv$), the uncertainty bands overlap well and decrease in size at higher orders, indicating that resummed perturbation theory is well behaved. Despite the fact that not all of the resummation ingredients at NNLL$'$ are known, the approximate aNNLL$'$ results still provide a substantial improvement compared to the previous NLL$'$ order. For the STXS bin with small $p_T^H$ (left panel) there is a gap between the NLL band and the others for large values of $\ptv$. This is of no concern, as this is the fixed-order region where the matching to fixed order becomes necessary.

\begin{figure} [t]
     \centering
     \begin{subfigure}{0.48\textwidth}
         \centering
         \includegraphics[width=\textwidth]{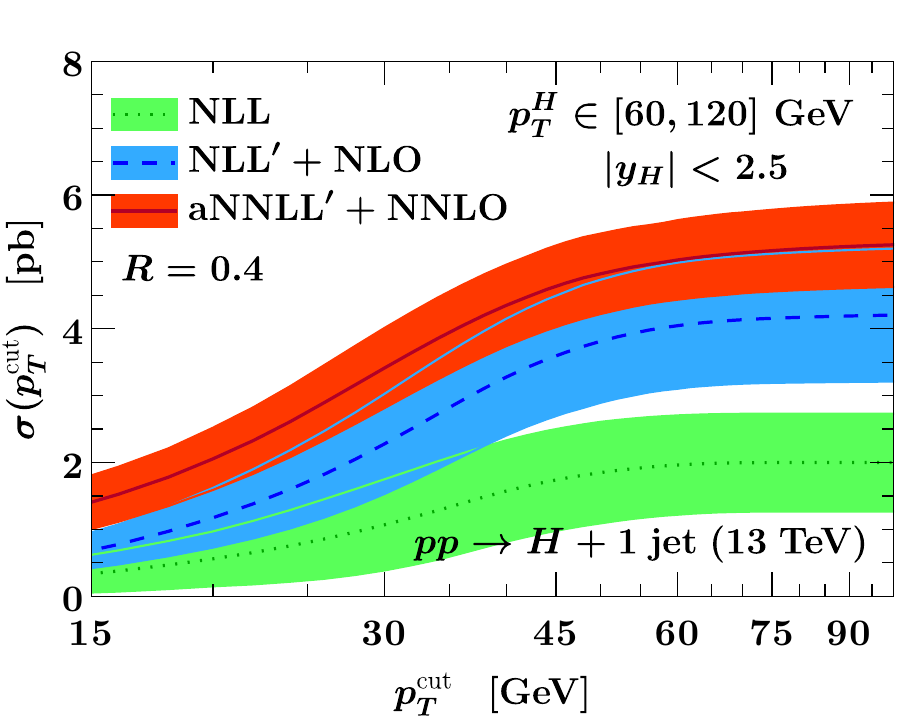}
     \end{subfigure}
     \hfill
     \begin{subfigure}{0.48\textwidth}
         \centering
         \includegraphics[width=\textwidth]{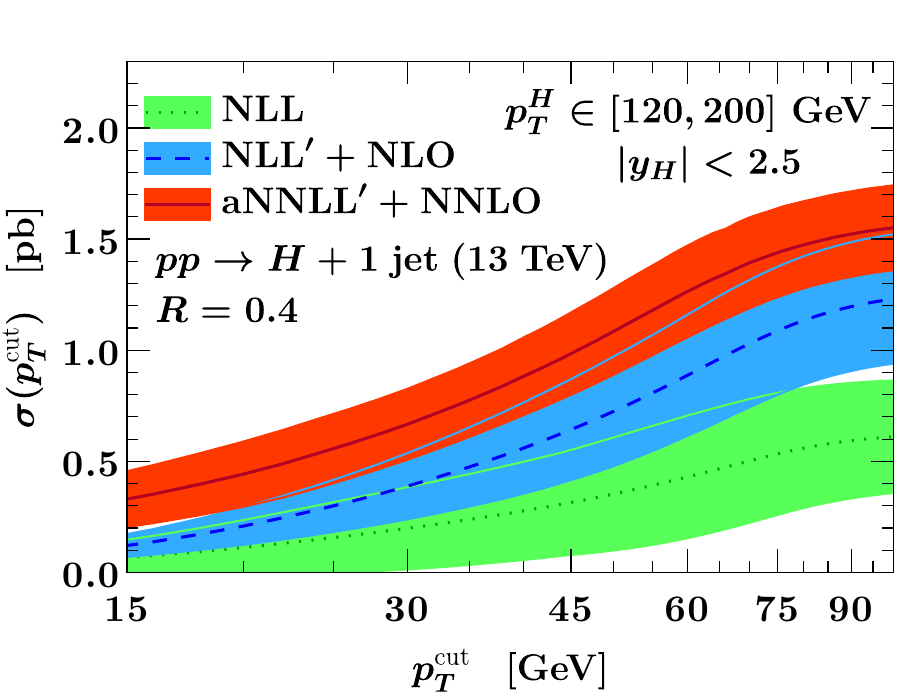}
     \end{subfigure}
    \hfill
\caption{Same as \fig{resum_convergence} but including the matching to fixed-order perturbation theory.} \label{fig:match_convergence}
\end{figure}

This brings us to \fig{match_convergence}, where compared to \fig{resum_convergence} the fixed-order matching is included. Including the nonsingular contribution increases the uncertainty band of the cross section. Nonetheless, the relative uncertainty at aNNLL$'$+NNLO compared to NLL$'$+NLO still shows a noticeable reduction. Explicit numerical results for two representative $\ptv$ values are given in \tab{numeric_vals}. The gap between the NLL and other orders for large values of $\ptv$ seen in the left panel of \fig{resum_convergence} becomes worse when the matching is included. This is again not a feature of our resummed calculation, but rather of the underlying inclusive Higgs+jet fixed-order cross section, for which it is known that the scale variations do not cover the increase from LO to NLO (see e.g. \refscite{Chen:2016zka, Becker:2020rjp}).

\begin{figure}[t]
     \centering
     \begin{subfigure}{0.48\textwidth}
         \centering
         \includegraphics[width=\textwidth]{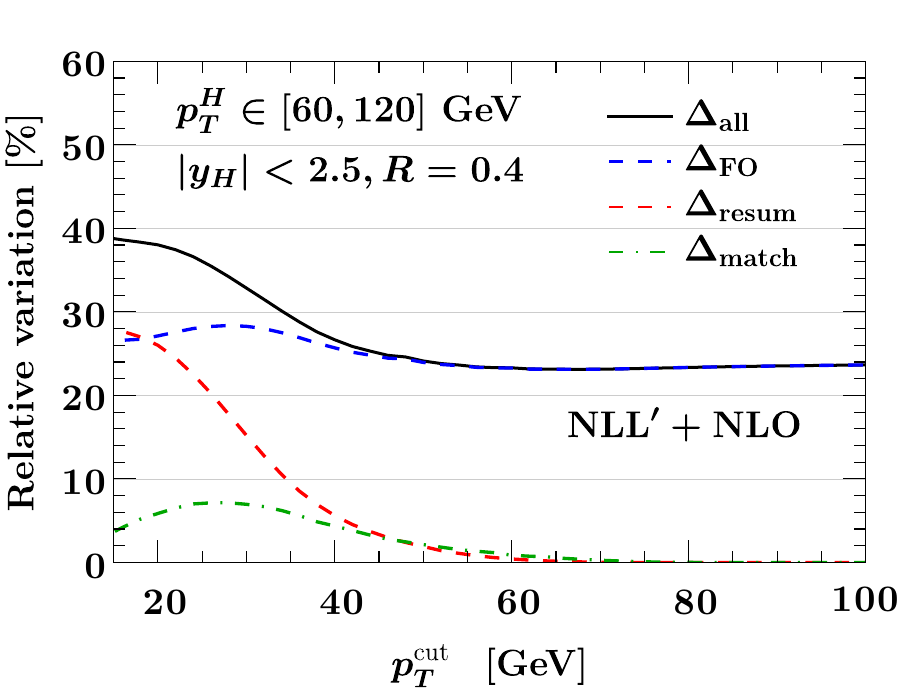}
     \end{subfigure}
     \hfill
     \begin{subfigure}{0.48\textwidth}
         \centering
         \includegraphics[width=\textwidth]{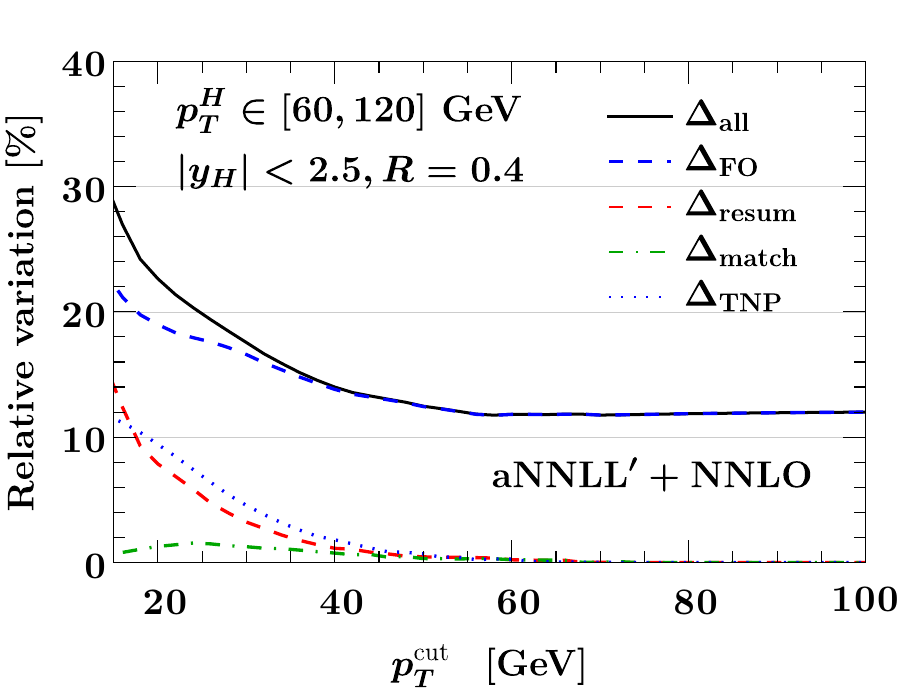}
     \end{subfigure}
    \hfill
    \begin{subfigure}{0.48\textwidth}
         \centering
         \includegraphics[width=\textwidth]{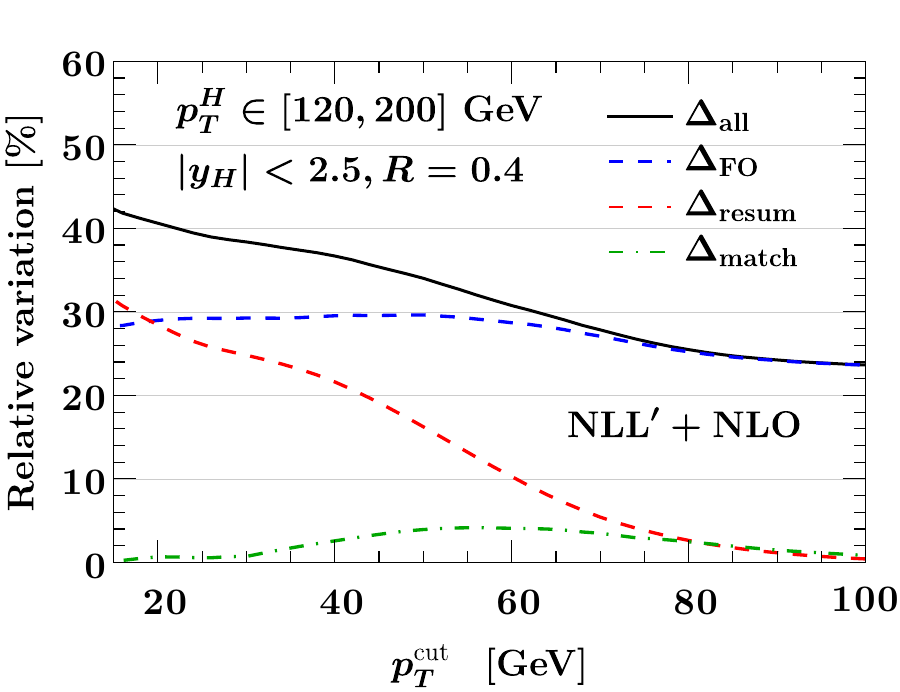}
     \end{subfigure}
     \hfill
     \begin{subfigure}{0.48\textwidth}
         \centering
         \includegraphics[width=\textwidth]{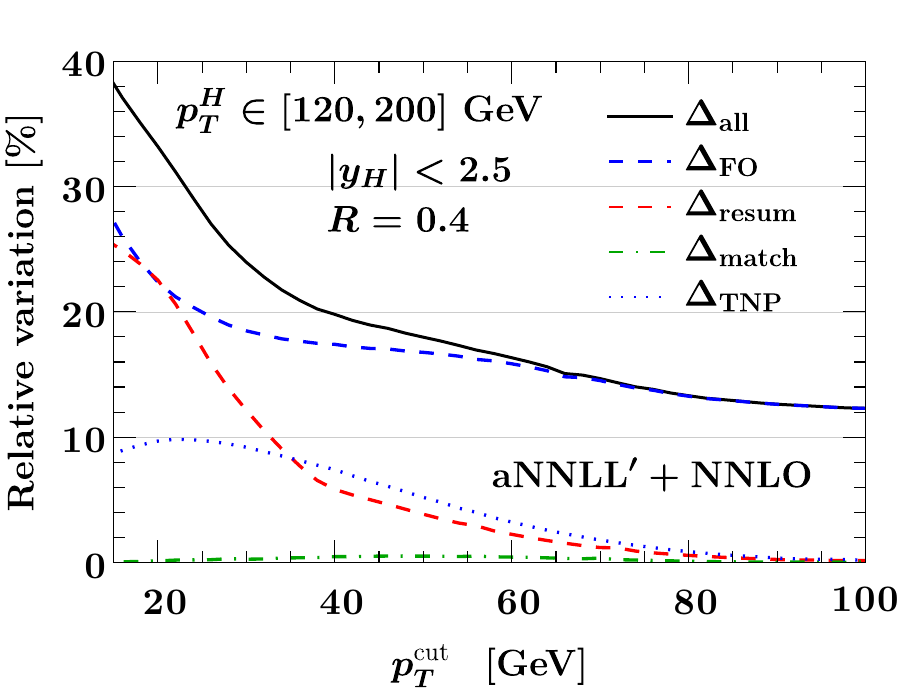}
     \end{subfigure}
    \hfill 
\caption{Breakdown of the total perturbative uncertainty as function of $\ptv$ for the low $p_T^H$ bin (top row) and the high $p_T^H$ bin (bottom row). For each bin, the uncertainties at NLL$'$+NLO (left) and aNNLL$'$+NNLO (right) are given. The relative uncertainties from the individual uncertainty
sources as discussed in \sec{TNP} are shown in blue dashed (fixed-order), red dashed (resummation), and green dashed (matching). At aNNLL$'$, the uncertainty due to TNPs is shown by the blue dotted curve. The total uncertainty (solid black) is obtained by adding these in quadrature.} \label{fig:impact_ptv}
\end{figure}

In \fig{impact_ptv} we show a detailed breakdown of the various contributions to our uncertainties at NLL$'$+NLO and aNNLL$'$+NNLO for the two STXS bins. As expected, for large values of $\ptv$, the uncertainty is simply the fixed-order uncertainty. For small values of $\ptv$ the uncertainty from resummation becomes larger and of similar size to the fixed-order uncertainty at NLL$'$+NLO. At aNNLL$'$+NNLO, the fixed-order uncertainty also grows for small values of $\ptv$ staying dominant for low $p_T^H$ and on par with the resummation uncertainty for the high $p_T^H$ STXS bin. The uncertainty from the matching procedure, governing the transition between the resummed and fixed-order calculation, is negligible compared to the other uncertainties. At aNNLL$'$+NNLO, there is also an uncertainty due to the TNPs. Its contribution is roughly on par with the resummation uncertainty for the small $p_T^H$ bin. For the high $p_T^H$ bin the TNP uncertainty tracks the resummation uncertainty for values of $\ptv>35$~GeV but then levels off at around $10\%$ for low $\ptv$ while the resummation uncertainty continues to grow.
\begin{figure} [t]
     \centering
     \begin{subfigure}{0.48\textwidth}
         \centering
         \includegraphics[width=\textwidth]{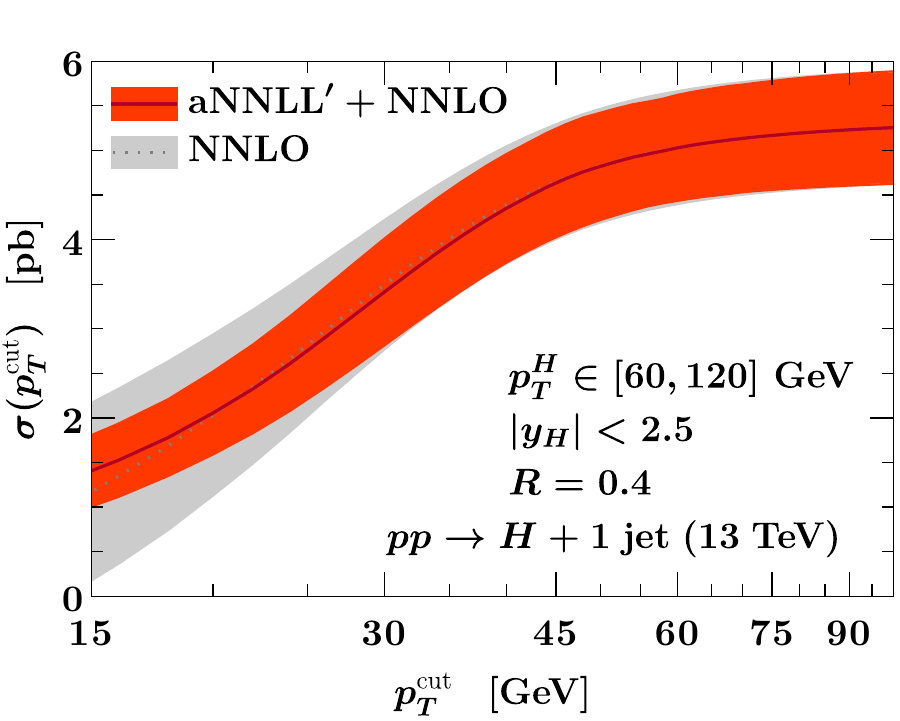}
     \end{subfigure}
     \hfill
     \begin{subfigure}{0.48\textwidth}
         \centering
         \includegraphics[width=\textwidth]{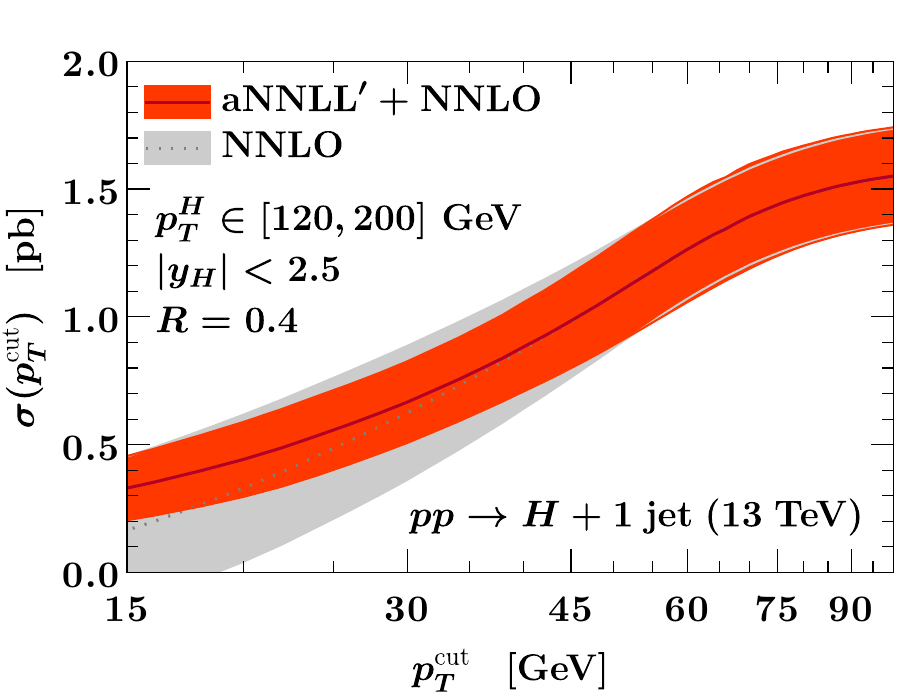}
     \end{subfigure}
    \hfill
\caption{Comparison of our aNNLL$'$+NNLO prediction (red) to NNLO (grey dotted) as function of $\ptv$, showing the reduction in uncertainty from the resummation. The uncertainty band on our prediction is obtained using the method described in \sec{TNP}, while the ST method is used for the uncertainty at fixed NNLO.} \label{fig:nnlo_v_resummed}
\end{figure}

We next compare our best result at aNNLL$'$+NNLO to the fixed NNLO prediction in \fig{nnlo_v_resummed}.
Here, the uncertainty at fixed NNLO is obtained using the ST method~\cite{Stewart:2011cf} (which is commonly used in this context to avoid artificially small scale variations due to accidental cancellation between the large K factor of the inclusive cross section and the large jet-veto logarithms). As expected, these predictions coincide for large values of $\ptv$, but the uncertainty of our resummed prediction becomes smaller than at fixed NNLO as $\ptv$ decreases. The improvement from resummation is more pronounced for the STXS bin in the right panel, since for fixed $\ptv$ the resummed logarithms increase with increasing $p_T^H$. The explicit numerical results for the two STXS bins at $\ptv=20$ GeV and $\ptv=30$ GeV are given in \tab{numeric_vals}. The uncertainties given correspond to the total uncertainty in \eq{total_unc} for the resummed results and the ST uncertainties for the fixed NNLO results. It should be noted that the individual sources of uncertainty between the two STXS bins should be considered correlated.
\begin{table}[t]
\centering
\begin{tabular}{r|cccc}
\hline\hline
\multicolumn{1}{l|}{}                   & \multicolumn{4}{c}{\textbf{Cross section} [pb]}                                                                                \\ \hline
\multicolumn{1}{c|}{STXS bin}           & \multicolumn{2}{c}{$p_T^H \in [60,120]$ GeV} & \multicolumn{2}{c}{$p_T^H \in [120,200]$ GeV} \\
\multicolumn{1}{c|}{$\ptv $ [GeV]} & $20$                      & $30$                      & $20$                           & $30$                          \\ \hline
NNLO                                     & $2.03\pm 44\%$            & $3.49\pm 21\%$            & $0.329\pm 87\%$                & $0.624\pm 42\%$               \\ \hline
NLL                                      & $0.56\pm 73\%$            & $1.07\pm 64\%$            & $0.112\pm 110\%$               & $0.198\pm 93\%$               \\
NLL$'$+NLO                               & $1.17\pm 38\%$            & $2.28\pm 33\%$            & $0.209\pm 41\%$                & $0.388 \pm 38\%$              \\
aNNLL$'$+NNLO                            & $2.05\pm 23\%$            & $3.41 \pm 18\%$           & $0.442\pm 33\%$                & $0.666\pm 24\%$               \\ \hline\hline
\end{tabular}
\caption{Sample numerical values for our resummed predictions. Both STXS bins use a cut $|y_H|<2.5$. The uncertainties correspond to the total uncertainty according to \eq{total_unc} for the resummed results
and to the ST uncertainties~\cite{Stewart:2011cf} for the NNLO results.}
\label{tab:numeric_vals}
\end{table}

\begin{figure} [t]
     \centering
     \begin{subfigure}{0.48\textwidth}
         \centering
         \includegraphics[width=\textwidth]{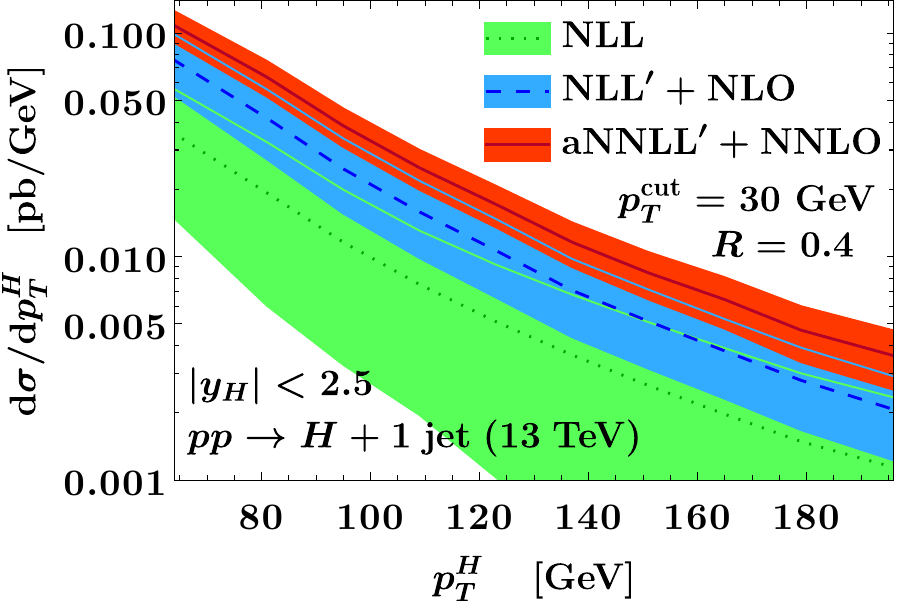}
     \end{subfigure}
     \hfill
     \begin{subfigure}{0.48\textwidth}
         \centering
         \includegraphics[width=\textwidth]{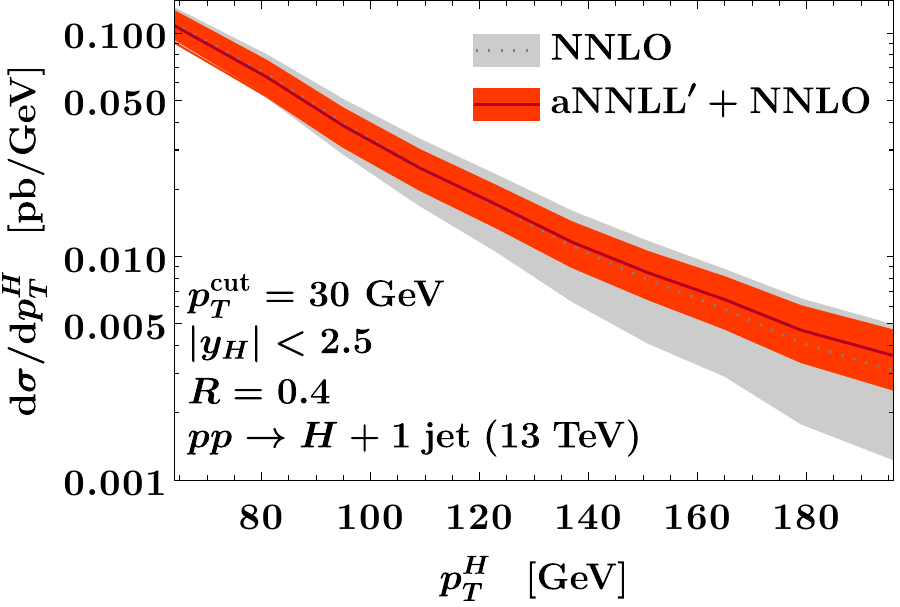}
     \end{subfigure}
    \hfill
\caption{Left panel: The convergence of resummed perturbation theory as function of $p_T^H$, comparing NLL (green dotted), NLL$'$+NLO (blue dashed), and aNNLL$'$+NNLO (red solid) for $\ptv = 30$ GeV, $|y_H| < 2.5$ and $R_J = 0.4$. The uncertainty bands are obtained using the method described in \sec{TNP}. Right panel: Comparison of our aNNLL$'$+NNLO prediction (red) to NNLO (grey dotted) as function of $p_T^H$ for $\ptv = 30$~GeV. The uncertainty on the NNLO is obtained using the ST method.} \label{fig:pth}
\end{figure}

\begin{figure} [t]
     \centering
     \begin{subfigure}{0.48\textwidth}
         \centering
         \includegraphics[width=\textwidth]{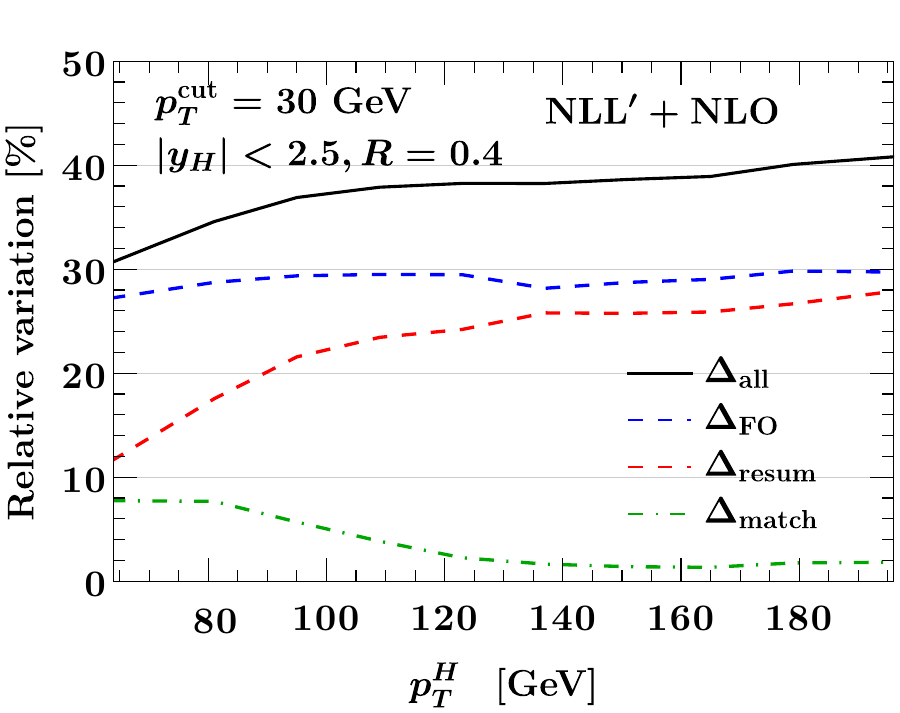}
     \end{subfigure}
     \hfill
     \begin{subfigure}{0.48\textwidth}
         \centering
         \includegraphics[width=\textwidth]{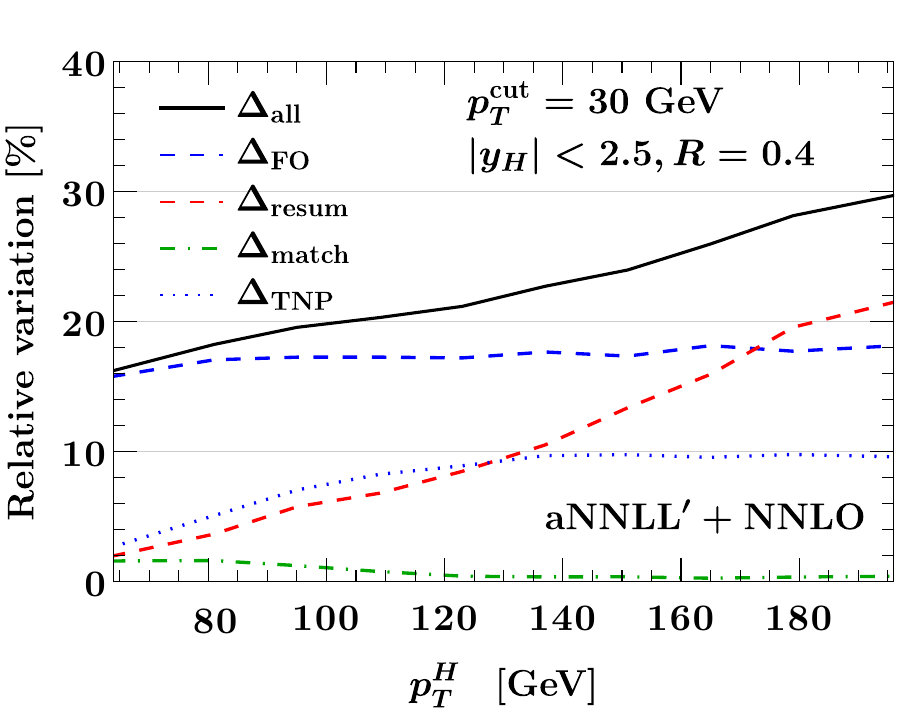}
     \end{subfigure}
    \hfill
\caption{Same as \fig{impact_ptv} but as function of $p_T^H$ for fixed $\ptv = 30$~GeV.} \label{fig:impact_pth}
\end{figure}

In the following set of plots we repeat the above results but explore the dependence on $p_T^H$ while keeping $\ptv = 30$ GeV fixed. In the left panel of \fig{pth}, we show the analogue of \fig{match_convergence}. We remind the reader that we work in the heavy-top approximation, and that for $p_T^H>200$ GeV finite top-quark mass effects become relevant~\cite{Jones:2018hbb,Czakon:2024ywb}. The overlapping bands and reduction of the bands at higher orders highlight the good convergence of resummed perturbation theory for the nominal value of the jet veto. In the right panel, we compare our aNNLL$'$+NNLO to the NNLO result. For small values of $p_T^H$, the hierarchy between $\ptv$ and $p_T^H$ reduces so we are effectively leaving the resummation region and approach the fixed-order region. Hence, the resummed and fixed-order results become equal. In contrast, for larger values of $p_T^H$ we are going deeper into the resummation region and the uncertainty on the aNNLL$'$+NNLO prediction reduces by a factor of $2$ or more compared to the NNLO prediction.
Finally, in \fig{impact_pth} we show the breakdown of the uncertainty as function of $p_T^H$ for fixed $\ptv = 30$ GeV analogous to \fig{impact_ptv}. As expected, as $p_T^H$ increases the resummation and TNP uncertainties increase while the matching uncertainty decreases.


\section{Conclusions}
\label{sec:conclusions}

In this work, we have provided state-of-the-art predictions for Higgs boson production in the exclusive 1-jet bin. Demanding exactly one hard jet requires the introduction of a veto scale, which in turn causes large logarithms of the ratio of that scale to the natural hard scale of the process to arise in the perturbative calculation. We have resummed these large logarithms to all orders in $\alpha_s$ using the framework of Soft-Collinear Effective Theory, extending a factorization formula first developed in ref.~\cite{Liu:2012sz}, and matched our results to a calculation at fixed order in perturbation theory. 

Our work has several novel features which provide important improvements over previous studies. First, we extend the formal accuracy to NNLL$'$ for the resummed component and push the fixed order part to NNLO using the numerical results of~\refscite{Chen:2016zka, Becker:2020rjp}. Though the lack of some of the requisite two-loop ingredients only allows us to reach an approximate NNLL$'$ accuracy in practice, we are able to quantify the additional perturbative uncertainty associated with these missing terms by variation of appropriately defined theory nuisance parameters. The total perturbative uncertainty is still significantly improved compared to NLL$'$. Second, we perform a refactorization of the soft sector which allows logarithms of the signal jet radius $\Rj$ to be resummed. This is achieved using a SCET$_+$ formalism, in which an additional soft-collinear mode is introduced. Third, we study the effect of power corrections in $\Rj$ on the soft function which multiply logarithms of $\ptv$. We have found that the inclusion of these power corrections is necessary to recover the correct singular behaviour in the small $\ptv$ limit, and that they in fact have a sizeable numerical effect in this region for phenomenologically relevant values of the jet radius $\Rj \sim 0.4$. Fourth, we include in an approximate form the effect of nonglobal logarithms. By exploiting a five-loop solution of the BMS equation which resums these terms at leading logarithmic order, we are able to include these subleading effects into our factorisation formula.

Our work can be extended in various ways. Most obviously, a full knowledge of the two-loop soft, soft-collinear and gluon jet functions would allow to further reduce the perturbative uncertainties pushing the formal accuracy to full NNLL$'$. Such a resummation (possibly also including the effects of nonglobal logarithms to higher orders) could then be used as the basis of an event generator matching NNLO calculations for $H+$jet to parton showers, in the same vein as \refcite{Gavardi:2023aco}. It would also be interesting to obtain resummed predictions for the STXS 1-jet bin at low $p_T^H < 60$ GeV. This scenario requires, however, a substantially different treatment, since the hierarchy of energy scales is altered. A suitable factorization formula for this setup still needs to be derived, likely necessitating the introduction of additional csoft modes in SCET$_+$. The work here can also be applied to the STXS 1-jet bins in VH production.
It would also be fairly straightforward to include the Higgs decay and cuts on the decay products
to obtain fully fiducial predictions for exclusive $H$+1-jet production.
Our approach could also be extended to cases with additional hard jets in the final state. This would not entail any changes to the collinear components of our calculation, but the hard function would need to be extracted from the (known) two-loop amplitudes and the relevant soft functions would need to be computed. In addition, this would introduce nontrivial colour structure to the resummation, which could be dealt with in a manner similar to that employed in e.g.\ \refscite{Ahrens:2010zv,Alioli:2021ggd,Gao:2023ivm}.

\acknowledgments
We are grateful to Piotr Pietrulewicz for early collaboration and to Simone Alioli for access to computing resources necessary for this work. MAL is indebted to Michael Penn, whose YouTube videos on logarithmic-trigonometric integrals were useful in the computation of the 1-loop soft function. This project has received
funding from the European Research Council (ERC) under the European
Union's Horizon 2020 research and innovation programme (Grant
agreement 101002090 COLORFREE). MAL is supported
by the UKRI guarantee scheme for the Marie Sk\l{}odowska-Curie
postdoctoral fellowship, grant ref. EP/X021416/1.

\appendix
\section{Clustering of collinear and soft-collinear emissions}
\label{app:cscclust}
A potential obstacle to the factorization in \eq{meas_full_fact_2} could be the clustering of jet-collinear and soft-collinear emissions. In this appendix, we demonstrate that this effect is in fact suppressed and contributes at the same order as subleading nonglobal logarithms, that is at $\mathcal{O}\bigl(\as^2 \ln (\ptj/\ptv)\bigr)$. 

We start by observing that we only have to concern ourselves with the effect of collinear emissions on the soft-collinear clustering, as soft-collinear emissions do not change the collinear clustering because they are much softer. 
If the jet consists of a single jet-collinear particle, the phase-space restriction of a soft-collinear emission is a fixed circle in $(\eta,\phi)$ with radius $R_J$. The effect of an additional jet-collinear emission is to shift the boundary, increasing the effective value of the radius $R_J \to k R_J$ (where $k \sim\mathcal{O}(1)$). The double logarithmic term appearing in the soft collinear function $L_R^2=\ln^2(\mu/(\ptv R_J))$ is then modified as
\begin{align}
L_R^2 \to L_R^2 - 2 L_R \ln k + \ln^2k\,,
\end{align}
inducing a change in the single logarithmic term. Since the effect is due to mixing between jet-collinear and soft-collinear sectors, we take $\mu=\ptj R_J$ and find that the contribution of this clustering effect on the cross section is $\propto \as^2\ln (\ptj/\ptv)$.

We still need to demonstrate that the integration over the phase space of the additional jet-collinear emission does not generate any further logarithms, otherwise the clustering could contribute at the same order as the leading nonglobal logarithms, which we do include. Concretely, we wish to show that, since the entire effect is given by a product of the one-loop jet and soft-collinear functions, the contribution from the jet function in the relevant phase space is a constant factor.

The  condition that both jet-collinear emissions are clustered into one jet leads to the following restriction
\begin{align}
\label{eq:cond1}
s \leq z(1-z) (p_T^J \Rj)^2
\end{align}
where $s$ is their invariant mass and $z$ the momentum fraction. The clustering effect described above only occurs when the distance between the most energetic of the two collinear particles and the jet boundary is smaller than the distance between the two collinear particles. The reason is that the most energetic particle will be clustered first, and the problematic effect only occurs if the first clustering is with a soft-collinear emission. For simplicity we may consider $z<1/2$ (the case $z>1/2$ is equivalent by considering $z\to 1-z$), in which case the condition for clustering to occur is given by
\begin{align}
\Rj - \sqrt{\frac{zs}{(1-z)(p_T^J)^2}} < \sqrt{\frac{s}{z(1-z)(p_T^J)^2}} 
\end{align}
with solution
\begin{align}
\label{eq:cond2}
s > \frac{z(1-z)}{(1+z)^2} (p_T^J \Rj)^2\,.
\end{align}
\Eqs{cond1}{cond2} and $z<1/2$ together determine the region of phase space $\mathcal{R}$ for which clustering occurs. Integrating over this region with a simplified integrand $1/(sz)$ is sufficient to capture the logarithmic structure. We find 
\begin{align}
\int_\mathcal{R} \frac{\mathrm{d}s}{s} \,\frac{\mathrm{d}z}{z} =- 2 \,\mathrm{Li}_2\left(-\frac{1}{2}\right)\,.
\end{align}
We observe that the integral is finite, and does not produce poles in $\epsilon$ (which when multiplying factors of $(p_T^J \Rj)^{\epsilon}$ would generate further logarithms). We thus conclude that clustering of jet-collinear and soft-collinear emissions contributes at most an $\mathcal{O}\left(\as^2 \ln (\ptj/\ptv)\right)$ effect.


\section{Calculation of the soft function at 1-loop}
\label{app:R2-corrections}

In this appendix, we revisit the calculation of the 1-loop soft function originally
described in ref.~\cite{Liu:2013hba}. We keep all logarithmic dependence on the jet radius
$\Rj$ explicit throughout, anticipating the refactorization into a global-soft and soft-collinear function.

We begin with the expression for the soft current integrated over phase space, given by
\begin{align}
S_{\kappa} =& -\frac{2g_s^2}{(2\pi)^{d-1}} \bigg(\frac{e^{\gamma_E}\mu^2}{4\pi} \bigg)^{\epsilon} \sum_{\substack{l<m\\ l,m\in\kappa}}\mathbf{T}_l\cdot \mathbf{T}_m  \,
\int \mathrm{d}^d k \, \delta(k^2) \frac{n_l\cdot n_m}{ (n_l\cdot k ) (n_m\cdot k)}\,\mathcal{M}_s,
\end{align}
where $l,m = \{a, b, j\}$ and we are using the parametrization 
\begin{align}
  k^\mu &= k_T(\cosh y,\dots, \cos\phi, \sin\phi, \sinh y)\,,\nn\\
  n^\mu_a &= (1,\dots,0,0,1)\,,\nn\\
  n^\mu_b &= (1,\dots,0,0,-1)\,,\nn\\
  n^\mu_j &= (\cosh y_J,\dots, 1, 0, \sinh y_J)
\,.\end{align}
The measurement function is given by
\begin{align}
  \label{eq:softmeasure}
  \mathcal{M}_{s,T}(k,\ptveto,\Rj)&=\Theta(\ptveto-k_T)+\Theta(k_T-\ptveto)\Theta(\Rj-\Delta R_{kJ})
\,,\end{align}
with $\Delta R_{kJ}=\sqrt{(y-y_J)^2 + \phi^2}$. The first term on the RHS is independent of the jet radius and contains all rapidity divergences, while the second
contains the complete $\Rj$ dependence.

We may  decompose the soft function according to colour structures, i.e.
\begin{align}
  \label{eq:softcolstruc}
  S_{\kappa}^{(1)}= \mathbf{T}_a\cdot\mathbf{T}_b\, S_{ab} + \mathbf{T}_a\cdot\mathbf{T}_j \,S_{aj} +\mathbf{T}_b\cdot\mathbf{T}_j \, S_{bj}
\,.\end{align}
We focus on the first term $S_{ab}$ in~\eq{softcolstruc}. Since this piece will suffer from rapidity divergences as $n_a \cdot k \to 0$ and $n_b \cdot k \to 0$, following ref.~\cite{Liu:2013hba} we introduce a regulator of the form
\begin{align}
|2k_{3}|^{-\eta}\nu^\eta \,
\xrightarrow{|y|\to \infty } k_T^{-\eta} \nu^{\eta} \exp(-\eta\, |y|)\,.
\end{align}
It is sufficient to retain only the asymptotic behaviour of the regulator, since divergences occur only in this limit and elsewhere we can safely set $\eta=0$. To leading order in $\Rj$, we may expand $\Theta(\Rj - \Delta {R_{kJ}}) = \mathcal{O}(\Rj^2)$, yielding 
\begin{align}
S_{ab} &=  -\frac{g_s^2 \Omega_{1-2\epsilon}}{2 (2\pi)^{d-1}} \bigg(\frac{e^{\gamma_E}\mu^2}{4\pi} \bigg)^{\epsilon}  \!\!
\int \mathrm{d} y \,\mathrm{d}\phi \,\mathrm{d}k_T^2 \, 
\, (k_T^2)^{-\epsilon-\eta/2 } \nu^\eta e^{ -\eta |y| } (\sin \phi)^{-2\epsilon}\
\frac{n_a\cdot n_b}{(n_a\cdot k) \,  (n_b\cdot k)}  \, \Theta (\ptveto- k_T) \nn\\
&= \frac{2 g_s^2  \, \Omega_{2-2\eps}}{(2\pi)^{3-2\eps}} \bigg(\frac{e^{\gamma_E}}{4\pi} \bigg)^{\eps}\frac{2}{(2\eps+\eta) \eta}\, \bigg(\frac{\nu}{\mu} \bigg)^ \eta \bigg(\frac{\ptveto}{\mu}\bigg)^{-2\eps-\eta} \nn \\
&=\frac{\alpha_s}{4\pi}\bigg\{\frac{1}{\eta}\bigg[\frac{8}{\epsilon}+16\ln\bigg(\frac{\mu}{\ptveto}\bigg)\bigg]-\frac{4}{\epsilon^2} + \frac{8}{\epsilon}\ln\bigg(\frac{\nu}{\mu}\bigg)+8\ln^2\bigg(\frac{\mu}{\ptveto}\bigg)\ln\bigg(\frac{\nu}{\mu}\bigg)\nn\\
& \qquad +16\ln\bigg(\frac{\mu}{\ptveto}\bigg)+\frac{\pi^2}{3}\bigg\}
\end{align}

We now turn to the computation of the $S_{aj}$ contribution. The rapidity divergence at $n_a \cdot k \to 0$ is now regulated by
\begin{align}
|2k_{3}|^{-\eta}\nu^\eta = 2^{-\eta}k_T^{-\eta}\nu^\eta|\sinh y|^{-\eta}
\xrightarrow{y\to \infty } k_T^{-\eta} \nu^{\eta} \exp(-\eta\, y\,\Theta(y) )\,.
\end{align}
We find
\begin{align}
S_{aj} &=  -\frac{g_s^2 \Omega_{1-2\epsilon}}{2 (2\pi)^{d-1}} \bigg(\frac{e^{\gamma_E}\mu^2}{4\pi} \bigg)^{\eps}  \,
\int \mathrm{d} y \,
\mathrm{d}\phi \,
\mathrm{d}k_T^2 \, 
\, (k_T^2)^{-\epsilon-\eta/2 } \nu^\eta e^{ -\eta\,  y\,  \Theta(y) } (\sin \phi)^{-2\eps}\
 \frac{n_a\cdot n_J}{(n_a\cdot k) \,  (n_J\cdot k)} \nn \\
 &\qquad \times  \Big[ \Theta (\ptveto- k_T)
+ \Theta(\Rj - \Delta {R_{kJ}}) \Theta( k_T - \ptveto)  \, \Big]\nn \\
&=  -\frac{g_s^2 \Omega_{1-2\epsilon} }{2 (2\pi)^{3-2\eps}} \bigg(\frac{e^{\gamma_E}\mu^2}{4\pi} \bigg)^{\eps}  \,
\int \mathrm{d} \dely \,
\mathrm{d}\phi \,
\mathrm{d}k_T^2 \, 
\, (k_T^2)^{-\epsilon-\eta/2 } \nu^\eta e^{ -\eta\, ( \dely + y_J)\,  \Theta(\dely + y_J) } (\sin \phi)^{-2\eps}\ \nn \\
 &\qquad \times e^{\dely} \frac{1}{k_T^2 (\cosh \dely-\cos \phi)}\Big[ \Theta (\ptveto- k_T)
  + \Theta(\Rj - \Delta {R_{kJ}}) \Theta( k_T - \ptveto)  \, \Big] \, \nn \\
&=\frac{g_s^2 \Omega_{1-2\epsilon}}{(2\pi)^{3-2\eps}} \bigg(\frac{e^{\gamma_E}}{4\pi} \bigg)^{\eps}  \frac{1}{2\eps+\eta} \bigg(\frac{\nu e^{-y_J}}{\mu} \bigg)^\eta  \bigg(\frac{\ptveto}{\mu}\bigg)^{-2\eps-\eta}  \, \nn\\ 
&\qquad \times \int \mathrm{d} \dely \, \mathrm{d}\phi \,
\,   e^{ -\eta\,  \dely \,  \Theta(\dely) } (\sin \phi)^{-2\eps}  \frac{e^{\dely} }{\cosh \dely-\cos \phi} \Theta(\Delta R_{kJ} - \Rj),
\label{eq:sajprelim}
\end{align}
where $\dely \equiv y - y_J$. In the above expression, the jet radius $\Rj$ serves to regulate a divergence associated with collinear emissions along the jet direction. Following ref.~\cite{Liu:2013hba}, we introduce the subtraction term
\begin{align}
{\cal I}_R &\equiv \,
(\sin \phi)^{-2\epsilon}\, 
 \frac{2\,e^{-\eta \dely\,\Theta(\dely)}}{\dely^2+\phi^2}\,,
\end{align}
which is obtained by taking the $\Rj\to 0$ limit of the integrand of \eq{sajprelim}. The $\ln \Rj$ dependence of this subtraction term can be computed exactly, while the difference of the subtraction term and the original integrand can be expanded in the limit $\Rj\to 0$. We thus write
\begin{align}
S_{aj}  &=  \frac{g_s^2\Omega_{1-2\epsilon} }{(2\pi)^{3-2\eps}} \bigg(\frac{e^{\gamma_E}}{4\pi} \bigg)^{\eps}  \frac{1}{2\eps+\eta} \bigg(\frac{\nu e^{-y_J}}{\mu} \bigg)^\eta  \bigg(\frac{\ptveto}{\mu}\bigg)^{-2\eps-\eta}  \, \nn\\ 
& \qquad \times  \bigg\{
 \int \mathrm{d} \dely \, \mathrm{d}\phi \,   (\sin \phi)^{-2\eps}  e^{ -\eta\,  \dely \,  \Theta(\dely) }   \bigg[ \frac{e^{\dely} }{ \cosh \dely-\cos \phi} -  \frac{2 }{ \Delta R_{kJ}^2} \bigg] \Theta(\Delta R_{kJ} - \Rj)\nn \\
& \qquad \quad + \int \mathrm{d} \dely \, \mathrm{d}\phi\,  (\sin \phi)^{-2\eps}   \frac{2}{\Delta R_{kJ}^2}  \Theta(\Delta R_{kJ} - \Rj)
\bigg\}
\label{eq:decomposition}
\end{align}

By replacing $\Theta(\Delta R_{kJ} - \Rj) = 1-\Theta(\Rj-\Delta R_{kJ})$ in the last line of \eq{decomposition}, we may identify the integrals
\begin{align}
S_{aj}^{\text{ln}\Rj} &\equiv- \frac{g_s^2 \Omega_{1-2\epsilon}}{(2\pi)^{3-2\eps}} \bigg(\frac{e^{\gamma_E}}{4\pi} \bigg)^{\eps} \,
\frac{1}{2\epsilon} \,
\left(\frac{\ptveto}{\mu}\right)^{-2\epsilon} \,
\int_{-\infty}^\infty \mathrm{d} \dely \,
\int_0^{\pi} \df\phi \, 
{\sin \phi}^{-2\epsilon}\, 
 \frac{2}{\Delta {R_{kJ}^2}}\,
 \Theta(\Rj - \Delta R_{kJ} )
\label{eq:sajlnR}
\end{align}
\begin{align}
S_{aj}^{\text{div}} &\equiv \frac{g_s^2 \Omega_{1-2\epsilon}}{(2\pi)^{3-2\eps}} \bigg(\frac{e^{\gamma_E}}{4\pi} \bigg)^{\eps} \,
\frac{1}{2\epsilon} \,
\left(\frac{\ptveto}{\mu}\right)^{-2\epsilon} \,
\int_{-\infty}^\infty \mathrm{d} \dely \,
\int_0^{\pi} \df\phi \, 
{\sin \phi}^{-2\epsilon}\, 
\frac{2}{\Delta {R_{kJ}^2}} \,.
\label{eq:sajdiv}
\end{align}
We now take the small $\Rj$ limit: writing $\dely=\rho\cos\chi$ and $\phi=\rho\sin\chi$, we keep the leading term in the $\rho \to 0 $ expansion. This amounts to
making the replacement $(\sin \phi)^{-2\epsilon}\to \phi^{-2\epsilon}$ in \eqs{sajlnR}{sajdiv}. We thus have 
\begin{align}
S_{aj,\Rj\to0}^{\text{ln}\Rj} &\equiv- \frac{g_s^2 \Omega_{1-2\epsilon}}{(2\pi)^{3-2\eps}} \bigg(\frac{e^{\gamma_E}}{4\pi} \bigg)^{\eps} \,
\frac{1}{2\epsilon} \,
\left(\frac{\ptveto}{\mu}\right)^{-2\epsilon} \,
\int_{-\infty}^\infty \mathrm{d} \dely \,
\int_0^{\pi} \df\phi \, 
{\phi}^{-2\epsilon}\, 
 \frac{2}{\Delta {R_{kJ}^2}}\,
 \Theta(\Rj - \Delta R_{kJ} )
\label{eq:sajlnRsmallR}
\end{align}
\begin{align}
S_{aj,\Rj\to0}^{\text{div}} &\equiv \frac{g_s^2 \Omega_{1-2\epsilon}}{(2\pi)^{3-2\eps}} \bigg(\frac{e^{\gamma_E}}{4\pi} \bigg)^{\eps} \,
\frac{1}{2\epsilon} \,
\left(\frac{\ptveto}{\mu}\right)^{-2\epsilon} \,
\int_{-\infty}^\infty \mathrm{d} \dely \,
\int_0^{\pi} \df\phi \, 
{\phi}^{-2\epsilon}\, 
\frac{2}{\Delta {R_{kJ}^2}}.
\label{eq:sajdivsmallR}
\end{align}

In the first line of \eq{decomposition}, we may repeat the same tactic as before and introduce a further subtraction term to isolate
the rapidity divergences of the integral for $\dely>0$. We define
\begin{align}
S^\eta_{aj} &=   \frac{g_s^2 \Omega_{1-2\epsilon} }{(2\pi)^{3-2\eps}} \bigg(\frac{e^{\gamma_E}}{4\pi} \bigg)^{\eps}  \frac{1}{2\eps+\eta} \bigg(\frac{\nu e^{-y_J}}{\mu} \bigg)^\eta  \bigg(\frac{\ptveto}{\mu}\bigg)^{-2\eps-\eta} \,
\int_{-\infty}^\infty \mathrm{d} \Delta y \,
\int_0^{\pi} \df\phi \, \,
2 (\sin {\phi})^{-2\epsilon}   e^{ -\eta\,  \dely } \nn \\
&\qquad \times  \Theta(\dely)\,\Theta(\Delta R_{kJ} - \Rj)
\label{eq:sajeta}
\end{align}
\begin{align}
S_{aj}^{\eta=0} &= \frac{g_s^2 \Omega_{1-2\epsilon}}{(2\pi)^{3-2\eps}}  \bigg(\frac{e^{\gamma_E}}{4\pi} \bigg)^{\eps} \,
\frac{1}{2\epsilon} \,
\left(\frac{\ptveto}{\mu}\right)^{-2\epsilon }\, 
\int_{-\infty}^\infty \mathrm{d} \dely \,
\int_0^\pi \df \phi \,(\sin \phi)^{-2\epsilon}  \Theta(\Delta R_{kJ} - \Rj) \nn \\
&\qquad \times\bigg\{ e^{ -\eta\,  \dely }   \bigg[ \frac{e^{\dely} }{ \cosh \dely-\cos \phi} -  \frac{2 }{ \Delta R_{kJ}^2} \bigg] - 2\Theta(\dely) \bigg\}\,,
\label{eq:sajeta0}
\end{align}
so that the whole expression in \eq{decomposition} can be written as
\begin{align}
S_{aj} &= S_{aj}^{\text{ln}\Rj}+S_{aj}^{\text{div}}+S_{aj}^{\eta}+S_{aj}^{\eta=0}\,.
\label{eq:decomposition2}
\end{align}
Since \eq{sajeta0} is free of rapidity divergences, we may set the rapidity regulator $\eta=0$ here. Again setting $\Theta(\Delta R_{kJ} - \Rj) = 1-\Theta(\Rj-\Delta R_{kJ})$ in \eqs{sajeta}{sajeta0}, the small $\Rj$ limit allows us to drop the second term in each case. We also recall that the second term in the square brackets of \eq{sajeta0} originates from \eq{sajlnR}, which necessitates the replacement $(\sin \phi)^{-2\epsilon}\to \phi^{-2\epsilon}$. We obtain
\begin{align}
S^\eta_{aj,\Rj\to0} &=   \frac{g_s^2 \Omega_{1-2\epsilon} }{(2\pi)^{3-2\eps}} \bigg(\frac{e^{\gamma_E}}{4\pi} \bigg)^{\eps}  \frac{1}{2\eps+\eta} \bigg(\frac{\nu e^{-y_J}}{\mu} \bigg)^\eta  \bigg(\frac{\ptveto}{\mu}\bigg)^{-2\eps-\eta} \,
\int_{-\infty}^\infty \mathrm{d} \Delta y \,
\int_0^{\pi} \df\phi \nn \\
&\qquad \times \,
2 (\sin {\phi})^{-2\epsilon}   e^{ -\eta\,  \dely }   \Theta(\dely)
\label{eq:sajetasmallR}
\end{align}
\begin{align}
S_{aj,\Rj\to0}^{\eta=0} &= \frac{g_s^2 \Omega_{1-2\epsilon}}{(2\pi)^{3-2\eps}}  \bigg(\frac{e^{\gamma_E}}{4\pi} \bigg)^{\eps} \,
\frac{1}{2\epsilon} \,
\left(\frac{\ptveto}{\mu}\right)^{-2\epsilon }\, 
\int_{-\infty}^\infty \mathrm{d} \dely \,
\int_0^\pi \df \phi \,   \nn \\
&\qquad \times\bigg\{ \frac{e^{\dely} (\sin \phi)^{-2\epsilon} }{ \cosh \dely-\cos \phi} -  \frac{2\phi^{-2\epsilon} }{ \Delta R_{kJ}^2}  - 2(\sin \phi)^{-2\epsilon}\Theta(\dely) \bigg\}\,.
\end{align}

Evaluating all integrals in the small $\Rj$ limit, we find
\begin{align}
S_{aj,\Rj\to0}^{\text{ln} \Rj} 
 &=  \frac{\alpha_s}{4\pi}\left[\frac{2}{\epsilon^2}+\frac{4}{\epsilon}\ln\left(\frac{\mu}{\ptveto \Rj}\right) + 4\ln^2\left(\frac{\mu}{\ptveto \Rj}\right)-\frac{\pi^2}{6}\right],
\end{align}
\begin{align}
 S_{aj,\Rj\to0}^{\text{div}}
 &=\frac{\alpha_s}{4\pi}\left[-\frac{2}{\epsilon^2} - \frac{4}{\epsilon}\ln\left(\frac{\mu}{2\pi \ptveto}\right) -4 \ln^2\left(\frac{\mu}{2\pi \ptveto }\right)+\frac{\pi^2}{2}\right], 
\end{align}
\begin{align}
  S^\eta_{aj,\Rj\to0} &=\frac{\alpha_s}{4\pi}\bigg\{\frac{1}{\eta}\left[\frac{4}{\epsilon}+8\ln\left(\frac{\mu}{\ptveto}\right)\right] - \frac{2}{\epsilon^2} + \frac{4}{\epsilon}\ln\left(\frac{\nu e^{-y_J}}{\mu}\right) + 4 \ln^2 \bigg(\frac{\mu}{\ptveto}\bigg)\nn\\
  &\qquad + 8 \ln\bigg(\frac{\mu}{\ptveto}\bigg)\ln \bigg(\frac{e^{-y_J} \nu}{\mu} \bigg) + \frac{\pi^2}{6} \bigg\},
\end{align}
\begin{align}
  S_{aj,\Rj\to0}^{\eta=0} &= \frac{\alpha_s}{4\pi} \bigg\{ \left[-\frac{4}{\epsilon} - 8\ln\left(\frac{\mu}{\ptveto}\right) + 8 \ln (2)\right] \ln (2 \pi) - 8\ln \bigg(\frac{\mu}{\ptveto}\bigg)\ln(2\pi) \nn \\
  &\qquad+4\ln^2(\pi) -\frac{\pi^2}{3} -4\ln^2(2)   \bigg\}\,,
\end{align}
thus completing the evaluation of $S_{aj}$ in the small $\Rj$ limit. The $S_{bj}$ contribution follows in exactly the same way and is related to the $S_{aj}$ contribution via
\begin{align}
  S_{bj}=S_{aj}\vert_{y_J\to-y_J}\,.
\end{align}
Inserting the expressions for the integrals, we therefore find that the soft function up to one-loop order is given by
\begin{align}
  S_{\kappa}^{(1)} &= (\mathbf{T}_a^2+\mathbf{T}_b^2)\left[-4\ln^2\left(\frac{\mu}{\ptveto}\right)-8\ln\left(\frac{\mu}{\ptveto}\right)\ln\left(\frac{\nu}{\mu}\right)-\frac{\pi^2}{6}\right] \nn \\
 & \qquad + 8 y_J(\mathbf{T}_a^2-\mathbf{T}_b^2)\ln\left(\frac{\mu}{\ptveto}\right)+\mathbf{T}_j^2\left[-4\ln^2\Rj+8\ln\left(\frac{\mu}{\ptveto}\right)\ln \Rj\right].
\end{align}

Including higher order terms from the expansion in $\rho$ of each of the integrals on the RHS of \eq{decomposition2}, we find that the soft function including power corrections in $\Rj$ is given by
\begin{align}
  S_{\kappa}^{(1)} &= (\mathbf{T}_a^2+\mathbf{T}_b^2)\left[-4\ln^2\left(\frac{\mu}{\ptveto}\right)-8\ln\left(\frac{\mu}{\ptveto}\right)\ln\left(\frac{\nu}{\mu}\right)-\frac{\pi^2}{6} + 2\Rj^2\ln\left(\frac{\mu}{\ptveto \Rj}\right) + \Rj^2\right] \nn \\
  & \qquad + 8 y_J(\mathbf{T}_a^2-\mathbf{T}_b^2)\ln\left(\frac{\mu}{\ptveto}\right)\nn\\
  & \qquad +\mathbf{T}_j^2\left[-4\ln^2\Rj+8\ln\left(\frac{\mu}{\ptveto}\right)\ln \Rj-\Rj^2\ln\left(\frac{\mu}{\ptveto \Rj}\right) + \frac{\Rj^2}{6}\right].
\end{align}

\section{Resummation ingredients}
\label{app:resummation}

The following functions enter in the evolution kernels in \sec{RGE}:
\begin{align}
K^i_\Gamma(\mu_0, \mu)
&=\int_{\as(\mu_0)}^{\as(\mu)} \f{\df \as}{\beta(\as)} \Gamma^i_\text{cusp}(\as)
 \int_{\as(\mu_0)}^{\as} \f{\df \as'}{\beta(\as')}
\,,\nn \\
\eta^i_\Gamma(\mu_0, \mu)
&=\int_{\as(\mu_0)}^{\as(\mu)} \f{\df \as}{\beta(\as)} \Gamma^i_\text{cusp}(\as)
\,,\nn \\
K_\gamma^i(\mu_0, \mu)
&=\int_{\as(\mu_0)}^{\as(\mu)} \f{\df \as}{\beta(\as)} \gamma^i(\as)
\,.\end{align}
Up to NNLL accuracy, their expressions are given by
\begin{align}
K^i_\Gamma(\mu_0, \mu) &= -\f{\Gamma^i_0}{4 \beta_0^2}\bigg\{\f{4\pi}{\as (\mu_0)} \bigg(1-\f{1}{r}-\ln r \bigg)
+\bigg(\f{\Gamma_1}{\Gamma_0}- \f{\beta_1}{\beta_0} \bigg)(1-r +\ln r ) \nn \\ & 
+ \f{\beta_1}{2 \beta_0} \ln^2 r  +\f{\as (\mu_0)}{4\pi} \bigg[\bigg( \f{\beta_1^2}{\beta_0^2} -\f{\beta_2}{\beta_0} \bigg)\bigg( \f{1-r^2}{2}+\ln r \bigg)  \nn \\ & + \bigg(\f{\beta_1 \Gamma_1 }{\beta_0 \Gamma_0}-\f{\beta_1^2}{\beta_0^2} \bigg)(1-r+r \ln r ) - \bigg(\f{\Gamma_2}{\Gamma_0} - \f{\beta_1 \Gamma_1 }{\beta_0 \Gamma_0}\bigg)\f{(1-r)^2}{2} \bigg]  \bigg\}
\,,\nn \\
\eta^i_\Gamma(\mu_0, \mu) &= -\f{\Gamma^i_0}{2\beta_0}\bigg[\ln r + \f{\as (\mu_0)}{4\pi} \bigg(\f{\Gamma_1}{\Gamma_0}-\f{\beta_1}{\beta_0} \bigg)(r-1)\nn \\ &
+\f{\as^2 (\mu_0)}{(4\pi)^2} \bigg(\f{\Gamma_2}{\Gamma_0}-\f{\beta_1 \Gamma_1}{\beta_0 \Gamma_0} +\f{\beta_1^2}{\beta_0^2}-\f{\beta_2}{\beta_0} \bigg) \f{r^2-1}{2}  \bigg]
\,,\nn \\
K^i_\gamma(\mu_0, \mu) &= -\f{\gamma^i_0}{2\beta_0}\bigg[\ln r + \f{\as (\mu_0)}{4\pi} \bigg(\f{\gamma^i_1}{\gamma^i_0}-\f{\beta_1}{\beta_0} \bigg)(r-1) \bigg],
\end{align}
 where we defined $r= \as(\mu)/ \as(\mu_0)$. 
The coefficients in the expansion of the cusp anomalous dimension in \eq{Gamma_exp} and $\beta$-function in \eq{beta_exp} that are needed up to NNLL$'$ are given by
\begin{align} \label{eq:beta_coef}
\beta_0
&= \frac{11}{3}\,C_A - \frac{4}{3}\,T_F\,n_f
\,, \nn \\
\beta_1
&= \frac{34}{3}\,C_A^2 - 2T_F\,n_f \Bigl(\frac{10}{3}\, C_A + 2 C_F\Bigr)
\,, \\
\beta_2
&= \frac{2857}{54}\,C_A^3
 + 2T_F\,n_f \Bigl(- \frac{1415}{54}\,C_A^2 - \frac{205}{18}\,C_F C_A + C_F^2  \Bigr)
 + 4T_F^2\,n_f^2 \Bigl(\frac{79}{54}\, C_A  + \frac{11}{9}\, C_F \Bigr)
\,,\nn \\
 \label{eq:Gamma_coef}
\Gamma_0^i &= 4 C_i
\,,\nn\\
\Gamma_1^i
&= 4 C_i \biggl[
   C_A \Bigl(\frac{67}{9} - 2\zeta_2 \Bigr)
   - \frac{20}{9}\, T_F\, n_f
\biggr]
\,,\nn\\
\Gamma_2^i
&= 4 C_i \biggl\{
   C_A^2 \Bigl(\frac{245}{6} - \frac{268}{9}\zeta_2 + \frac{22}{3}\zeta_3 + 22\zeta_4 \Bigr)
\nn \\ & \qquad\quad
   + 2T_F\,n_f \biggl[
      C_A \Bigl(-\frac{209}{27} + \frac{40}{9}\zeta_2 - \frac{28}{3}\zeta_3 \Bigr)
      + C_F \Bigl(-\frac{55}{6} + 8\zeta_3 \Bigr)
   \biggr]
   - \frac{16}{27}\, T_F^2\, n_f^2
\biggr\}
\,.\end{align}
We sometimes use the colour-stripped coefficients $\Gamma_n=\Gamma_n^i/C_i$ for $n\leq 2$.


\addcontentsline{toc}{section}{References}
\bibliographystyle{jhep}
\bibliography{refs}

\end{document}